\documentclass[rmp,aps,reprint,groupedaddress,eqsecnum,longbibliography]{revtex4-1}
\usepackage{bm,graphicx}
\usepackage{epstopdf}
\usepackage{enumerate}
\usepackage{amsmath}

\begin{document}


\title{An introduction to the Ginzburg-Landau theory of phase transitions and nonequilibrium patterns}


\author{P. C. Hohenberg}
\affiliation{Department of Physics, New York University, New York, NY 10012 USA}
\author{A. P. Krekhov}
\affiliation{Max Planck Institute for Dynamics and Self-Organization, 37077 G\"ottingen, Germany}


\date{\today}

\begin{abstract}
This paper presents an introduction to phase transitions and critical phenomena on the one hand, and nonequilibrium patterns on the other, using the Ginzburg-Landau theory as a unified language.
In the first part, mean-field theory is presented, for both statics and dynamics, and its validity tested self-consistently.
As is well known, the mean-field approximation breaks down below four spatial dimensions, where it can be replaced by a scaling phenomenology.
The Ginzburg-Landau formalism can then be used to justify the phenomenological theory using the renormalization group, which elucidates the physical and mathematical mechanism for universality.
In the second part of the paper it is shown how near pattern forming linear instabilities of dynamical systems, a formally similar Ginzburg-Landau theory can be derived for nonequilibrium macroscopic phenomena.
The real and complex Ginzburg-Landau equations thus obtained yield nontrivial solutions of the original dynamical system, valid near the linear instability.
Examples of such solutions are plane waves, defects such as dislocations or spirals, and states of temporal or spatiotemporal (extensive) chaos.
\end{abstract}


\maketitle

\tableofcontents

%
%

\section{\label{sec:1} Introduction: systems, models, phenomena}
This paper describes two classes of physical phenomena, continuous phase transitions and nonequilibrium patterns, using a unified theoretical approach, the so-called Ginzburg-Landau theory.
We will show that a rich variety of observable phenomena can be usefully unified and understood using this approach, which emphasizes important physical principles and seeks to avoid excessive technical complications.
%

\subsection{\label{sec:1A} Phase transitions and critical phenomena in bulk thermodynamic systems}
\begin{figure}[hbt]
\includegraphics[width=7.5cm]{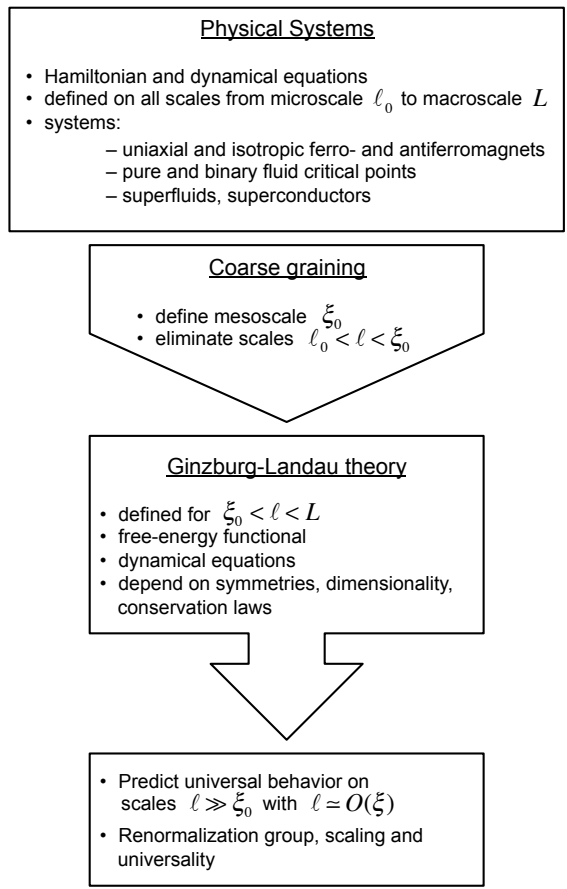}
\caption{\label{fig:schema1}
Schematic structure of the Ginzburg-Landau theory of continuous phase transitions.}
\end{figure}

We begin by considering thermodynamic systems undergoing continuous phase transition from a `symmetric' state to a more `ordered' state.
Examples are fluids or fluid mixtures at their critical point, uniaxial and isotropic ferro- and antiferromagnets, superfluids and superconductors.
The systems are defined on the microscale $\ell_0$ (which is generally an atomic dimension) by their Hamiltonian and classical or quantum dynamics.
These quantities control the behavior from the microscale $\ell_0$ all the way to the macroscale $L$, which we think of as being the scale of experiments (typically from millimeters to meters), but which will also be considered to go to infinity in the so-called `thermodynamic limit'.
The systems we are considering all undergo a continuous phase transitions at a temperature $T =T_c$, from a high-temperature symmetric phase to a low-temperature ordered phase in which some symmetry is broken.
The notion of equilibrium phases of matter is fundamental to thermodynamics and statistical mechanics.
Each phase can be characterized by its symmetries and conserved variables, from which specific hydrodynamic modes follow at long wavelengths and long times.
For example, a fluid supports sound waves whose velocity is exactly related to the compressibility, an equilibrium thermodynamic quantity.
In the solid crystalline phase the system displays additional transverse sound modes, reflecting the broken translational symmetry, in addition to the (longitudinal) compression mode already present in the fluid.
All of these modes exist generally for classical or quantum systems, quite independent of the specific atomic or molecular details of the constituents.
This generality motivates a theoretical description in terms of coarse-grained variables, i.e. local averages in which the short-scale properties have been eliminated in favor of densities varying slowly in space and time.
As explained below, the most powerful theoretical description of thermodynamic phases is in terms of a coarsening operation, the Wilson renormalization group, in which short-scale fluctuations are progressively eliminated.
This is most easily visualized in an abstract space whose elements are different system Hamiltonians.
The coarsening operation is then represented by a trajectory in this space, whose endpoint or fixed point describes the system properties at the longest scales and thus serves to characterize the thermodynamic phase.
We show below that this general renormalization group framework introduced by K.~G.~Wilson in 1968-72 and elaborated by others, not only serves to illuminate the physics of thermodynamic phases but it also leads to powerful theoretical methods for understanding critical phenomena at continuous phase transitions quite generally.
The renormalization group fixed points represent different phases of matter at low and high temperatures, respectively, as well as distinct universality classes of critical behavior at the transition point between the two phases: different physical systems flowing to the same fixed point belong to the same universality class.
To be more specific, let us return to a consideration of a system undergoing a continuous phase transition from a high-temperature symmetric phase to a low-temperature ordered phase in which some symmetry is broken.
Prior to the 1960s the most general and accurate description of such transitions was the Landau mean-field theory, based on defining a local order parameter $\psi$ whose average value controlled the thermodynamic phase.
The theory was in qualitative agreement with experiment, especially in the prediction of long-range spatial correlations of the order parameter over a length $\xi$ which diverges at the phase transition. As explained below, at this point the system displays
a separation of scales in which the microscopic details can be averaged over (to define $\psi$) and the long-range properties are associated with a fixed-point of the renormalization group trajectory.
The quantitative features of the high- and low-temperature phases and of the mean-field phase transitions, as reflected in the properties of the respective fixed points, could be largely determined by arguments based on dimensional analysis, symmetry and analyticity in appropriate variables.
By the 1960s, however, it was understood that while mean-field theory worked well for the high- and low-temperature fixed points, it was quantitatively inaccurate at the phase transition, and many improvements and corrections were devised, as discussed below.
It is the singular achievement of K. G. Wilson to have linked these departures from mean-field theory to the behavior of the renormalization group trajectories near the critical fixed point, and to have devised theoretical methods for arriving at systematic quantitative results, later elaborated by many workers.
Specifically, in contrast to the mean-field fixed points which can be fully characterized in terms of the local order parameter that embodies the dominant short-range fluctuations, Wilson argued that at the critical fixed point fluctuations on all scales, from microscopic to order $\xi$, make non-negligible contributions to the renormalization group trajectories and these must be accounted for to determine the quantitative critical behavior.
The first part of the present paper provides an introductory treatment of continuous phase transitions using the so-called Ginzburg-Landau theory as a convenient general language to describe both the mean-field theory and the renormalization group framework.
As mentioned above, we begin with a microscopic Hamiltonian and note that according to statistical mechanics, thermodynamic quantities and correlation functions are all derivable from a free energy which can be expressed in terms of the microscopic Hamiltonian as a sequence of integrals over all scales from the microscale $\ell_0$ to the macroscale $L$ (and out to infinity) [see Eq.~(\ref{eq:Z})].
We now introduce the mesoscale $\xi_0 =k_0^{-1}$, which is intermediate between the microscale $\ell_0$ and the macroscale $L$, $\ell_0 \ll \xi_0 \ll L$, and note that the correlation length $\xi$ extends from $\xi_0$ to $L$ ($\xi_0 < \xi < L$), and it diverges at the transition.
Since near the transition the properties of interest involve fluctuations on the varying scale $\xi$, a fundamental assumption of the Ginzburg-Landau approach is that the scales extending from the microscale to the mesoscale ($\ell_0 < \ell < \xi_0$) are unimportant, and may be averaged over [see Eq.~(\ref{eq:exp_Phi})].
One is then left with a model derived in a precise way from the microscopic Hamiltonian, but involving only scales extending from the mesoscale $\xi_0$ to the macroscale $L$.
This is so-called Ginzburg-Landau free energy function $\Phi[\psi]$, which is a general functional of the coarse-grained order parameter $\psi(x,t)$ [see Eqs.~(\ref{eq:Phi_psi}) and (\ref{eq:phi_psi})].
This functional can be a complicated nonlinear and nonlocal functional of the field $\psi(x,t)$, but it no longer involves the microscopic details of the system under study.
It only reflects general features of the system such as the dimensionality of space and the symmetry of the ordered state, i.e., the number $n$ of relevant components of the order parameter $\psi(x,t)$.
In this way, even before attempting to analyze the precise behavior of the thermodynamics and correlation functions near the transition, we have achieved a considerable level of {\em universality}: different physical systems, with different Hamiltonians, will lead to the same Ginzburg-Landau free energy functional, provided they have the same spatial dimension and order parameter symmetry.
In this representation the microscopic details of the original system are summarized by the values of the parameters in the Ginzburg-Landau free energy, e.g., the values of $T_c$, $\xi_0$, etc.
Starting from the Ginzburg-Landau free energy function we focus on the long-wavelength region $\ell \gg \xi_0 $ with $\ell \simeq O(\xi)$ (i.e. both $ \ell< \xi$ and $\ell >\xi$ are considered), where $\xi \gg \xi_0$ is the diverging correlation length.
These are the degrees of freedom that will control the renormalization group trajectories and universal behavior near $T_c$.
Up to now we have been discussing thermodynamic functions and static (time independent) correlations.
In order to investigate dynamic properties such as transport coefficients or dynamical modes, we must carry out a similar coarse-graining (i.e., averaging) on the dynamical equations, eliminating the microscopic modes involving the scales $\ell_0 < l < \xi_0$.
The remaining modes then describe the time dependence of the order parameter, which slows down near the transition, and the time dependence of any conserved densities that remain coupled to the order parameter at long wavelengths ($k \xi \approx 1$), as the transition is approached ($T \to T_c$).
In this way one obtains a dynamic generalization of the Ginzburg-Landau free energy, whose long-wavelength modes are precisely those of the original systems near $T_c$.
The important difference between statics and dynamics, which is already apparent from the Ginzburg-Landau theory itself, is that a single static universality class (given spatial dimension $d$ and order parameter symmetry $n$) will correspond to a multiplicity of dynamic universality classes, depending on the long-wavelength dynamics of the order parameter and of the conserved densities that couple to it.
A schematic representation of the above description of the Ginzburg-Landau theory is shown in Fig.~\ref{fig:schema1}.
%

%
%
%

\subsection{\label{sec:1B} Nonequilibrium patterns near linear instabilities}
We now turn to a different application of the Ginzburg-Landau theory, namely the study of nonequilibrium pattern formation in systems undergoing linear instabilities at a nonzero length and/or time scale.
We should say at the outset that whereas in the case of continuous phase transitions the most interesting properties near the transition are fully captured by the Ginzburg-Landau approach, for nonequilibrium patterns this is not the case.
It is only because the validity of our methods is confined to the vicinity of the linear instability that we focus on this regime.
Thus the analogy between phase transitions and nonequilibrium patterns is formal, rather than physical.
On the other hand, it should be mentioned that much less is known in general about systems far from equilibrium than about equilibrium and near-equilibrium phenomena and our treatment does provide nontrivial results for certain far from equilibrium systems, so we believe this more limited theory does make a contribution.
We consider a nonequilibrium system defined by dynamical equations, typically by a set of partial differential equations.
The system is subjected to a constant external drive, represented by a control parameter $R$, which vanishes in equilibrium.
We imagine that for sufficiently small $R >0$, the solutions $u(x,t)$ of the dynamical equations are `simple' nonequilibrium steady states which we represent by a constant $\bar{u}$.
At a critical value $R =R_c$ of the control parameter, the steady state $\bar{u}$ becomes unstable, and a mode with wave vector $q_0$ and frequency $\omega_0$ (length scale $q_0^{-1}$ and time scale $\omega_0^{-1}$) is the one that grows most rapidly.
In analogy with the situation of continuous phase transitions we now define the `microscale' as $\ell_0 =q_0^{-1}$ (or some other length scale associated with the linear instability) and note that the starting dynamical equations, though they may originate from some physical macroscopic theory, can from a formal point of view be considered as a `microscopic model', valid from the {\em microscale} $\ell_0$ to the macroscale $L$.
We then introduce the reduced control parameter $r =(R-R_c)/R_c$, and define a {\em mesoscale} $\xi =\ell_0 |r|^{-1/2} =q_0^{-1}|r|^{-1/2}$, which sufficiently close to the instability ($|r| \ll 1$) provides a scale separation between the micro- and macroscales ($\ell_0 \ll \xi$), with $\xi \lesssim L$.
Note that since the starting equations are themselves physically macroscopic, we do not need the coarse graining step employed in the phase transition case, and we here define the microscale $\ell_0$ to be what we called $\xi_0$ previously (see Fig.~\ref{fig:schema1}).
The Ginzburg-Landau equations are only valid in the critical (or
\lq universal') region $|r| \ll 1$ ($\xi \gg \ell_0$) and it describes scales $\ell \simeq O(\xi)$.
We now represent the solution of the original dynamical system as
\begin{eqnarray}
\label{eq:u_approx}
u(x,t) = u_0 \left[ A(x,t) e^{i (q_0 x - \omega_0 t)} + \textrm{c.c.} \right] \,,
\end{eqnarray}
where $u_0$ is a function related to the linear instability, and c.c. signifies complex conjugate.
Then sufficiently close to the linear instability, it can be shown that solutions $u(x,t)$ of the starting dynamical system are given by Eq.~(\ref{eq:u_approx}), provided $A(x,t)$ satisfies the so-called real or complex Ginzburg-Landau equations given by Eqs.~(\ref{eq:ampl}) and (\ref{eq:CGLE1d}).
For this case we have thus reduced the problem of finding solutions of a general dynamical system to analysis of a relatively simple nonlinear partial differential equation.
We also demonstrate thereby that at least sufficiently close to the linear instability the behavior is entirely determined by the parameters of that instability, so that vastly different systems can thus admit a universal description, as long as they have similar linear instabilities.
Of course, as mentioned above, this universality is confined to the vicinity of the linear instability, which is not necessarily the most interesting physical regime, in contrast to thermodynamic phase transitions where the vicinity of the critical point is of primary physical relevance.
Nonequilibrium systems undergoing pattern forming linear instabilities include Rayleigh-B\'enard convection, convection in fluid mixtures, Taylor-Couette flow, oscillatory chemical reactions and reaction-diffusion dynamics in neural systems and heart muscle, to cite only a few.
Solutions of the Ginzburg-Landau equations can then be found for $0 < r \ll 1$, and they constitute nontrivial solutions of the original dynamical system via Eq.~(\ref{eq:u_approx}).
For example a continuum of stationary or traveling plane wave solutions can be constructed and their stability investigated.
More complicated solutions, which we refer to as `defects', can be found and their dynamics investigated.
These are then bona fide solutions of the starting dynamical system and they are observed in experiments on a variety of systems.
One of the most interesting aspects of pattern forming nonequilibrium systems is the phenomenon of chaos, and the complex Ginzburg-Landau equation [see Eq.~(\ref{eq:CGLE})] provides an excellent example, where the transition from temporal to spatiotemporal chaos as the system size is increased can be vividly illustrated both numerically and experimentally.
%

%
%
\begin{figure}[ht]
\includegraphics[width=7.5cm]{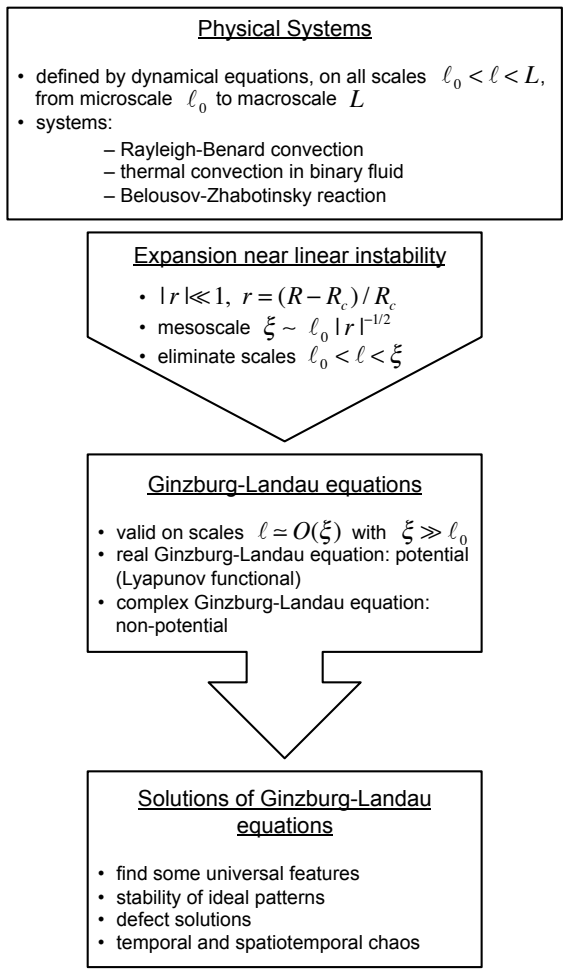}
\caption{\label{fig:schema2}
Schematic structure of the Ginzburg-Landau theory of pattern formation and chaos.}
\end{figure}
A schematic structure of the Ginzburg-Landau theory of pattern formation and chaos is shown in Fig.~\ref{fig:schema2}.
%

\subsection{\label{sec:1C} Nature of the presentation}
This paper is designed to introduce the reader to critical phenomena and nonequilibrium pattern formation using a unified language, that of the Ginzburg-Landau theory.
It is by no means intended to be a full survey of these fields even for the first portion (phase transitions) and certainly not for the second portion (patterns).
Rather, the Ginzburg-Landau theory is presented as a convenient and transparent {\em language} with which to highlight the essential principles that govern the behavior.
There is little emphasis on calculational techniques or on detailed experimental developments, and the historical aspects of the field are treated rather superficially.
The authors consider those items to be adequately treated in the existing literature, to which references can be found in the various reviews and monographs referred to in the bibliography.
It is hoped that by tying together the two primary applications of the Ginzburg-Landau equations, phase transitions and nonequilibrium patterns, which are usually discussed separately, this paper will lead to a unified conceptual understanding of cooperative equilibrium and nonequilibrium behavior.
A word about references.
In accordance with the introductory nature of the discussion, we have not provided citations for the occasional references to historical materials.
These can be found in the textbooks, monographs and review articles that appear in our bibliography.
%

\section{\label{sec:2} Mean-field theory: statics}
%
\subsection{\label{sec:2A} Order parameters and broken symmetries: the Landau expansion}
Continuous (also known as second-order) phase transitions occur when a new state of reduced symmetry emerges continuously from the disordered or symmetric phase as the temperature is reduced.
The ordered phase at low temperature has a lower symmetry than the disordered phase at high temperature.
There are a multiplicity of equivalent states (equal free energy) in the ordered phase, sometimes an infinite number.
These states are macroscopically different, so fluctuations do not connect them in the macroscopic ($L \to \infty$) limit, also known as the thermodynamic limit.
The ordered phases are described by a phenomenological order parameter $\psi(T)$ which is nonzero below the transition point $T_c$ and vanishes at and above $T_c$, in equilibrium.
\underline{The Landau expansion:}\\
For spatially uniform systems the free energy for given value of the order parameter is analytic in $\psi$ and $T$. Near the transition it thus takes the form
\begin{eqnarray}
\label{eq:free_energy}
\Phi(P,T) = \Phi_0(T) + V \left[ a(T) \psi^2 + b(T) \psi^4 + \dots \right] \,, \qquad
\end{eqnarray}
where $\Phi_0$ is smooth at $T_c$.
For the coefficients $a$ and $b$ we have
\begin{eqnarray}
\label{eq:ab}
&& a(T) = a_0 \tau + \dots \,, \; a_0 > 0 \,,
\nonumber \\
&& b(T) = b_0 + \dots \,, \; b_0 > 0 \,,
\end{eqnarray}
where the reduced temperature $\tau$ is defined by
\begin{eqnarray}
\label{eq:tau_def}
\tau = (T - T_c)/T_c \,.
\end{eqnarray}
The equilibrium condition (Landau equation) is given by minimization of the Landau free energy $\Phi$ with respect to $\psi$
\begin{eqnarray}
\label{eq:min_psi}
\frac{\partial \Phi}{\partial \psi} = 0 \,:\Rightarrow \;
2 a \psi + 4 b \psi^3 = 0 \,.
\end{eqnarray}
The solutions $\bar{\psi}$ of Eq.~(\ref{eq:min_psi}) are given by
\begin{eqnarray}
\label{eq:min_psi_sol}
\bar{\psi} =
\begin{cases}
0 \,, & \tau >0 \; (a > 0) \\
\pm \left( \frac{-a}{2 b} \right)^{1/2} =
\pm \left( \frac{-a_0 \tau}{2 b_0} \right)^{1/2} \,, & \tau < 0 \; (a < 0)
\end{cases}
\qquad
\end{eqnarray}
Substituting $\psi = \bar{\psi}$ into the free energy given by Eq.~(\ref{eq:free_energy}) one obtains
\begin{eqnarray}
\label{eq:Phi_psi1}
\Phi = \Phi_0 - V \frac{a^2}{4 b} = \Phi_0 - V \frac{a_0^2 \tau^2}{4 b_0} \,.
\end{eqnarray}
The specific heat $C_p$ is given by
\begin{eqnarray}
\label{eq:C_p}
C_p = - \frac{T}{V} \frac{\partial^2 \Phi}{\partial T^2} =
\begin{cases}
C_0 \;, & \tau > 0 \\
C_0 + \frac{a_0^2 T}{2 b_0 T_c^2} \,, & \tau < 0
\end{cases}
\end{eqnarray}
One has a jump $\Delta C_p = a_0^2/(2 b_0 T_c)$ at the transition temperature $T = T_c$.

In addition to temperature one introduces an external field $h$, which couples linearly to the order parameter.
The free energy contains an additional term
\begin{eqnarray}
\label{eq:free_energy_h}
\tilde{\Phi}(P,T,h) = \Phi(P,T) - V \psi h \,,
\end{eqnarray}
where $\Phi(T,P)$ is given by Eq.~(\ref{eq:free_energy}) and $h$ is the external field.
The equilibrium value of $\psi$ is determined by minimization of $\tilde{\Phi}(P,T,h)$
\begin{eqnarray}
\label{eq:min_psi_h}
\frac{\partial \tilde{\Phi}}{\partial \psi} = 0 \,:\Rightarrow \;
2 a \psi + 4 b \psi^3 = h \,.
\end{eqnarray}
The susceptibility is the derivative $\chi = (\partial \psi /\partial h)_{T, h \to 0}$.
Differentiation of Eq.~(\ref{eq:min_psi_h}) gives
\begin{eqnarray}
\label{eq:chi_inv}
\chi^{-1} = \frac{\partial h}{\partial \psi} = 2 a + 12 b \psi^2 \,.
\end{eqnarray}
Then one obtains for $h \to 0$ in the disordered phase
\begin{eqnarray}
\label{eq:chi_h0_dis}\
\tau > 0 \,, \; \psi^2 = \bar{\psi}^2 = 0 \; \textrm{and} \; \chi^{-1} = 2 a \,,
\end{eqnarray}
and in the ordered phase
\begin{eqnarray}
\label{eq:chi_h0_ord}
\tau < 0 \,, \; \psi^2 = \bar{\psi}^2 = -a/(2 b) \; \textrm{and} \; \chi^{-1} = -4 a \,. \qquad
\end{eqnarray}
Thus the susceptibility diverges at the transition point $T \to T_c$ ($a = a_0 \tau \to 0$).
For nonzero external field $h \ne 0$ at the transition point $\tau =0$ ($a =0$), the order parameter is
\begin{eqnarray}
\label{eq:psi_h}
\psi = \left( \frac{h}{4 b} \right)^{1/3} \,.
\end{eqnarray}
\noindent It is the minimization with respect to $\psi$ which turns the smooth free energy \eqref{eq:free_energy_h} into one having a singularity at $T=T_c$ and $h=0$.
\underline{First-order phase transitions:}\\
We assume a free energy in the form
\begin{eqnarray}
\label{eq:free_energy_1st}
\Phi(P,T) = \Phi_0(T) + V \left[ a \psi^2 + e \psi^3 + b \psi^4 + \dots \right] \,. \qquad
\end{eqnarray}
In the presence of a cubic term ($e \ne 0$) one has metastability, for example at a solid -- liquid phase transition (melting, freezing) one has
\begin{eqnarray}
\label{eq:metastab}
& T = T_{sol} \,: & \quad \Phi_{sol} > \Phi_{liq} \,, \;
\psi = \psi_{liq} \,,
\nonumber \\
& T > T_{meta} \,: & \quad \Phi_{sol} = \Phi_{liq} \,, \;
\psi_{sol} = \psi_{liq}  \;,
\nonumber \\
& T < T_{meta} \,: & \quad \Phi_{sol} < \Phi_{liq} \,, \;
\psi = \psi_{sol} \,.
\end{eqnarray}
Thus the order parameter $\psi$ jumps at $T = T_{meta}$.
One also has a first-order phase transition for a free energy of the form
\begin{eqnarray}
\label{eq:free_energy_1st_6}
\Phi(P,T) = \Phi_0(T) + V \left[ a \psi^2 - b \psi^4 + f \psi^6 + \dots \right] \,, \qquad
\end{eqnarray}
with $b >0$, $f >0$.
Note that the expansion (\ref{eq:free_energy_1st}) is only valid if the transition is weakly first-order, i.e. $|\psi_{liq} - \psi_{sol}| \ll |\psi_{liq}| + |\psi_{sol}|$.
%

\subsection{\label{sec:2B} Spatial variations and fluctuations: the Ginzburg-Landau free energy}

Let us consider spatially nonuniform systems, i.e., we allow the order parameter to be spatially dependent, $\psi = \psi(x)$.
The free energy is now a {\em functional} of $\psi(x)$, and in the presence of an external field $h$ it has the following form
\begin{eqnarray}
\label{eq:free_energy_x}
&& \tilde{\Phi}[P,T,\psi(x),h] = \Phi_0(T)
\nonumber \\
&& + \int d^3x \left[ a \psi^2(x) + b \psi^4 (x) + c (\nabla \psi)^2 - h \psi(x) \right] \,. \qquad
\end{eqnarray}
This expression is for historical reasons referred to as the Ginzburg-Landau free energy, though it was introduced by Landau before the appearance of the Ginzburg-Landau paper (1950).
The probability of a fluctuation $\psi(x)$ is
\begin{eqnarray}
\label{eq:prob}
&& {\cal P}[\psi(x)] = Z^{-1} \exp\left\{ -\beta \tilde{\Phi}[\psi(x)] \right\} \,,
\nonumber \\
&& \beta = \frac{1}{k_B T} = \frac{1}{T} \,,
\end{eqnarray}
where $Z$ is the partition function (normalization) obtained by integration over all possible configurations ${\cal D}\psi(x)$ of the order parameter
\begin{eqnarray}
\label{eq:Z1}
Z = \int {\cal D}\psi(x) \exp\left\{ -\beta \tilde{\Phi}[\psi(x)] \right\} \,.
\end{eqnarray}
Knowing the probabilities as in Eq.~(\ref{eq:prob}) one can write in general the average value of some function of the order parameter $A(\psi)$ as
\begin{eqnarray}
\label{eq:A_avr}
\langle A(\psi) \rangle = Z^{-1}
\int {\cal D}\psi A(\psi) \exp\left\{ -\beta \tilde{\Phi}[\psi(x)] \right\} \,. \qquad
\end{eqnarray}
The average value of the order parameter is given by
\begin{eqnarray}
\label{eq:psi_avr}
\langle \psi \rangle = \frac{Z^{-1}}{V} \frac{\partial Z}{\partial h} = \frac{T}{V} \frac{\partial \ln Z}{\partial h} \,,
\end{eqnarray}
and the susceptibility can be found from the linear response
\begin{eqnarray}
\label{eq:chi}
\chi = \frac{\partial \langle \psi \rangle}{\partial h} =
\frac{T}{V} \frac{\partial^2 \ln Z}{\partial h^2} \,.
\end{eqnarray}
Let us now relate the partition function to the correlation function.
Introduce Fourier modes
\begin{eqnarray}
\label{eq:Fourier}
&& \psi(x) = \frac{V}{(2\pi)^3} \int d^3q \, \psi(q) e^{i \bm q \cdot \bm x} \,,
\nonumber \\
&& \psi(q) = \frac{1}{V} \int d^3x \, \psi(x) e^{-i \bm q \cdot \bm x} \,.
\end{eqnarray}
Define the correlation function
\begin{eqnarray}
\label{eq:corr}
C(x) = \langle\langle \psi(x) \psi(0) \rangle\rangle  \,,
\end{eqnarray}
where the double bracket is defined as
$\langle ( \psi(x) - \langle \psi \rangle )
( \psi(0) - \langle \psi \rangle ) \rangle$.
Using Fourier modes from Eq.~(\ref{eq:Fourier}) the free energy can be written (for $h =0$) as
\begin{eqnarray}
\label{eq:free_energy_Four}
\Phi = \int d^3q \, \left[ a \psi(q)\psi(-q) + c q^2 \psi(q)\psi(-q) + \dots \right] \,. \qquad
\end{eqnarray}
The fluctuation-response relation (fluctuation dissipation theorem) is
\begin{eqnarray}
\label{eq:fluc_rel}
\chi = \frac{1}{T} \int d^3x \, C(x) \,,
\end{eqnarray}
a relation which is valid for weak fluctuations since linear response was assumed.
More generally, for nonzero wave vector we have
\begin{eqnarray}
\label{eq:chi_q}
\chi(q) = \frac{V}{T} C(q) \,,
\end{eqnarray}
which relates the response $\chi$ and the correlations $C$ (or fluctuations).
According to the free energy given by Eq.~(\ref{eq:free_energy_Four}) the coefficient $a$ in the Landau expansion is replaced by $a + c q^2$ in the Ginzburg-Landau expansion.
Thus one has for the susceptibility in Fourier space [compare with Eqs.~(\ref{eq:chi_h0_dis}) and (\ref{eq:chi_h0_ord})]
\begin{eqnarray}
\label{eq:chi_q_exp}
\chi^{-1}(q) =
\begin{cases}
2 (a + c q^2) \;, & \tau >0 \\
2 (-2 a + c q^2) \;, & \tau < 0 \,.
\end{cases}
\end{eqnarray}
Using the relation between susceptibility and correlation function given by Eq.~(\ref{eq:chi_q}) one finds after Fourier transformation
\begin{eqnarray}
\label{eq:corr_x}
C(x) = \frac{T}{(2\pi)^3}\int d^3q \, \chi(q) e^{i \bm q \cdot \bm x} = \frac{T}{8\pi c x} e^{-x/\xi} \,, \qquad
\end{eqnarray}
where the so-called correlation length $\xi$ is given by
\begin{eqnarray}
\label{eq:xi2}
\xi^2 = 2 c \, \chi_{q=0} =
\begin{cases}
c/a = c/(a_0 \tau) \;, & \tau >0 \\
-c/(2 a) = -c/(2 a_0 \tau) \;, & \tau <0 \,.
\end{cases}
\qquad
\end{eqnarray}
The correlation length $\xi \propto 1/\sqrt{\tau}$ diverges when the transition point is approached (Ornstein and Zernike).
%

\subsection{\label{sec:2C} Continuous broken symmetries}
Up to now the order parameter $\psi$ was considered to be a real scalar.
The ordered state has the broken symmetry $\psi \leftrightarrow -\psi$ (discrete broken symmetry).
The more general case is a vector order parameter ($n$-vector model):
\begin{eqnarray}
\label{eq:psi_vec}
\bm \psi(x) = \left\{ \psi_1(x), \dots, \psi_n(x) \right\} \;,
\end{eqnarray}
and one has in Eq.~(\ref{eq:free_energy_x})
\begin{eqnarray}
\label{eq:scal_vec}
&& \psi^2 \to |\bm \psi |^2 = \sum_{i=1}^n \psi_i^2 \;,
\nonumber \\
&& (\nabla \psi)^2 \to |\nabla \bm \psi|^2
= \sum_{i=1}^n (\partial_x \psi_i)^2 \,.
\end{eqnarray}
The scalar case corresponds to $n =1$.
An external field is now also a vector and $h \psi \to \bm h \cdot \bm \psi$.
In the ordered state the order parameter is equal to $\psi_1$, say, but it could be equal to any other component $\psi_i$, i.e., there is an $n$-fold degeneracy.
In this case we speak of a {\em continuous} broken symmetry.
The free energy for a spatially uniform system in the presence of an external field is given by
\begin{eqnarray}
\label{eq:Phi_tilde}
&& \tilde{\Phi} = \Phi(|\bm \psi|^2) - V \bm h \cdot \bm \psi \;,
\nonumber \\
&& \Phi(|\bm \psi|^2) = V \left[ a |\bm \psi|^2 + b |\bm \psi|^4 \right] \,.
\end{eqnarray}
The equilibrium state is determined by minimization of $\tilde{\Phi}$
\begin{eqnarray}
\label{eq:Phi_tilde_min}
\frac{\partial \tilde{\Phi}}{\partial \psi_i} = 0 \,:\Rightarrow \;
2 \psi_i \Phi' = V h_i \,,
\end{eqnarray}
where $\Phi'$ means the derivative of $\Phi$ with respect to its argument $|\bm \psi|^2$.
What is now the susceptibility?
We introduce the matrix
\begin{eqnarray}
\label{eq:chi_ij}
\chi_{i j} = \frac{\partial \psi_i}{\partial h_j}
\qquad \textrm{and} \qquad
\chi^{-1}_{i j} = \frac{\partial h_i}{\partial \psi_j} \,.
\end{eqnarray}
We consider the field to be applied either along the vector order parameter or transverse to it, with corresponding susceptibilities $\chi_{\parallel}$ and $\chi_{\perp}$, respectively.
The susceptibility matrix is
\begin{eqnarray}
\label{eq:chi_pp}
\chi_{i j} = \chi_{\parallel} \hat{h}_i \hat{h}_j
+ \chi_{\perp} (\delta_{i j} - \hat{h}_i \hat{h}_j) \,,
\end{eqnarray}
where $\hat{\bm h} = \bm h/|\bm h|$ is a unit vector along the external field.
Similarly, for the inverse susceptibility we have
\begin{eqnarray}
\label{eq:chi_inv_pp}
\chi^{-1}_{i j} = \chi^{-1}_{\parallel} \hat{h}_i \hat{h}_j
+ \chi^{-1}_{\perp} (\delta_{i j} - \hat{h}_i \hat{h}_j) \,.
\end{eqnarray}
Taking into account Eq.~(\ref{eq:Phi_tilde_min}) and differentiating with respect to $\psi_j$ one finds for the inverse susceptibility
\begin{eqnarray}
\label{eq:V_chi_inv}
V \chi^{-1}_{i j} = 2 \delta_{i j} \Phi' + 4 \psi_i \psi_j \Phi''
= 2 \delta_{i j} \Phi' + 4 \hat{h}_i \hat{h}_j |\bm \psi|^2 \Phi'' \,. \qquad
\end{eqnarray}
Adding and subtracting the term $2 \hat{h}_i \hat{h}_j \Phi'$ to the right hand side of Eq.~(\ref{eq:V_chi_inv}) one finds
\begin{eqnarray}
\label{eq:V_chi_inv_2}
V \chi^{-1}_{i j} = \hat{h}_i \hat{h}_j (2 \Phi' + 4 |\bm \psi|^2 \Phi'') + (\delta_{i j} - \hat{h}_i \hat{h}_j) 2 \Phi' \,. \qquad
\end{eqnarray}
Comparing with Eq.~(\ref{eq:chi_inv_pp}) one obtains for the longitudinal and transverse inverse susceptibilities
\begin{eqnarray}
\label{eq:chi_inv_pp_2}
&& V \chi^{-1}_{\parallel} =  2 \Phi' + 4 |\bm \psi|^2 \Phi'' \,,
\nonumber \\
&& V \chi^{-1}_{\perp} = 2 \Phi' \,.
\end{eqnarray}
For $\Phi(|\bm \psi|^2)$ given by Eq.~(\ref{eq:Phi_tilde}) one finds
\begin{eqnarray}
\label{eq:Phi_diff}
\Phi' = V \left[ a + 2 b |\bm \psi|^2 \right] \,, \;
\Phi'' = V 2 b \,,
\end{eqnarray}
and substituting into Eqs.~(\ref{eq:chi_inv_pp_2}) obtains
\begin{eqnarray}
\label{eq:chi_inv_pp_3}
&& \chi^{-1}_{\parallel} =  2 a + 12 b |\bm \psi|^2 \,,
\nonumber \\
&& \chi^{-1}_{\perp} = 2 a + 4 b |\bm \psi|^2 \,.
\end{eqnarray}
Then one finds in the disordered phase
\begin{eqnarray}
\label{eq:chi_pp_dis}
\tau > 0 \,, \; |\bm \psi|^2 = 0 \; \textrm{and} \;
\chi^{-1}_{\parallel} = \chi^{-1}_{\perp} = 2 a \,, \qquad
\end{eqnarray}
and in the ordered state
\begin{eqnarray}
\label{eq:chi_pp_ord}
\tau < 0 \,, \; |\bm \psi|^2 = -a/(2 b) \; \textrm{and} \;
\chi^{-1}_{\parallel} = -4 a \,, \; \chi^{-1}_{\perp} = 0 \,. \;\;\qquad
\end{eqnarray}
Since $\chi^{-1}_{\perp} = 0$ for all $\tau < 0$ one has a divergence of the transverse susceptibility not only at the critical point but throughout the ordered phase.
The significance of this result is apparent when one looks at a spatially dependent vector order parameter.
The free energy will contain an additional square gradient term
\begin{eqnarray}
\label{eq:Phi_int}
\Phi = \int d^3x \left[ a |\bm \psi|^2 + b |\bm \psi|^4
+ c |\nabla \bm \psi|^2 \right] \,.
\end{eqnarray}
The same structure occurs in Fourier space and again the coefficient $a$ is replaced by $a + c q^2$.
One can then write
\begin{eqnarray}
\label{eq:chi_inv_pp_q}
&& \tau >0 \;, \;
\chi^{-1}_{\parallel} = \chi^{-1}_{\perp} = 2 (a + c q^2) \,,
\nonumber \\
&& \tau <0 \,, \;
\chi^{-1}_{\parallel} = 2 (-2 a + c q^2) \,, \;
\chi^{-1}_{\perp} = 2 c q^2 \,.
\end{eqnarray}
Using the relation between susceptibility and correlation function given by Eq.~(\ref{eq:chi_q}), one finds after Fourier transformation for the longitudinal correlation function $C_{\parallel}(x)$
\begin{eqnarray}
\label{eq:C_par}
C_{\parallel}(x) = \frac{T}{(2\pi)^3}\int d^3q \, \chi_{\parallel}(q) e^{i \bm q \cdot \bm x} = \frac{T}{8\pi c x} e^{-x/\xi_{\parallel}} \,. \qquad
\end{eqnarray}
However for the transverse correlation function $C_{\perp}(x)$ in the ordered phase one finds
\begin{eqnarray}
\label{eq:C_perp}
C_{\perp}(x) = \frac{T}{(2\pi)^3}\int d^3q \, \frac{e^{i \bm q \cdot \bm x}}{2 c q^2}
= \frac{T}{8\pi c x} \,.
\end{eqnarray}
Thus one has a power-law decay of correlations for all $T < T_c$, rather than an exponential, i.e., there is an infinite correlation length $\xi_{\perp} \to \infty$.
A continuous broken symmetry possesses a kind of critical behavior not only at the critical point but along the whole ordered (condensed) phase at zero field.
Such behavior is referred to as a `soft mode', even though it occurs in the static (time-independent) correlations.
Let us consider a vector order parameter with planar order ($n =2$).
Suppose the symmetry is broken in a certain way and one has
\begin{eqnarray}
\label{eq:psi_2_example}
\bm \psi(x) = \{ \psi_1, \psi_2 \} = \bar{\psi} e^{i \theta(x)} \;.
\end{eqnarray}
Since the free energy depends only on $|\bm \psi|^2$, changing the phase $\theta$ in Eq.~(\ref{eq:psi_2_example}) does not change the free energy.
Although there is no barrier in the free energy when the direction of $\bm \psi$ is changed, there is a so-called finite `stiffness'.
Consider the square gradient term in Eq.~(\ref{eq:Phi_int}) in the ordered state $\tau < 0$ with the order parameter given by Eq.~(\ref{eq:psi_2_example}); we have
\begin{eqnarray}
\label{eq:sg_psi_example}
|\nabla \bm \psi|^2 = |i \nabla \theta \bar{\psi} e^{i \theta}|^2
= \bar{\psi}^2 (\nabla \theta)^2 \,.
\end{eqnarray}
Then the free energy Eq.~(\ref{eq:Phi_int}) can be rewritten in the form
\begin{eqnarray}
\label{eq:Phi_int_example}
\Phi = a \bar{\psi}^2 + b \bar{\psi}^4
+ \frac{\rho_s}{2} \int v_s^2 \,,
\end{eqnarray}
where we have introduced
\begin{eqnarray}
\label{eq:vs_rhos}
v_s = \nabla \theta \,, \;
\rho_s = 2 c \bar{\psi}^2 \,,
\end{eqnarray}
and the coefficient $\rho_s$ is called the stiffness.
The free energy $\Phi$ is independent of $\theta$ (continuous degeneracy), but it depends on the gradient of $\theta(x)$.
%

\subsection{\label{sec:2D} Physical systems}
Let us briefly describe the most commonly studied physical systems in which continuous phase transitions occur.
%

\subsubsection{\label{sec:2D1} Uniaxial magnet}
\begin{figure}[ht]
(a)\includegraphics[height=5cm]{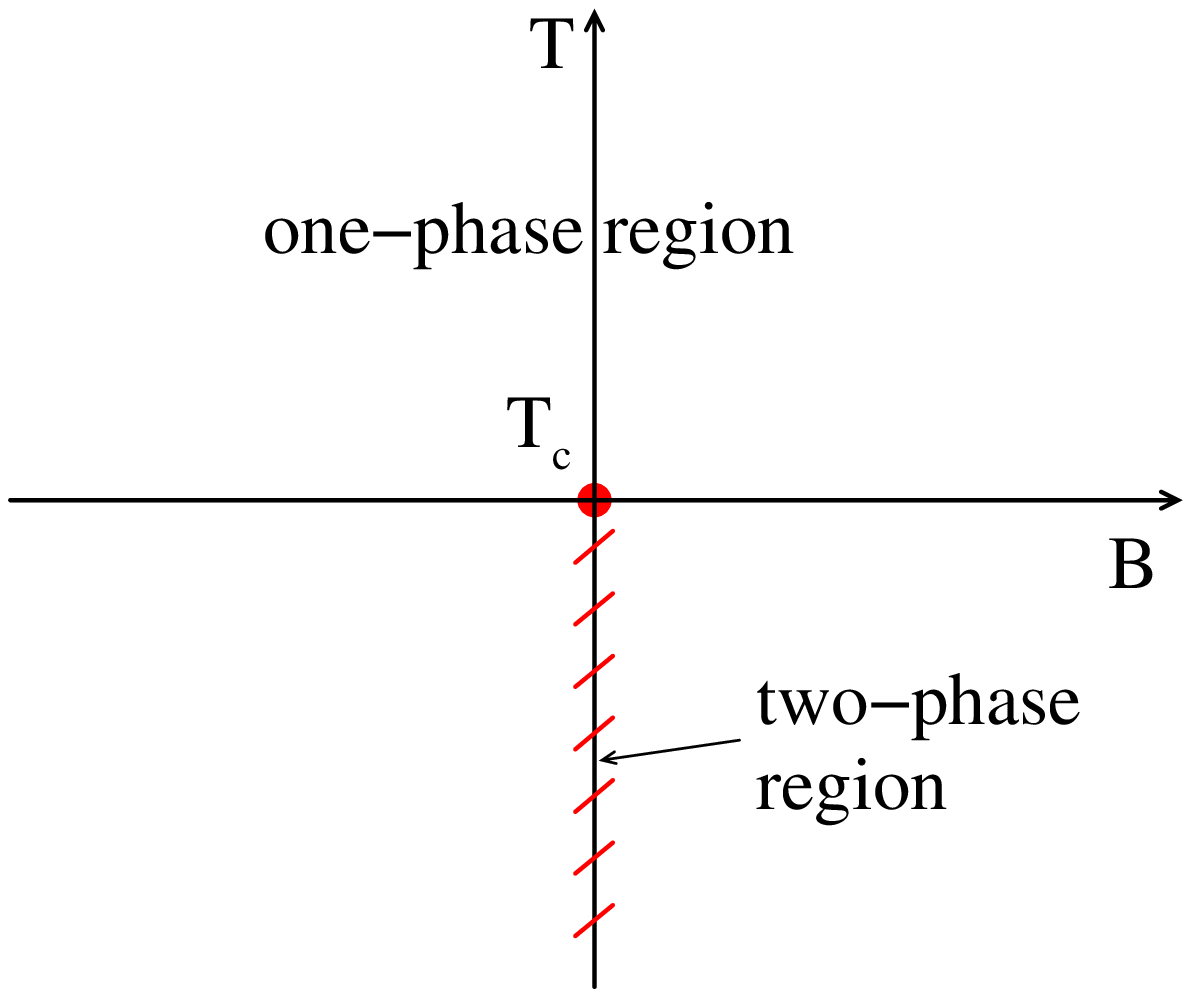}
(b)\includegraphics[height=5cm]{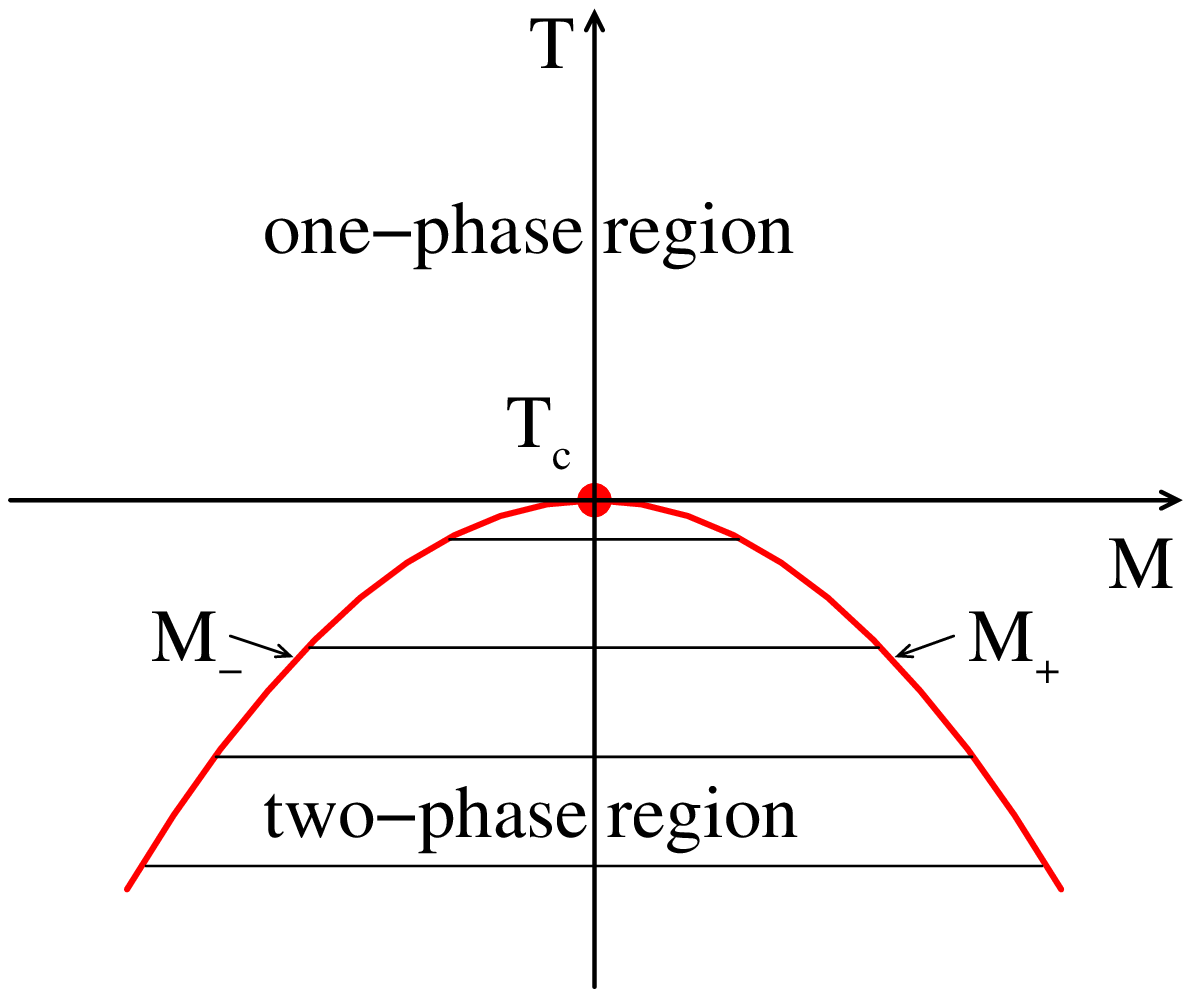}
\caption{\label{fig:ising}
Phase diagram of Ising model ($n =1$):
(a) Field variable: temperature $T$ vs. external field $B$, ordered state below $T_c$;
(b) Density variable: temperature $T$ vs. magnetization $M$, one phase region above $T_c$ and two phase region below $T_c$.
For the gas--liquid critical point, $B$ is replaced by $\mu = \mu_L - \mu_G$ and $M$ is replaced by $\rho = \rho_L - \rho_G$, so that $M_{+}=\rho_L$, $M_{-}=\rho_G$.
For the binary fluid $M$ is replaced by $\psi =c_A - c_B$ the difference of concentrations, and $B =h =\mu_A - \mu_B$.}
\end{figure}

This is the simplest physical system since it is  described by a scalar order parameter ($n =1$).
In the case of a ferromagnet $\psi \sim M$ is the magnetization and $h \sim B$ a magnetic induction, and in the ordered phase we have $\psi = \pm \bar{\psi}$.
For antiferromagnets $\psi \sim M_s$ is the so-called staggered or sub-lattice magnetization.
Considering a lattice of spins there will be an `up-lattice' and a `down-lattice' and $\psi$ characterizes each sub-lattice.
The external field $h$ is the staggered field that acts on each sub-lattice separately.
The simplest model for a uniaxial magnet is the Ising model ($n =1$).
On the microscale (lattice spacing $\ell_0$) the Hamiltonian is
\begin{eqnarray}
\label{eq:H_Ising}
{\cal H} = -J \sum_{\langle i, j \rangle} S_i S_j \,,
\end{eqnarray}
where $\langle i, j \rangle$ means the sum over nearest neighbors, and $S_i =\pm 1$ is a classical spin.
For $J >0$ one has a ferromagnet and for $J <0$ an antiferromagnet.
The phase diagram can be written in terms of a field variable, the temperature $T$ vs. the external field $B$, or alternatively, in terms of a density variable, $T$ vs. $M$.
In the latter case one has a one-phase region above $T_c$ and a two-phase region below $T_c$ [see Fig.~\ref{fig:ising}].
%

%
%
%

\subsubsection{\label{sec:2D2} Pure fluid: liquid-gas critical point}
For a pure fluid the order parameter is the difference between the liquid and gas densities, $\psi = \rho_L - \rho_G$, and the external field is the difference between the liquid and gas chemical potentials, $h = \mu_L - \mu_G$.
The symmetry $\psi \to -\psi$ is true only asymptotically as $T \to T_c$ ($\tau \to 0$).
The liquid-gas transition can also be described by an Ising model (lattice gas model).
%

\subsubsection{\label{sec:2D3} Binary fluid}
For a fluid mixture the order parameter is the difference between the concentrations of the two components, $\psi = c_A - c_B$, the external field is the difference between the chemical potentials of the two components, $h = \mu_A - \mu_B$.
This system can also be represented by an Ising model ($n =1$).
%

\subsubsection{\label{sec:2D4} Planar magnet}
This system is also known as an easy-plane magnet.
It is a magnetic system in which the ordered state is characterized by a vector isotropic in a plane, say the $x-y$ plane.
The order parameter now has two components ($n =2$), $\bm \psi = (M_x, M_y)$, and the external field is $\bm h = (B_x, B_y)$.
The orthogonal components, $M_z$ and $B_z$ do not enter the static description, only the dynamics (see below).
%

%
%
\begin{figure}[ht]
\includegraphics[width=6cm]{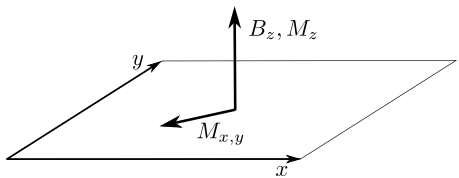}
\caption{\label{fig:easyplane}
Easy-plane magnet.}
\end{figure}
The microscopic model for this system is given by
\begin{eqnarray}
\label{eq:H_planar}
{\cal H} = -J_{x y} \sum_{\langle i, j \rangle} \left( S^x_i S^x_j
+ S^y_i S^y_j \right)
+ J_z \sum_{\langle i, j \rangle} S^z_i S^z_j \,, \qquad
\end{eqnarray}
where the coefficients $|J_{x y}| > |J_z|$ depend on the details of the lattice.
For $J_{x y} >0$ one has an easy-plane ferromagnet and for $J_{x y} <0$ an easy-plane antiferromagnet.
%

\subsubsection{\label{sec:2D5} Isotropic ferromagnet or antiferromagnet}
The ordered state is characterized by a vector isotropic in space, i.e., $n =3$.
In a ferromagnet one has $\bm \psi = \bm M$ where $\bm M =(M_x, M_y, M_z)$ is the uniform magnetization and the field $\bm h = \bm B$ where $\bm B = (B_x, B_y, B_z)$ is the magnetic induction.
In an antiferromagnet the order parameter is the staggered magnetization $\bm M_s$ and $\bm h$ is the staggered field $\bm B_s$.
The model on the microscale is the Heisenberg model
\begin{eqnarray}
\label{eq:H_Heisenberg}
{\cal H} = -J \sum_{\langle i, j \rangle} \bm S_i \cdot \bm S_j
- \bm h \cdot \sum_i \bm S_i \,.
\end{eqnarray}
For $J >0$ one has a ferromagnet and for $J <0$ an antiferromagnet.
%

\subsubsection{\label{sec:2D6} Superfluid}
The superfluid or Bose-fluid is described by an $n =2$ order parameter $\psi$, which is the complex superfluid `wave function'.
It comes from the off-diagonal density matrix $\rho(r,r')$ of a Bose-fluid,
\begin{eqnarray}
\label{eq:denmat}
\rho(r, r') = \langle a^{\dagger}(r) a(r') \rangle \,,
\end{eqnarray}
where $a^{\dagger}$, $a$ are the quantum creation and annihilation operators of the Bose-fluid and the bracket mean a thermal average.
The complex order parameter $\psi(r)$ is defined as
\begin{eqnarray}
\label{eq:psi_superfluid}
\lim_{|r-r'| \to \infty} \rho(r,r') = \psi(r) \psi^{*}(r') \,.
\end{eqnarray}
If the off-diagonal density matrix does not decay to zero at large distances, then $\psi \ne 0$ and one has a Bose condensate.
For example $^4$He has such a Bose condensation at $T \le T_{\lambda}$ (the lambda-temperature).
Since the order parameter $\psi$ is complex one has a phase degeneracy ($n =2$).
The field $h$ is a `source of quantum particles' and is not physically realizable.
Thus $\psi$ and $h$ are not directly measurable in liquid helium.
However they are coupled to physical quantities such as temperature $T$, entropy $S$, pressure $P$, and density $\rho$.
So the effect of $\psi$ on thermodynamic quantities can be measured, e.g., $C_p$ and the stiffness $\rho_s$ (also known as the superfluid density) can be measured.
%

\subsubsection{\label{sec:2D7} Superconductor}
Another system with quantum condensation is a superconductor.
It is also described by an $n =2$ order parameter $\psi$, which is the complex `pair wavefunction'.
This case is like Bose condensation but instead of the quantum creation and annihilation operators of the Bose-fluid $a$ and $a^{\dagger}$ one has for Fermi particles, pairs operators
\begin{eqnarray}
\label{eq:op_pairs}
a, a^{\dagger} \;\Rightarrow \; a a, a^{\dagger} a^{\dagger} \,.
\end{eqnarray}
The superconducting order parameter is related to an appropriate two-particle density matrix
\begin{eqnarray}
\label{eq:denmat2}
\rho_2(r,r')
= \langle a(r) a(r) a^{\dagger}(r') a^{\dagger}(r') \rangle \,,
\end{eqnarray}
by
\begin{eqnarray}
\label{eq:psi_cond}
\lim_{|r-r'| \to \infty} \rho_2(r,r') = \psi(r) \psi^{*}(r') \,.
\end{eqnarray}
The order parameter was introduced phenomenologically by Ginzburg and Landau in 1950 via Eq.~(\ref{eq:psi_cond}), without knowledge of the microscopic quantum relations Eq.~(\ref{eq:denmat2}) for the density matrix.
The field $h$ is again not physically realizable.
In superconductors the important new element from the point of view of physics is the coupling to electromagnetic fields since the electrons are charged.
The square gradient term in the Ginzburg-Landau free energy takes the form
\begin{eqnarray}
\label{eq:grad_psi_A}
|\nabla \psi|^2 \; \to  \;
|(\nabla - \frac{i e^{*}}{\hbar c} {\bm A})\psi|^2 \,,
\end{eqnarray}
where ${\bm A}(r,t)$ is the vector potential and $e^{*}$ is the charge associated with the `particles' which are actually pairs, i.e., $e^{*} = 2 e$.
This coupling leads to many important physical consequences, such as: \\
(i) the Meissner effect, an expulsion of a magnetic field from a superconductor below the transition to the superconducting state;\\
(ii) interfaces between the normal and superconducting states;\\
(iii) at nonzero magnetic field the Abrikosov instability leading to patterns of vortices of supercurrent with finite wavenumber $q_0 \sim 1/\xi$, where $\xi$ is the Ginzburg-Landau correlation length. \\
Note that in zero field one has the same expression for the free energy as for a superfluid, namely Eq.~(\ref{eq:Phi_int_example}).
%

\section{\label{sec:3} Dynamics: hydrodynamic modes}
%
\subsection{\label{sec:3A} Relaxational dynamics: conserved and non-conserved order parameter}
In terms of the Ginzburg-Landau description an equilibrium state is determined by the relation 
\begin{eqnarray}
\label{eq:equilib}
\frac{\partial \Phi}{\partial \psi} = 0 \,,
\end{eqnarray}
so away from equilibrium the simplest dynamics is relaxational
\begin{eqnarray}
\label{eq:reldyn}
\frac{\partial \psi}{\partial t} = -\frac{\Lambda}{V} 
\frac{\partial \Phi}{\partial \psi} \,,
\end{eqnarray}
i.e., $\psi$ decays to equilibrium, and the proportionality constant $\Lambda$ is called a `kinetic coefficient'.
In the spirit of the Ginzburg-Landau expansion, for $\psi$ near equilibrium (and near the phase transition) one finds [see Eq.~(\ref{eq:free_energy_h})]
\begin{eqnarray}
\label{eq:dPhidpsi}
&& \frac{\partial \tilde{\Phi}}{\partial h} = 0 \,:\Rightarrow \;
\frac{\partial \Phi}{\partial h} = V \psi \;,
\nonumber \\
&& \frac{\partial \Phi}{\partial \psi} 
= \frac{\partial \Phi}{\partial h} \frac{\partial h}{\partial \psi}
= V \psi \chi^{-1} \,.
\end{eqnarray}
The relaxational dynamics is then given by
\begin{eqnarray}
\label{eq:reldyn2}
\partial_t \psi = -\Gamma \psi \,,
\end{eqnarray}
where $\Gamma = \Lambda/\chi$ is the `relaxation rate'.
In the ordered phase ($\tau <0$) where $\psi = \bar{\psi}$ one has
\begin{eqnarray}
\label{eq:reldyn2psi}
\partial_t (\psi - \bar{\psi}) = -\Gamma (\psi - \bar{\psi}) \,.
\end{eqnarray}
Let us introduce the notion of a conserved order parameter, namely 
\begin{eqnarray}
\label{eq:conserved}
\partial_t \int d^3x \psi(x,t) = 0 \,,
\end{eqnarray}
or in Fourier space
\begin{eqnarray}
\label{eq:conserved_k}
\partial_t \psi(k=0,t) = 0 \,.
\end{eqnarray}
If the order parameter is conserved it implies
\begin{eqnarray}
\label{eq:Lambda_cons}
\Lambda \to \lambda \nabla^2 \,,
\end{eqnarray}
where $\lambda$ is known as a `transport coefficient'.
For a conserved order parameter one finds in Fourier space
\begin{eqnarray}
\label{eq:reldyn_k}
\partial_t \psi(k) = \Gamma(k) \psi(k) \,, \;
\Gamma(k) = \frac{\lambda}{\chi} k^2 = D k^2 \,.
\end{eqnarray}
Here $D = \lambda/\chi$ is the diffusion coefficient and the relaxation rate $\Gamma(k)$ goes to zero as $k \to 0$.
The expression $D =\lambda/\chi$ is known as an `Einstein relation'.
When the order parameter is not conserved we have $\Gamma(k=0)=\Gamma_0 \ne 0$ and $\psi(t)$ decay to equilibrium at a finite rate for $k \to 0$: $\psi(t) - \bar{\psi} \sim e^{-\Gamma_0 t}$.
%

\subsection{\label{sec:3B} Coupling to conserved densities: propagating modes}
Let us consider a situation where a non-conserved order parameter [$\Gamma(k=0)=\Gamma_0 \ne 0$] is coupled to a conserved density.
The model we consider is the planar magnet (see Fig.~\ref{fig:easyplane}).
The system has rotational symmetry around the $z$-axis and $\bm \psi$ lies in the $x-y$ plane.
There is no temporal symmetry of the dynamics of $\bm \psi$ in the plane, therefore $\bm \psi$ is not conserved, whereas $m_z$ is conserved, and $\partial_t m_z \sim \nabla^2 m_z$.
In this model $\bm \psi$ is coupled to $m_z$, and the normal component of the field, $h_z$, generates rotations of $\bm \psi$.
Classically we have the Poisson bracket 
\begin{eqnarray}
\label{eq:PB}
\left\{ \bm \psi, m_z \right\}_{PB} = i \bm \psi \,,
\end{eqnarray}
or quantum mechanically, in terms of the commutator we can write 
\begin{eqnarray}
\label{eq:commut}
\left[ S_x + i S_y, S_z \right] \propto i (S_x + i S_y) \,.
\end{eqnarray}
We consider an extension of the simple relaxational model of Sec.~\ref{sec:3A} to this case of a non-conserved order parameter $\psi = M_x + i M_y$ coupled to a conserved density $m_z$.
The free energy, which now depends on $\psi$, $m_z$, and $h_z$, takes the form
\begin{eqnarray}
\label{eq:fe_psi_mz}
\Phi(\psi, m_z, h_z) &=& \int d^3x \left[ 
a |\psi|^2 + b |\psi|^4 + c |\nabla \psi|^2 \right.
\nonumber \\
&& \qquad \left. + \frac{1}{2} \chi_m^{-1} m_z^2 - h_z m_z \right] \,.
\end{eqnarray}
The dynamics of $\psi$ and $m_z$ is then given by
\begin{eqnarray}
\label{eq:psi_t_mz_t}
&& \partial_t \psi = -\frac{\Lambda_{\psi}}{V} 
\frac{\partial \Phi}{\partial \psi}
- i g_0 \frac{\psi}{V} \frac{\partial \Phi}{\partial m_z} \,,
\nonumber \\
&& \partial_t m_z = \frac{\lambda_m}{V} 
\nabla^2 \frac{\partial \Phi}{\partial m_z}
+ \frac{2 g_0}{V} 
\textrm{Im} \left[ \psi^{*} \frac{\partial \Phi}{\partial \psi^{*}} \right] \,.
\end{eqnarray}
This is model E in the classification of \textcite{Hohenberg:1977}.
For the disordered phase, $\tau >0$, the cross coupling is negligible since $|\psi| \to 0$.
In the ordered phase, $\tau <0$, one has $\psi = \bar{\psi} e^{i \theta}$ and [see Eq.~(\ref{eq:vs_rhos})]
\begin{eqnarray}
\label{eq:fe_grad}
\Phi \sim \frac{\rho_s}{2} \int |\nabla \theta|^2 \,, \;
\rho_s = 2 c \bar{\psi}^2 \,.
\end{eqnarray}
Then in lowest order, the dynamics of $\theta$, $m_z$ is given by
\begin{eqnarray}
\label{eq:theta_t_mz_t}
&& \partial_t \theta = g_0 \chi_m^{-1} m_z \,,
\nonumber \\
&& \partial_t m_z = g_0 \rho_s \nabla^2 \theta \,.
\end{eqnarray}
Going into Fourier space, $\nabla^2 \to -k^2$, we find for the dynamical modes $\theta \sim \exp(i \omega_{\theta} t)$, $m_z \sim \exp(i \omega_m t)$
\begin{eqnarray}
\label{eq:omega_theta_m}
\omega_{\theta}(k) = \omega_m(k) = \pm c_s k \,, \;
c_s^2 = g_0^2 \rho_s / \chi_m \,.
\end{eqnarray}
Thus a non-conserved order parameter relaxes at a nonzero rate for $\tau >0$, but it is coupled to a conserved density ($m_z$) for $\tau <0$, due to the broken continuous symmetry.
This leads to a {\em propagating} `Goldstone' (spin wave) mode with $\omega = \pm c_s k$ and $c_s \propto \sqrt{\rho_s}$.
At the critical point one has $\rho_s = 2 c \bar{\psi}^2 \to 0$ and the velocity of the Goldstone mode goes to zero.
As we will see below, this result is directly related to the superfluid model with $\rho_s$ as the superfluid density.
%

\subsection{\label{sec:3C} Physical systems}
%
\subsubsection{\label{sec:3C1} Liquid-gas critical point}
This is an example of a system where a conserved order parameter is coupled to a conserved momentum current.
As mentioned above, in the static description the order parameter is the difference between the liquid and gas densities, $\psi = \rho_L - \rho_G$, the external field is the difference between the liquid and gas chemical potentials, $h = \mu_L - \mu_G$, and $\chi_{\psi} = \partial \rho / \partial \mu$ is the compressibility.
In the dynamics the order parameter $\psi$ is proportional to the entropy density $s = \varepsilon - (\bar{\mu} + T_c \bar{s})\rho$, where $\varepsilon$ is the energy density, and $\rho$ is the mass density.
The field $h_{\psi}$ is $T$ and $\chi_{\psi} = \partial s / \partial T|_{p} = C_p$.
The order parameter couples to the transverse momentum $\bm j_T$, a conserved current, with diffusion coefficient proportional to the viscosity $\bar{\eta}$: $D_j = \bar{\eta} / \rho$.
Note that a fluid in a porous medium does not obey momentum conservation so that both the sound mode and the viscous diffusion mode disappear at long wavelengths.
For this system one can also write a Ginzburg-Landau model [model H of \textcite{Hohenberg:1977}].
The relevant dynamical modes [see \textcite{FluidMech:1987}] are the thermal diffusion (Rayleigh) and viscous diffusion modes:
\begin{eqnarray}
\label{eq:Ra_m}
&& \textrm{Rayleigh} : \quad
\omega_{\psi} = i D_T k^2 \,, \; D_T = \lambda / C_p \,,
\nonumber \\
&& \textrm{viscous} : \quad
\omega_j = i D_j k^2 \,, \; D_j = \bar{\eta} / \rho \;,
\end{eqnarray}
where $\lambda$ is the thermal conductivity and $\bar{\eta}$ the viscosity.
There also are modes related to sound waves, the so-called Brillouin modes
\begin{eqnarray}
\label{eq:Br_m}
\textrm{Brillouin} : \quad
\omega_{B} = \pm c k \,, \; 
c^2 \propto (\partial \rho / \partial p)_s^{-1} \,,
\end{eqnarray}
but these are not important near $T_c$.
%

\subsubsection{\label{sec:3C2} Isotropic magnets}
The dynamics of the isotropic Heisenberg antiferromagnet ($n =3$) can be mapped onto the planar magnet (model E).
One has for the non-conserved order parameter $\bm \psi \sim \bm M_s$, the staggered (or sublattice) magnetization, which is mapped to the components $M_{x, y}$ in the planar magnet model.
The average total magnetization $\bm M$ is conserved and it is mapped onto the orthogonal component $m_z$ of the planar magnet model.
Thus $\bm M$ generates rotations of $\bm M_s$.
The dynamical modes for $\tau >0$ are
\begin{eqnarray}
\label{eq:af_modes1}
&& \omega_{\psi} \sim i \Gamma_0 = i \frac{\Lambda}{\chi_{\psi}} \,,
\nonumber \\
&& \omega_M \sim i \frac{\lambda}{\chi_M} k^2 \,,
\end{eqnarray}
where $\chi_M$ is the magnetic susceptibility.
In the ordered phase, $\tau <0$, the staggered and the total magnetization are coupled and one has
\begin{eqnarray}
\label{eq:af_modes2}
\omega_{\psi} = \omega_M = \pm c_s k \,, \;
c_s^2 \sim \rho_s \chi_M^{-1} \,,
\end{eqnarray}
which is a linear spin wave mode.
The isotropic ferromagnetic case is similar but there we have an $n =3$ {\em conserved} vector order parameter $\bm \psi \sim \bm M$ (Bloch equations, Landau-Lifshitz equations).
The dynamical modes are given for $\tau >0$ by
\begin{eqnarray}
\label{eq:fm_modes1}
\omega_{\psi} \sim i D_s k^2 = i \frac{\lambda}{\chi_{\psi}} k^2 \,,
\end{eqnarray}
which corresponds to spin diffusion.
This is in contrast to the antiferromagnet where for $\tau >0$ the order parameter just decays at a finite rate.
In the ordered phase, $\tau <0$, the different components of $\bm \psi$ are coupled and one has
\begin{eqnarray}
\label{eq:fm_modes2}
\omega_{\psi} = \pm b k^2 \;,
\end{eqnarray}
which describes the propagation of spin waves with quadratic wave vector dependence, and $b$ is again given by pure thermodynamics, $b = \rho_s / \bar{\psi}$, where $\bar{\psi}$ is the magnitude of the order parameter, and $\rho_s$ is the stiffness.
%

\subsubsection{\label{sec:3C3} Superfluids}
As mentioned above, the Bose fluid is described by an $n =2$ order parameter $\psi$.
We first consider a simple model of helium in a porous medium, i.e., no velocity diffusion (no momentum conservation), which makes the hydrodynamics simpler.
In analogy with the planar magnet we can use model E
\begin{eqnarray}
\label{eq:hp}
&& \psi \sim M_x + i M_y \to \textrm{Bose wave function} \; \psi
\nonumber \\
&& m_z \to \rho \,, \; h_z \to \mu \,.
\end{eqnarray}
For the dynamical modes for $\tau >0$ one has a non-conserved order parameter $\psi$ and a conserved (mass) density $\rho$
\begin{eqnarray}
\label{eq:hp_modes1}
&& \omega_{\psi} \sim i \Gamma_0 = i \frac{\Lambda}{\chi_{\psi}} \,,
\nonumber \\
&& \omega_{\rho} \sim i \frac{\lambda}{\chi_{\rho}} k^2 \,, \;
\chi_{\rho} = \partial \rho / \partial \mu \,,
\end{eqnarray}
describing relaxation of the order parameter and diffusion of density with transport coefficient $\lambda$.
In the ordered phase, $\tau <0$, the order parameter and density modes are coupled and one has a propagating mode with linear dispersion relation 
\begin{eqnarray}
\label{eq:hp_modes2}
\omega_{\psi} = \omega_{\rho} = \pm c_s k \,, \;
c_s^2 = \rho_s / \chi_{\rho} \,.
\end{eqnarray}
In the normal (disordered) phase there is no sound propagation.
However, when Bose condensation happens, one gets a propagating sound mode appearing as a result of the continuous broken symmetry.
In a porous medium this mode is known as `fourth sound' and it has been observed experimentally.
There is also a mode called 'third sound', which describes propagation of sound in thin films of superfluid.
Pure helium is more complicated.
For $\tau >0$ it is essentially the same model as for a pure (normal) fluid critical point and one has the Rayleigh mode for the conserved entropy density $s$ and a decaying mode for the non-conserved Bose order parameter $\psi$
\begin{eqnarray}
\label{eq:he_modes1}
&& \omega_{s} = i D_T k^2 \,, \; D_T = \frac{\lambda}{C_p} \,,
\nonumber \\
&& \omega_{\psi} = i \Gamma_0 = i \frac{\Lambda_{\psi}}{\chi_{\psi}} \,.
\end{eqnarray}
In the ordered phase, $\tau <0$, there is a contribution to the free energy $\sim (\rho_s/2) \int d^3x |\nabla \theta|^2$ as in the planar magnet where we had [see Eq.~(\ref{eq:theta_t_mz_t})]
\begin{eqnarray}
\label{eq:theta_t}
\partial_t \theta = g_0 \chi_m^{-1} m_z = g_0 h_z \,.
\end{eqnarray}
This equation expresses the fact that $m_z$ generates rotations of $\bm \psi$ (changes in the phase $\theta$ of the complex order parameter $\psi$).
In the case of a superfluid $h_z \to \mu$ and taking into account the units for the chemical potential $\mu$ one can write
\begin{eqnarray}
\label{eq:theta_t_sf}
\partial_t \theta = \mu / \hbar \,,
\end{eqnarray}
which represents the Josephson relation between changes of the phase of the order parameter and the chemical potential.
In the context of the Ginzburg-Landau description it just expresses the generation of rotations of the order parameter by the field $h_z$ in the planar magnet.
One can also define a superfluid velocity $v_s$ by
\begin{eqnarray}
\label{eq:vs}
&& v_s = \frac{\hbar}{m} \nabla \theta \,,
\nonumber \\
&& \partial_t v_s = \frac{1}{m} \nabla \mu \,.
\end{eqnarray}
This is the Landau equation for superfluid hydrodynamics, which can be obtained by taking the gradient of Eq.~(\ref{eq:theta_t_sf}).
Equation (\ref{eq:vs}) was derived by Landau in 1941 without any reference to Bose condensation, only on the basis of symmetry arguments.
Finally one finds for the modes in the ordered phase
\begin{eqnarray}
\label{eq:sf_modes2}
\omega_{s} = \omega_{\psi} = \pm c_2 k \,, \;
c_2^2 = \rho_s / C_p \,. 
\end{eqnarray}
This mode is known as `second sound', which is the new mode that appears in a superfluid, and its velocity $c_2 \to 0$ when approaching $T_c$.
The Brillouin mode also exists and it is given by 
\begin{eqnarray}
\label{eq:sf_modes2_br}
\omega_{B} = \pm c_1 k \,,
\end{eqnarray}
which is known as 'first sound'.
It represents ordinary compression of the fluid and has only a weak singularity at the transition to the superfluid state.
%

%
%
\begin{figure}[ht]
\includegraphics[width=7cm]{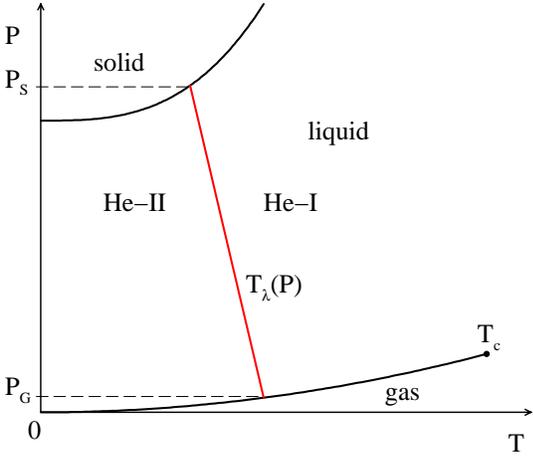}
\caption{\label{fig:4He_PT}
Phase diagram of $^4$He.
The gas-liquid coexistence curve ends at the critical point.
The liquid-solid melting line is shown at high pressure, and the so-called $\lambda$-line $T_{\lambda}(P)$, the transition from normal $^4$He-I to superfluid $^4$He-II depends on pressure in the range $P_G < P < P_S$.}
\end{figure}
All of these results are dramatic predictions for superfluids, all within the mean-field theory.
The phase diagram of helium ($PT$-diagram) is shown in Fig.~\ref{fig:4He_PT}.
One can separately measure the second sound velocity $c_2$, the superfluid density $\rho_s$ (by measuring the transverse response $\chi_{\perp}$), the specific heat $C_p$, and for example check the exact relation given by Eq.~(\ref{eq:sf_modes2}).
%

\subsection{\label{sec:3D} Phase transitions in dynamics: mean-field or conventional theory}
The discussion of dynamics thus far in Sec.~\ref{sec:3}, based as it is on mean-field theory, is nevertheless {\em exact} in the long-wavelength limit, away from the phase transition, since it refers either to the low-temperature or the high-temperature fixed point.
This is because mean-field theory correctly captures the symmetries and couplings that determine the long-wavelength hydrodynamics.
At the phase transition, we do not expect mean-field theory to be any more accurate for the dynamics than for the statics.
In that approximation the modes will reflect the behavior of the thermodynamic quantities $\chi_{\psi} \sim |\tau|^{-1}$ and $\rho_s \sim |\bar{\psi}|^2 \sim |\tau|$, and all the singularities (jump in $C_p$, correlation length $\xi \sim |\tau|^{-1/2}$) come from the vanishing of $\tau$ at the transition.
In particular, this so-called `conventional theory' assumes that all transport and kinetic coefficients $\Gamma_0$, $\Lambda$, $\lambda$, are non-singular (smooth).
Thus $\chi_{\psi}^{-1} \to 0$, $\rho_s \to 0$ and since they enter into the mode frequencies one has also $\omega_{\psi} \to 0$ at the transition.
This phenomenon is known as critical slowing down: for example in relaxational dynamics we have $\Gamma \sim \chi_{\psi}^{-1} \to 0$.
Let us consider for example a pure fluid for which $\omega_{\psi} \sim C_p^{-1}$ and $C_p \sim \chi_{\psi}$, thus $\omega_{\psi} \to 0$ at $|\tau| \to 0$.
In the isotropic antiferromagnet $\omega_{\psi} = \omega_M \sim c_s$ and $c_s \sim \rho_s^{1/2} \sim |\tau|^{1/2}$ ($\chi_M$ is non-singular).
For the isotropic ferromagnet $\omega_{\psi} \sim b k^2$ where $b \sim \rho_s \sim |\tau|$.
For the superfluid $\omega_{s} = \omega_{\psi} \sim c_2 k$ and $c_2 \sim \rho_s^{1/2} \sim |\tau|^{1/2}$, in the ordered phase, but $\omega_{\psi} \sim \chi_{\psi}^{-1} \sim \tau$ above $T_c$.
%

\section{\label{sec:4} Phenomenology of critical behavior: scaling and universality}
We shall follow the historical order and introduce scaling and universality phenomenologically before discussing the renormalization group, even though this reverses the logical order.
%

\subsection{\label{sec:4A} Statics}
As noted earlier, in mean-field theory $a = a_0 \tau$ and we have for the order parameter $\bar{\psi} \sim \sqrt{-a}$ for $\tau <0$ and $\bar{\psi} =0$ for $\tau \ge 0$.
For the specific heat one has $C_p = C_0$ for $\tau >0$ and $C_p = C_0 + \Delta C_p$ for $\tau <0$.
Finally for the susceptibility one has $\chi_{\psi} \sim |\tau|^{-1}$ for all $\tau$.
These lead to the following critical exponents in the disordered phase
\begin{eqnarray}
\label{eq:crit_exp1}
\tau >0 \,: \qquad
&& \chi_{\psi} \sim \tau^{-1} = \tau^{-\gamma} \,, \; \gamma =1 \,,
\nonumber \\
&& C_p \sim const = \tau^{-\alpha} \,, \; \alpha =0 \,.
\end{eqnarray}
Along the critical isochore we have
\begin{eqnarray}
\label{eq:isochore}
\tau =0 \,, \; h \ne 0: \qquad
h(\psi) \sim \psi^3 = \psi^{\delta} \,, \; \delta =3 \,,
\end{eqnarray}
and at the critical point we have
\begin{eqnarray}
\label{eq:crit_exp2}
C(x) = \langle \psi(x)\psi(0) \rangle \sim x^{-(1+\eta)} \,, \; \eta =0 \,.
\end{eqnarray}
In the ordered phase one has
\begin{eqnarray}
\label{eq:crit_exp3}
\tau <0 \,: \qquad
&& \chi_{\psi} \sim |\tau|^{-1} = |\tau|^{-\gamma} \,, \; \gamma =1 \,,
\nonumber \\
&& \bar{\psi} \sim |\tau|^{1/2} = |\tau|^{\beta} \,, \; \beta =1/2 \,,
\nonumber \\
&& C(x) \sim x^{-1-\eta} e^{-x / \xi} \,, \; \eta =0 \,,
\nonumber \\
&& \xi \sim |\tau|^{-1/2} = |\tau|^{-\nu} \,, \; \nu =1/2 \,.
\end{eqnarray}
Note that for models with $n >1$ (continuous symmetry breaking) one has two correlation lengths [see Eqs.~(\ref{eq:C_par}) and (\ref{eq:C_perp})]: $\xi_{\parallel} \sim |\tau|^{-1/2}$ and $\xi_{\perp} = \infty$.
The corresponding critical exponent is $\nu_{\parallel} = 1/2$ and $\nu_{\perp}$ is undefined.
These six critical exponents $\alpha$, $\beta$, $\gamma$, $\delta$, $\eta$, and $\nu$ are universal in the sense that they are the same for all $n$ (except for the difference between $\nu_{\parallel}$ and $\nu_{\perp}$) and all space dimensions
\begin{eqnarray}
\label{eq:univ_exp}
\begin{array}{cccccc}
\alpha & \beta & \gamma & \delta & \eta & \nu \,,
\\
0 & \frac{1}{2} & 1 & 3 & 0 & \frac{1}{2} \,.
\end{array}
\end{eqnarray}
As is well known, however, experiments and approximate calculations of exponents show that the mean-field theory is not quantitatively correct, as regards values for the exponents and the fact that the values depend on the system.
During 1960s a highly successful phenomenological theory was developed, which we call scaling and universality.
It is based on the idea that the diverging correlation length $\xi$ controls all the singularities in the thermodynamics and correlation functions.
Specifically, one assumes for the free energy of the system in the vicinity of the critical point ($\tau \to 0$, $h \to 0$, $\xi^{-1} \to 0$)
\begin{eqnarray}
\label{eq:fe_hom}
\tilde{\Phi}(T,h) = \Phi_{reg}(\tau,h) + V \tilde{\phi}(\tau,h) \,,
\end{eqnarray}
where $\Phi_{reg}$ represents the regular part, and the function $\tilde{\phi}(\tau,h) \to \tilde{\phi}(\xi,h)$ is expressed in terms of the correlation length $\xi(\tau,h)$ as a homogeneous function of $\xi$ and $h$
\begin{eqnarray}
\label{eq:phi_hom}
\tilde{\phi}(\xi,h) = \xi^{-y} f_{\pm}(h / \xi^{-w}) \,.
\end{eqnarray}
For the correlation function one also assumes homogeneity:
\begin{eqnarray}
\label{eq:corr_hom}
C(\xi,h,x) = x^{-(d-2+\eta)} g_{\pm}(x / \xi, h / \xi^{-w}) \,.
\end{eqnarray}
Now from the fluctuation-dissipation theorem, Eq.~(\ref{eq:fluc_rel})
\begin{eqnarray}
\label{eq:fluc_diss}
\chi = \frac{1}{T} \int d^dx \, C(x) \,,
\end{eqnarray}
we obtain a relation between $y$, $w$, and $\eta$ which leaves two independent exponents.
From Eqs.~(\ref{eq:phi_hom}) and (\ref{eq:corr_hom}) one can calculate the exponents $\alpha$, $\beta$, $\gamma$, $\delta$, and $\nu$ as they are defined in Eqs.~(\ref{eq:crit_exp1})-(\ref{eq:crit_exp3}), just re-expressing them in terms of $y$, $w$, and $\eta$.
One finds
\begin{eqnarray}
\label{eq:yw}
y = d \,, \; w = \beta \delta / \nu \,,
\end{eqnarray}
and
\begin{eqnarray}
\label{eq:rel_exp}
&& 2 - \alpha = 2 \beta + \gamma = d \nu \,,
\nonumber \\
&& \gamma = \beta (\delta - 1) = (2 - \eta) \nu \,.
\end{eqnarray}
These 4 relations between the 6 exponents (known as 'scaling laws') allow all static exponents to be expressed in terms of 2 independent ones, say, $\nu$ and $\eta$.
This follows directly from the homogeneity assumptions Eqs.~(\ref{eq:phi_hom}) and (\ref{eq:corr_hom}).
One now assumes that $\nu$ and $\eta$ depend only on the order parameter dimension $n$ and the space dimension $d$, as suggested by experimental data.
This is known as universality, namely that within a universality class, defined by $d$ and $n$, the exponents are the same:\\
$n =1$, $d =3 :$ liquid-gas critical point = uniaxial magnet (Ising model);\\
$n =2$, $d =3 :$ superfluid = planar magnet;\\
$n =3$, $d =3 :$ isotropic magnet (ferro- and antiferromagnet).\\
As explained in the next section the validity of the phenomenological theory turns out to be justified by the renormalization group.
Finally, let us consider the special case of a continuous symmetry, where in the mean-field theory one has for the correlation lengths $\xi_{\parallel} \ne \xi_{\perp}$ for $\tau <0$ and $n >1$.
In the scaling theory we have assumed a single $\xi$.
The simplest way to do this is to define the transverse correlation function in $d$ dimensions in terms of the Fourier transform of Eq.~(\ref{eq:chi_inv_pp_q}) as follows:
\begin{eqnarray}
\label{eq:corr_perpd}
C_{\perp}(x) = \frac{T \bar{\psi}^2}{\rho_s x^{d-2}} \sim
T \bar{\psi}^2 \left( \frac{\xi_{\perp}}{x} \right)^{d-2} \,,
\end{eqnarray}
which also defines $\xi_{\perp} \propto \xi_{\parallel}$ and thus $\nu_{\parallel} = \nu_{\perp}$.
It implies that
\begin{eqnarray}
\label{eq:rho_xi_scale}
\xi_{\perp}^{2-d} \sim \rho_s \,.
\end{eqnarray}
In $d =3$ Eq.~(\ref{eq:corr_perpd}) agrees with Eq.~(\ref{eq:C_perp}) and we have $\rho_s \sim \xi^{-1} \sim |\tau|^{\nu}$, a relation which is sometimes associated with the name of Josephson, although it was understood earlier.
%

\subsection{\label{sec:4B} Dynamics}
Is there a phenomenology for dynamics?
As we saw in Sec.~\ref{sec:3A} the simplest dynamics is relaxational, where for a non-conserved order parameter one has
\begin{eqnarray}
\label{eq:relax}
\partial_t \psi = -\Gamma_0 (\psi - \bar{\psi}) \,,
\end{eqnarray}
and $\Gamma_0 = \Lambda / \chi_{\psi}$ is the relaxation rate.
For a conserved order parameter the condition $\partial_t \int d^dx \, \psi = 0$ is achieved by $\Lambda \to -\lambda \nabla^2$ and
\begin{eqnarray}
\label{eq:relax_conserved}
\partial_t \psi = D \nabla^2 \psi \,,
\end{eqnarray}
where $D = \lambda / \chi_{\psi}$ is the diffusion constant.
In Fourier space one can write
\begin{eqnarray}
\label{eq:relax_fourier}
\partial_t \psi(k) = \Gamma(k) \psi(k) \,,
\end{eqnarray}
where $\Gamma(k) = \Gamma_0$ for a non-conserved order parameter and $\Gamma(k) = D k^2$ for a conserved order parameter, respectively.
We have seen that in mean-field theory different characteristic frequencies $\omega_{\psi}$, $\omega_m$ go to zero with different exponents for $\tau >0$, $\tau <0$, and with different exponents for different coupled densities.
Hydrodynamics is different for $\tau >0$ and $\tau <0$.
The first assumption of the phenomenological scaling theory is that because of the divergence of the correlation length $\xi$, the breakdown of hydrodynamics is controlled by $\xi$ alone in all modes.
We can discuss hydrodynamics by considering the time dependent correlation function for the order parameter
\begin{eqnarray}
\label{eq:corr_xt}
C(x,t) = \langle (\psi(x,t) - \bar{\psi})
(\psi(0,0) - \bar{\psi}) \rangle \,,
\end{eqnarray}
which can be Fourier transformed to get $C(k,\omega)$, whose time dependence is controlled by modes $\omega(k)$.
$C(k,\omega)$ is characterized by either decay or propagation for different modes.
Similar definitions apply for the conserved densities entering the hydrodynamics.
The second assumption of the phenomenological theory is the homogeneity of characteristic frequencies $\omega = \omega(k,\xi)$ whose form depends on the dynamic universality class defined by the hydrodynamics [dynamic scaling, Halperin and Hohenberg (1967)].
%

%
%
\begin{figure}[ht]
\includegraphics[width=7cm]{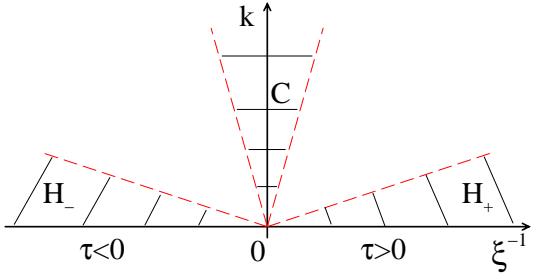}
\caption{\label{fig:hydrocorr}
Hydrodynamic regimes H$_{+}$, H$_{-}$, and critical dynamics C shown on a plot of the wave vector $k$ vs. the temperature $T$ measured by $\xi^{-1}$, with the critical point at $\xi^{-1} =0$.
$\xi_{+} \sim \tau^{\nu}$, $\xi_{-} \sim |\tau|^{\nu}$.}
\end{figure}
In Fig.~\ref{fig:hydrocorr} a schematic diagram of the hydrodynamic regimes is shown.
In the region H$_{+}$, $k \xi \ll 1$, we have hydrodynamics for $\tau >0$.
In the region H$_{-}$, $k \xi \ll 1$, we have hydrodynamics for $\tau <0$.
In the region C, $\tau \approx 0$, $k \xi \gg 1$, we have critical dynamics and no hydrodynamic laws.
The third assumption of the phenomenological theory is that near $T_c$ the link between regimes is also controlled by the correlation length $\xi$.
Thus, for the characteristic frequency of the order parameter, for example, one assumes a homogeneous function
\begin{eqnarray}
\label{eq:omega_hom}
\omega_{\psi}(k,\xi) = k^z \Omega_{\pm}(k \xi) \,,
\end{eqnarray}
where $z$ is a new `dynamic' exponent.
Any density that couples to the order parameter has a characteristic frequency with a similar functional form and the same dynamic exponent $z$, but a different scaling function,
\begin{eqnarray}
\label{eq:omegam_hom}
\omega_m(k,\xi) = k^z \Omega_{\pm}^m(k \xi) \,.
\end{eqnarray}
Since at nonzero $k$, the frequency $\omega$ should remain finite at $T_c$, we have in the critical dynamics regime, $\omega \sim k^z$.
From these quite general assumptions one can already draw an important conclusion.
Since the dispersion relation of propagating hydrodynamic modes can be expressed entirely in terms of static (equilibrium) quantities, the dynamic exponent $z$ of Eq.~(\ref{eq:omega_hom}) is always exactly related to static exponents.
It is only in cases where the order parameter relaxes that new dynamic exponents appear, relating to kinetic and transport coefficients.
Consider relaxational dynamics for a non-conserved order parameter, where
\begin{eqnarray}
\label{eq:omega_ncon1}
\omega_{\psi} = i \Gamma_0 = i \frac{\Lambda}{\chi_{\psi}}
\sim \frac{\xi^{x_{\lambda}}}{\xi^{2-\eta}} \,,
\end{eqnarray}
where we have introduced $x_{\lambda}$ for the scaling of $\Lambda$.
According to the dynamic scaling assumption Eq.~(\ref{eq:omega_hom}) one can also write
\begin{eqnarray}
\label{eq:omega_ncon2}
\omega_{\psi} \sim k^z (k \xi)^{-z} \,,
\end{eqnarray}
which gives for the dynamic exponent
\begin{eqnarray}
\label{eq:z_noncon}
z = 2 - \eta - x_{\lambda} \,.
\end{eqnarray}
In the case of a conserved order parameter one has
\begin{eqnarray}
\label{eq:omega_con1}
\omega_{\psi} = i D k^2 = i \frac{\lambda}{\chi_{\psi}} k^2
\sim \frac{\xi^{x_{\lambda}}}{\xi^{2-\eta}} k^2 \,,
\end{eqnarray}
which can be written in the form of a homogeneous function
\begin{eqnarray}
\label{eq:omega_con2}
\omega_{\psi} \sim k^z (k \xi)^{-(2 - \eta - x_{\lambda})} \,,
\end{eqnarray}
yielding for the dynamic exponent
\begin{eqnarray}
\label{eq:z_con}
z = 4 - \eta - x_{\lambda} \,.
\end{eqnarray}
%

\subsubsection{\label{sec:4B1} Planar magnet}
Consider now the planar magnet where the non-conserved order parameter is coupled to a conserved density.
In the region H$_{+}$ ($\tau >0$, $k \xi \ll 1$, see Fig.~\ref{fig:hydrocorr}) the dynamics of $\psi$ is relaxational, $\omega_{\psi} \sim i \Gamma_0$ decays and the dynamic exponent is given by Eq.~(\ref{eq:z_noncon}).
For the conserved density $m_z$ the frequency is given by
\begin{eqnarray}
\label{eq:omegam_neg}
\omega_m = i \frac{\lambda_m}{\chi_m} k^2
\sim \frac{\xi^{x_{\lambda_m}}}{\xi^{0}} k^2
= k^{2 - x_{\lambda_m}} (k \xi)^{x_{\lambda_m}}\,,
\end{eqnarray}
which results in the dynamic exponent
\begin{eqnarray}
\label{eq:zm_neg}
z = 2 - x_{\lambda_m} \,.
\end{eqnarray}
In the region H$_{-}$ ($\tau <0$, $k \xi \ll 1$) one has propagating modes for $\psi$ and $m_z$ with frequencies $\omega_{\psi} = \omega_m = \pm c_s k$ where $c_s^2 \sim \rho_s / \chi_m$.
According to Eq.~(\ref{eq:rho_xi_scale}) one has $\rho_s \sim \xi^{2 - d}$ with $d =3$ and taking into account that $\chi_m \sim \xi^0$ one finds for the frequency scaling
\begin{eqnarray}
\label{eq:omega_pos}
\omega_{\psi} = \omega_m \sim \xi^{-1/2} k = k^{3/2} (k \xi)^{-1/2} \,,
\end{eqnarray}
with dynamic exponent
\begin{eqnarray}
\label{eq:z_pos}
z = \frac{3}{2} \,.
\end{eqnarray}

Now we assume that since the $\omega_{\psi}$ and $\omega_m$ modes agree for $\tau <0$, the same dynamic scaling assumption (with the same exponent $z$) holds for $\omega_m$ for $\tau >0$.
Then we have
\begin{eqnarray}
\label{eq:xlm}
x_{\lambda_m} = 2 - z = \frac{1}{2} \,.
\end{eqnarray}
%

\subsubsection{\label{sec:4B2} Pure fluid}
For fluids the order parameter does not have propagating modes so the dynamic exponent $z$ is not related to static exponents.
One does, however predict the $k$-dependence of $\omega_{\psi}(k,\xi)$, which can be checked by inelastic light scattering (Rayleigh scattering) to extract the dynamic exponent $z$ (Swinney and Henry).
%

\subsubsection{\label{sec:4B3} Isotropic magnets}
The isotropic antiferromagnet can be mapped to the planar magnet case, for which $z =3/2$.
This can be verified by measurements of $\omega_{\psi}(k, \xi)$ by neutron scattering.
In the case of ferromagnets ($n =3$, $d =3$) one has for $\tau <0$ propagating spin waves with $\omega_{\psi} = \pm b k^2$ where $b = \rho_s/\bar{\psi}$ and
\begin{eqnarray}
\label{eq:omega_ferro}
\omega_{\psi} \sim \frac{\rho_s}{\bar{\psi}} k^2
\sim \frac{\xi^{-1}}{\xi^{-\beta / \nu}} k^2
= k^z (k \xi)^{\beta / \nu - 1} \,,
\end{eqnarray}
which gives for the dynamic exponent
\begin{eqnarray}
\label{eq:z_ferro}
z = 3 - \frac{\beta}{\nu} \,.
\end{eqnarray}
Taking into account the static critical exponents $\beta$, $\nu$ for isotropic ferromagnets one finds $z \approx 5/2$.
In the disordered phase, $\tau >0$, the dynamic mode is given by $\omega_{\psi} = i D k^2$ and similar to Eqs.~(\ref{eq:omega_con1})-(\ref{eq:omega_con2}) the dynamic exponent  is given by Eq.~(\ref{eq:z_con}), with $z$ determined by Eq.~(\ref{eq:z_ferro}), yielding $x_{\lambda} = 1 - \eta + \beta/\nu \approx 3/2 - \eta$.
These predictions have also been confirmed experimentally.
%

\subsubsection{\label{sec:4B4} Superfluid}
The case of helium in pores is analogous to the planar magnet (Sec.~\ref{sec:4B1}) and the dynamic exponent is $z =3/2$, yielding  $x_{\lambda_m} =1/2$.
For pure helium the specific heat singularity enters and one has the following scaling:
\begin{eqnarray}
\label{eq:chiCp}
\chi_m \sim C_p \sim \xi^{\alpha / \nu} \,.
\end{eqnarray}
For $\tau >0$ one has
\begin{eqnarray}
\label{eq:omega_s1}
\omega_m \sim \omega_s \sim \frac{\lambda}{C_p} k^2 \,,
\end{eqnarray}
and for $\tau <0$ (propagating modes) we find
\begin{eqnarray}
\label{eq:omega_s2}
\omega_s = \omega_{\psi} = \pm c_s k \,.
\end{eqnarray}
Assuming again the same dynamic exponent for $\omega_{\psi}$ for $\tau >0$ as well, one finds
\begin{eqnarray}
\label{eq:z_helium}
&& z = \frac{3}{2} + \frac{\alpha}{2 \nu} \;,
\nonumber \\
&& x_{\lambda} = \frac{1}{2} + \frac{\alpha}{2 \nu} \;.
\end{eqnarray}
In this way the dynamic exponent $z$ is evaluated in terms of static exponents, yielding the dramatic prediction by Ferrell et al. and by Halperin and Hohenberg in 1967 for the divergence of the thermal conductivity at the superfluid transition.
This prediction was verified experimentally by Ahlers in 1968.
To summarize, the Landau or mean-field theory is universal in that all thermodynamic  properties (critical exponents) are the same in all systems.
The scaling theory assumes universality classes, i.e., that critical exponents and scaling functions are the same for all systems belonging to the same class, but different for different classes.
For static phenomena the classes depend on $d$ (dimension of space) and $n$ (dimension of the order parameter).
For dynamic phenomena the classes depend also on the form of the hydrodynamics.
Thus a single static class ($d, n$) splits up into different dynamic universality classes, depending on the form of the hydrodynamic modes.
We list below the principal dynamic universality classes, along with the corresponding Ginzburg-Landau model defined by \textcite{Hohenberg:1977}.
\begin{tabular}{lll}
$n =1$: & Relaxation: & non-conserved $\psi$ (model A) \\
 & Diffusion: & conserved $\psi$ (model B) \\
 & Fluid: & conserved $\psi$ coupled to \\
 & & conserved transverse \\
 & & current $\bm j_T$ (model H) \\ \\
$n =2$: & Relaxation: & non-conserved $\psi$ (model A) \\
 & Diffusion: & conserved $\psi$ (model B) \\
 & Planar magnet, & $h_z =0$, $\chi_m \sim const$ (model E) \\
 & Helium in pores: & $z =\frac{3}{2}$ (model E) \\
 & Planar magnet, & $h_z \ne 0$, $\chi_m \sim \xi^{\alpha / \nu}$ (model F) \\
 & pure helium: &  $z =\frac{3}{2} + \frac{\alpha}{2 \nu}$, $x_{\lambda} =\frac{1}{2} + \frac{\alpha}{2 \nu}$ (model F) \\ \\
$n =3$: & Relaxation: & non-conserved $\psi$ (model A) \\
 & Diffusion: & conserved $\psi$ (model B) \\
 & Antiferromagnet: & $z =\frac{3}{2}$ (model G) \\
 & Ferromagnet: & $z =3 - \frac{\beta}{\nu}$ (model J)
\end{tabular}
%

\section{\label{sec:5} Effects of thermal fluctuations: renormalization group}
The mean-field theory neglects the effects of thermal fluctuations on the thermodynamic functions, even though it predicts divergent fluctuations via the correlation function $C(x)$ and response $\chi(k)$ as $\tau, h \to 0$.
It is thus not self-consistent.
However the Ginzburg-Landau theory can be used to determine the domain of validity (self-consistency) of mean-field theory, and also to calculate the corrections to mean-field theory.
For this it is sufficient to take into account the effects of thermal noise.
%

\subsection{\label{sec:5A} The `Ginzburg-Landau-Wilson' model}
For illustration, let us consider the Ising model on a lattice as a starting point for a microscopic description over the whole range of scales $\ell_0 < l < L$.
The Hamiltonian is given by [see Eq.~(\ref{eq:H_Ising})]
\begin{eqnarray}
\label{eq:Ising}
{\cal H} = -J \sum_{\langle i, j \rangle} S_i S_j \,,
\end{eqnarray}
where $\langle i, j \rangle$ means the sum over nearest neighbors, $S_i =\pm 1$ are classical spins and the lattice spacing is $\ell_0$.
The Gibbs free energy and the partition function are
\begin{eqnarray}
\label{eq:Gibbs}
&& \Omega = -T \ln Z \,,
\\
\label{eq:Z_Gibbs}
&& Z = \sum_{\{S_i\}} \exp\left[ -{\cal H}/T \right] \,,
\end{eqnarray}
where the sum in Eq.~(\ref{eq:Z_Gibbs}) signifies a sum over all configurations of the $S_i$ on the lattice.
Define the Fourier transform
\begin{eqnarray}
\label{eq:spins}
S_k = \sum_{i} S_i e^{-i k x_i} \,, \;
0 \le k \le \ell_0^{-1} \,,
\end{eqnarray}
and take the system volume to be $V = L^{d}$.
Then the partition function $Z$ can be rewritten in terms of $S_k$ as
\begin{eqnarray}
\label{eq:Z}
Z = \int\limits_{L^{-1} < k < \ell_0^{-1}} {\cal D}S_k \, \exp\left\{ -{\cal H}[S_k]/T \right\} \,,
\end{eqnarray}
where ${\cal D}S_k \equiv d^d S_{k_1} d^d S_{k_2} \dots d^d S_{k_n}$, with $k_1 =L^{-1}$ and $k_n =\ell_0^{-1}$, i.e., we have discretized the modes for clarity.
In the thermodynamic (continuum) limit ($L \to \infty$), $k_1 \to 0$ and the number of modes diverges.
We can divide the integral in Eq.~(\ref{eq:Z}) into two parts: 
$L^{-1} < k < k_0$ and $k_0 < k < \ell_0^{-1}$, where we have introduced the `mesoscale' wave vector $k_0 =\xi_0^{-1}$.
Then for the partition function we can write
\begin{eqnarray}
\label{eq:Z2}
Z = \int\limits_{L^{-1} < k < k_0} {\cal D}S_k \, \exp\left\{ -\Phi[S_k] \right\} \,,
\end{eqnarray}
with the definition
\begin{eqnarray}
\label{eq:exp_Phi}
\exp\left\{ -\Phi[S_k] \right\} \equiv \int\limits_{k_0 < p < \ell_0^{-1}} 
{\cal D}S_p \exp\left\{ -{\cal H}(S_k, S_p)/T \right\} \,. \qquad
\end{eqnarray}
For $L^{-1} < k < k_0$ we define $\psi_k = S_k$, and going back to $x$ (inverse Fourier transform), we have
\begin{eqnarray}
\label{eq:psi_x}
\psi(x) = \sum_{L^{-1} < k < k_0} \psi_k e^{i k x} \,.
\end{eqnarray}
The field $\psi(x)$ thus represents not the full spin but a `coarse-grained spin', since only the modes $L^{-1} < k < k_0$ are taken into account in Eq.~(\ref{eq:psi_x}).
Now $\Phi[S_k]$ becomes a functional of $\psi(x)$
\begin{eqnarray}
\label{eq:Phi_psi}
\Phi[\psi(x)] &&=
\int\limits_{k_0^{-1} < x < L} d^dx \left[
a |\psi|^2 + b |\psi|^4 + \dots 
+ c |\nabla \psi|^2 + \dots
\right]
\nonumber \\
&&\equiv \int\limits_{k_0^{-1} < x < L} d^dx \phi[\psi(x)] \,.
\end{eqnarray}
The free energy Eq.~(\ref{eq:Phi_psi}) is referred to as the Ginzburg-Landau-Wilson model.
It is related to the exact partition function by Eq.~(\ref{eq:Z2}) and its general form has in principle an infinite number of terms.
It was popularized in the west by Wilson in 1968-1972, but it was first introduced by Landau as part of a general formulation of critical phenomena in 1958 [see footnote in Sec.~147 in \textcite{StatPhys:1994}, and \textcite{Patashinskii:1964}].
The mean-field theory corresponds to a saddle-point (or steepest descent) approximation of the functional integral in Eq.~(\ref{eq:Z2}), i.e, to the `stationary phase' condition 
\begin{eqnarray}
\frac{\delta \Phi}{\delta \psi} = 0 \,.
\end{eqnarray}
We now wish to study the fluctuation corrections to mean-field theory.
%

\subsection{\label{sec:5B} Effects of fluctuations: the Levanyuk-Ginzburg criterion}
It is important to test the self-consistency of the mean-field theory and of the Ginzburg-Landau expansion to see where they might break down.
This was first done by \textcite{Levanyuk:1959} but it was reformulated by \textcite{Ginzburg:1960} and it is often referred to as the Ginzburg criterion.
We shall refer to it as the `Levanyuk-Ginzburg criterion'.
As mentioned above, we can use the Ginzburg-Landau theory to estimate the fluctuations approximately from the correlation function in mean-field theory.
For self-consistency we require the fluctuations of the order parameter over a volume $v =\xi^d$ to be less than the average value of the order parameter over that volume
\begin{eqnarray}
\label{eq:criterion}
\langle (\psi(x) - \bar{\psi})^2 \rangle_v = 
\langle (\Delta \psi)^2 \rangle_v 
\ll \langle \psi \rangle_v^2 = \bar{\psi}^2 \,.
\end{eqnarray}
Let us evaluate the fluctuations $\langle (\Delta \psi)^2 \rangle_v$ for $\tau >0$ and assume that the answer is comparable for $\tau <0$ when expressed in terms of $\xi$.
From Eq.~(\ref{eq:corr_x}) we have in three dimensions
\begin{eqnarray}
\label{eq:corr_Gc}
C(x) = \langle (\psi(x) - \bar{\psi})(\psi(0) - \bar{\psi}) \rangle =
\frac{T}{8\pi c x} e^{-x/\xi} \,, \qquad
\end{eqnarray}
and then
\begin{eqnarray}
\label{eq:Dpsi}
\langle (\Delta \psi)^2 \rangle_v = 
\frac{1}{v} \int\limits_{x < \xi} d^3x \, C(x) \approx 
\frac{T}{c \xi} \,.
\end{eqnarray}
For the average value of the order parameter one has
\begin{eqnarray}
\label{eq:barpsi2}
\bar{\psi}^2 = \frac{a}{2 b} \,,
\end{eqnarray}
and Eq.~(\ref{eq:criterion}) takes the form
\begin{eqnarray}
\label{eq:criterion2}
\frac{T}{c \xi} \ll \frac{a}{2 b} \,.
\end{eqnarray}
Taking into account $\xi^2 = c/a$ for $\tau >0$ [see Eq.~(\ref{eq:xi2})] and $a = a_0 \tau$ we can rewrite Eq.~(\ref{eq:criterion2}), expressing the validity of mean-field theory in the vicinity of $T_c$ as 
\begin{eqnarray}
\label{eq:Ginzcrit}
\tau \gg \frac{4 T_c b^2}{a_0 c^3} = \tau_{LG} \,,
\end{eqnarray}
where $\tau_{LG}$ denotes 'Levanyuk-Ginzburg' (not Landau-Ginzburg!).
In $d$-dimensions we have
\begin{eqnarray}
\label{eq:Dpsi_d}
\langle (\Delta \psi)^2 \rangle_v = 
\xi^{-d} \int d^dx \, \frac{e^{-x / \xi}}{c x^{d - 2}} \approx 
\frac{T}{c \xi^{d - 2}} \,,
\end{eqnarray}
and Eq.~(\ref{eq:Ginzcrit}) becomes 
\begin{eqnarray}
\label{eq:Ginzcrit_d}
\tau \gg \left( \frac{4 T_c b^2}{a_0^{4-d} c^d} \right)^{1/(4 - d)} =
\tau_{LG} \,,
\end{eqnarray}
or
\begin{eqnarray}
\label{eq:Ginzcrit_d2}
\tau^{4 - d} \gg \frac{4 T_c b^2}{a_0^{4-d} c^d} = \tau_{LG}^{4-d} \,.
\end{eqnarray}
For dimensions $d >4$ one has $\tau^{4 - d} \to \infty$ as $\tau \to 0$ and the Levanyuk-Ginzburg criterion is satisfied as $T \to T_c$.
For dimensions $d <4$ the Levanyuk-Ginzburg criterion breaks down at $|\tau| \approx \tau_{LG}$.
The case of $d =4$ is marginal or border line.
In the case of long-range forces $c \sim R_0^2$, where $R_0 \gg \ell_0$ is the range of the forces.
Then one has in $d$-dimensions
\begin{eqnarray}
\label{eq:longrf}
\tau_{LG} = \left( \frac{4 T_c b^2}{a_0^{4-d} c^d} \right)^{1/(4 - d)} \propto 
\left( \frac{1}{R_0} \right)^{2 d/(4 - d)} \;.
\end{eqnarray}
If $d <4$ then $\tau_{LG} \to 0$ for $R_0 \to \infty$ and the Levanyuk-Ginzburg criterion is satisfied closer and closer to $T_c$ as $R_0$ grows.
For superconductors one has $\xi = \xi_0 \tau^{-1/2}$, where $\xi_0$ is the pair size and $\xi_0 k_F \sim (E_F/T_c)$.
Then one has
\begin{eqnarray}
\label{eq:tauLG_supercond}
\tau_{LG} \sim \left( \frac{T_c}{E_F} \right)^{2(d - 1)/(4 - d)} \,.
\end{eqnarray}
Typically for superconductors $E_F/T_c \sim 10^3 - 10^4$ and in three dimensions one has
\begin{eqnarray}
\label{eq:tauLG_supercond2}
\tau_{LG} \sim \left( \frac{T_c}{E_F} \right)^{4} \sim 10^{-14} \ll 1 \,.
\end{eqnarray}
Thus the Levanyuk-Ginzburg criterion (as well as the Ginzburg-Landau theory) is satisfied up to very small $|\tau|$ close to $T_c$.
Note that in high-$T_c$ superconductors the ratio $E_F/T_c \sim 1 - 10$ is not large, so fluctuations become important.
%

\subsection{\label{sec:5C} Static critical phenomena: dimensional analysis}
Let us carry out dimensional analysis of the general Ginzburg-Landau-Wilson model.
The free energy functional in $d$ dimensions is given by 
\begin{eqnarray}
\label{eq:GLW_fe}
&& \Phi[\psi] = \int d^dx \, \phi[\psi] \,,
\nonumber \\
&& \phi[\psi] = a |\psi|^2 + b |\psi|^4 + c |\nabla \psi|^2 
- h \psi + \dots \,.
\end{eqnarray}
How do the different terms in $\phi[\psi]$ scale?
We introduce the following notation for the scaling dimension: if some quantity $A$ scales as $A \sim l^{-d_A}$ we define the dimension of $A$ as $[A] = d_A$.
Assume now that the total free energy $\Phi$ has no scale, i.e., $\Phi \sim l^0$ and $[\Phi] = 0$.
This means that the free energy density $\phi$ scales as $\phi \sim l^{-d}$ and $[\phi] = d$.
Let us first determine the 'naive dimensions' applicable to mean-field theory, based on the assumption that each term in the Landau expansion Eq.~(\ref{eq:GLW_fe}) has the same dimension.
We have some freedom in the definition of the dimension of $\psi$ and to fix it we choose the dimension of the coefficient of the square gradient term $[c]$ to be zero.
With these conventions we can find the dimension of $\psi$ by looking at the square gradient term in Eq.~(\ref{eq:GLW_fe})
\begin{eqnarray}
\label{eq:dimsgt}
[c |\nabla \psi|^2] = 0 + 2 + 2 d_{\psi} = [\phi] = d \,,
\end{eqnarray}
since $[\nabla] =1$, and thus the dimension of $\psi$ is
\begin{eqnarray}
\label{eq:dimpsi}
[\psi] = d_{\psi} = \frac{d - 2}{2} \,.
\end{eqnarray}
Similarly we can find the dimensions of $h$, $a$, and $b$ from the assumption that the terms in Eq.~(\ref{eq:GLW_fe}) all scale in the same way:
\begin{eqnarray}
\label{eq:dimhab}
&& [h] = d_h = \frac{d + 2}{2} \,,
\nonumber \\
&& [a] = d_a = 2 \,,
\nonumber \\
&& [b] = d_b = 4 - d \,.
\end{eqnarray}
For the dimension of $\chi$ one has
\begin{eqnarray}
\label{eq:dimchi}
[\chi] = d_{\chi} = d_{\psi} - d_h = -2 \,.
\end{eqnarray}
Equations (\ref{eq:dimpsi})-(\ref{eq:dimchi}) yield what we call the naive dimensions.
In the critical regime, on the other hand, we will assume phenomenological scaling (Sec.~\ref{sec:4A}).
All dimensions are supposed to be controlled by the correlation length $\xi$.
We want to know the scaling dimensions, also known as 'anomalous dimensions', of the various quantities, determined by their dependence on $\xi$.
The quantity $a$ scales as $a \sim a_0 \tau \sim \xi^{-1/\nu}$, so $[\tau] =\nu^{-1}$.
The dimension of $\psi$ follows from Eq.~(\ref{eq:corr_hom}), since $C(x) \sim \psi^2$ so
\begin{eqnarray}
\label{eq:d_psi}
2 d_{\psi} = d - 2 + \eta \,.
\end{eqnarray}
Similarly, from Eq.~(\ref{eq:corr_hom}) we see that $h$ scales as $\xi^{-w}$ so $d_h = w = (d + 2 - \eta)/2$, and from Eq.~(\ref{eq:dimchi}) we obtain $d_{\chi} = \eta - 2$.
The naive and anomalous dimensions are summarized in Table~\ref{tab:dim}.
\begin{table}[htb]
\caption{\label{tab:dim}
Comparison between naive and anomalous dimensions.}
\begin{ruledtabular}
\begin{tabular}{c|c|c}
Quantity & Naive dimension & Anomalous dimension \\ \hline
$\Phi$ & $0$ & $0$ \\
$\phi$ & $d$ & $d$ \\
$c$ & $0$ & $0$ \\
$\psi$ & $\frac{d - 2}{2}$ & $\frac{d - 2 + \eta}{2}$ \\
$a \sim \tau$ & 2 & $\nu^{-1}$ \\
$h$ & $\frac{d + 2}{2}$ & $\frac{d + 2 - \eta}{2}$ \\
$\chi$ & $-2$ & $\eta - 2$ \\
$\xi$ & $-1$ & $-1$ \\
$b$ & $4 - d$ & ? \\
\end{tabular}
\end{ruledtabular}
\end{table}
The renormalization group provides a calculation or a schema for understanding these anomalous dimensions.
%

\subsection{\label{sec:5D} The renormalization group: statics}
Let us now describe the renormalization group transformation which explains how the phenomenological scaling theory emerges near the critical point.
To see how this comes about we start from the general Ginzburg-Landau-Wilson free energy, as defined by the partition function given in Eq.~(\ref{eq:Z2}) which we rewrite as 
\begin{eqnarray}
\label{eq:Z_psi}
Z = \int\limits_{0 < k < k_0} {\cal D}\psi_k \exp\left\{ -\Phi[\psi_k] \right\} \,,
\end{eqnarray}
with a free energy density $\phi$, Eq.~(\ref{eq:Phi_psi}) in the general form
\begin{eqnarray}
\label{eq:phi_psi}
\phi[\psi] = \sum_i \mu_i O_i \,, \;
O_{n m} = |\psi|^n |\nabla \psi|^m \,.
\end{eqnarray}
In Eq.~(\ref{eq:phi_psi}) we have introduced the generalized fields $\mu_i = \mu_{m n}$.
We want to study the renormalization group, which is a transformation of the free energy density ${\cal R}[\phi] = \phi'$, defined as follows:\\
(i) Integrate out wave numbers in the momentum shell $k_0/s < k < k_0$ in Eq.~(\ref{eq:Z_psi}), with $s >1$.\\
(ii) Change the length scale so that $k_0/s \to k_0$, i.e., for the length $l \to l/s$.\\
(iii) Renormalize the order parameter as $\psi \to s^{d_{\psi}} \psi $.
Then the partition function has once more the form Eq.~(\ref{eq:phi_psi}), but with $\phi \to \phi'$ and
\begin{eqnarray}
\label{eq:phi_prime}
\phi' = \sum_i \mu_i' O_i \,.
\end{eqnarray}
In other words, one can write ${\cal R}$ as a transformation of the fields $\mu_i$, ${\cal R}[\phi] \equiv {\cal R}[\{ \mu_i \}]$, because $\phi$ is entirely defined by these fields: 
\begin{eqnarray}
\label{eq:rt}
{\cal R}[\phi] \equiv {\cal R}[\{ \mu_i \}] = \phi' \equiv \{ \mu_i' \} \,.
\end{eqnarray}
We can thus consider the renormalization group to be a transformation of the huge vector $\{ \mu_i \}$ to $\{ \mu_i' \}$,
\begin{eqnarray}
\label{eq:rg}
{\cal R} : \{ \mu_i \} \to \{ \mu_i' \} \,,
\end{eqnarray}
which is a highly nonlinear and a very complicated function, e.g., $\mu_1' = M_1(\mu_1, \mu_2, \dots , \mu_n)$ and so on.
%

%
%
\begin{figure}[ht]
\includegraphics[width=7.5cm]{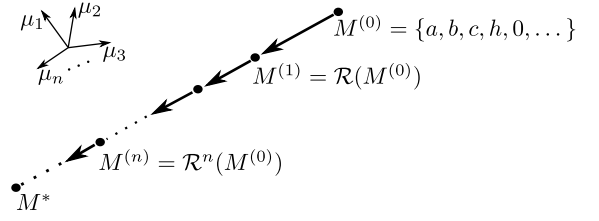}
\caption{\label{fig:muspace}
Representation of the renormalization group in $\mu$-space.
The transformation ${\cal R}[\{\mu_i\}]$ is represented as ${\cal R}(M^{(n)}) =M^{(n+1)}$.}
\end{figure}
We can consider $\bm M$ to be a vector in an $n$-dimensional $\mu$-space of fields $\{ \mu_i \}$ with $n \to \infty$.
Thus each $\bm M$ is a point in $\mu$-space that corresponds to some free energy density $\phi$ and therefore to some free energy $\Phi$.
The transformation ${\cal R}$ can be thought of as a trajectory in $\mu$-space.
The topology of $\mu$-space is shown schematically in Fig.~\ref{fig:muspace}.
We start with some point that we call $\bm M^{(0)}$ which is the original Ginzburg-Landau free energy Eq.~(\ref{eq:phi_psi}).
Applying the transformation ${\cal R}(\bm M^{(0)})$ one arrives at another point $\bm M^{(1)}$.
Applying the transformation ${\cal R}(\bm M^{(1)})$ once again one arrives at the point $\bm M^{(2)}$ and so on.
Thus we have ${\cal R}^n(\bm M^{(0)}) = \bm M^{(n)}$ and ${\cal R}$ has a group property ${\cal R}^{n+m} = {\cal R}^n {\cal R}^m$, whence the name `renormalization group'.
It is actually not a group but only a semi-group because the transformation is not reversible.
For further information on the renormalization group see the textbooks by \textcite{Pfeuty:1977} and by \textcite{Goldenfeld:1992}.
We now state the so-called `Wilson conjectures' for the behavior of the renormalization group transformation near a continuous transition.
RG~1: There exists a fixed point $M^{*}$ or $\phi^{*}$ defined by  $\lim_{n \to \infty} {\cal R}^n(\bm M) = \bm M^{*}$ or $\lim_{n \to \infty} {\cal R}^n(\phi) = \phi^{*}$.
RG~2: For $\phi$ near the fixed point $\phi^{*}$ one can {\em linearize} the transformation ${\cal R}$, i.e., one can represent the very complicated nonlinear function $\{ \mu_i' \} = {\cal M}(\{ \mu_i \})$ by a linear function.
Let us write for $\phi$ near $\phi^{*}$
\begin{eqnarray}
\label{eq:diffphi}
\phi - \phi^{*} = \sum_{i} \mu_i O_i \;,
\end{eqnarray}
and apply the transformation ${\cal R}$ to it
\begin{eqnarray}
\label{eq:Rdiffphi}
{\cal R}(\phi - \phi^{*}) = \phi' - \phi^{*} 
= \sum_{i j} A_{i j} \mu_j O_i \,,
\end{eqnarray}
which yields linear relations $\mu_i' = \sum_{j} A_{i j} \mu_j$ via the matrix $A_{i j}$.
We can diagonalize this matrix and introduce eigenvalues $\Lambda_i$ (corresponding to `eigenfields' $\{ g_i \}$) and eigenfunctions $\tilde{O}_i$ (`eigenoperators')
\begin{eqnarray}
\label{eq:eigen}
{\cal R}(\tilde{O}_i) = \Lambda_i \tilde{O}_i \,.
\end{eqnarray}
Then the transformation can be rewritten as
\begin{eqnarray}
\label{eq:Rdiffphi2}
&& \phi - \phi^{*} = \sum_{i} g_i \tilde{O}_i \;,
\nonumber \\
&& {\cal R}(\phi - \phi^{*}) = \phi' - \phi^{*} 
= \sum_i g_i \Lambda_i \tilde{O}_i 
= \sum_i g_i' \tilde{O}_i \,. \qquad
\end{eqnarray}
Thus near the fixed point one has eigenfields and eigenoperators and the transformation is linear.
Let us write $\Lambda_i = s^{\lambda_i}$ where $s >1$ is the scale chosen in the transformation steps.
If $\Lambda_i >1$, i.e., $\lambda_i >0$, every time the transformation is repeated, $g_i' = s^{\lambda_i} g_i$, the corresponding $g_i$ grows near the fixed point.
Such a $g_i$ is called a {\em relevant} field.
If $\Lambda_i <1$, i.e., $\lambda_i <0$, one has $g_i \to 0$ when the transformation is repeated.
In this case $g_i$ is called an {\em irrelevant} field.
If $\lambda =1$, i.e., $\lambda_i =0$, the corresponding $g_i$ is called {\em marginal}.
The third Wilson conjecture is:
RG~3: There are only {\em two} relevant fields (and two relevant operators), namely, $g_1 =h$ and $g_2 =a \propto \tau$ with the positive exponents $\lambda_1$ and $\lambda_2$.
All other fields scale to zero.
The corresponding relevant operators are $\tilde{O}_1 =\psi$ and $\tilde{O}_2 =|\psi|^2$.
This assumption is necessary from the very definition of a critical point.
Finally we have: 
RG~4: A universality class is defined by its fixed point.
All systems that flow to the same fixed point have the same exponents and belong to the same universality class.
The consequences of these renormalization group conjectures are the following:
According to the definitions of the transformation we have 
\begin{eqnarray}
\label{eq:phi_g}
\phi'(g_i) = \phi(g_i') \,.
\end{eqnarray}
Each time one renormalizes $\phi$ (whose dimension is $d$) by a factor $s$, one gets
\begin{eqnarray}
\label{eq:phi_ren}
\phi'(g_i) = s^{d} \phi(g_i) \,,
\end{eqnarray}
and therefore
\begin{eqnarray}
\label{eq:phi_ren2}
\phi(g_i) = s^{-d} \phi'(g_i) = s^{-d} \phi(g_i') = s^{-d} \phi(\Lambda_i g_i) \,, \qquad
\end{eqnarray}
so that
\begin{eqnarray}
\label{eq:phi_scale}
\phi(g_i) = s^{-d} \phi(s^{\lambda_i} g_i) \,.
\end{eqnarray}
This is the scaling relation which follows from the linearization of ${\cal R}$ close to the fixed point.
For most fields $g_i$ the corresponding $\lambda_i$ is negative and such $g_i$ are irrelevant.
By our assumption, as one goes near the fixed point there are only two relevant fields, $g_1 =h$ and $g_2 = a =a_0 \tau$.
Let us write $s = \xi$ and $\xi^{\lambda_2} g_2 =1$.
Then $\tau \sim \xi^{-\lambda_2}$ with $\lambda_2 = 1 / \nu$.
Near the fixed point Eq.~(\ref{eq:phi_scale}) can be rewritten as
\begin{eqnarray}
\label{eq:phi_scale2}
\phi(g_1, g_2) = \xi^{-d} \phi(\xi^{\lambda_1} g_1, \xi^{\lambda_2} g_2) \,,
\end{eqnarray}
and thus
\begin{eqnarray}
\label{eq:phi_scale3}
\phi(h, \tau) = \xi^{-d} \phi(\xi^{\lambda_1} h, 1) 
= \xi^{-d} f_{\pm}(h/\xi^{-\lambda_1}) \,,
\end{eqnarray}
with the sign $\pm$ for positive and negative $\tau$, respectively, 
which is just the homogeneity relation Eq.~(\ref{eq:phi_hom}), and there are only two exponents $\lambda_1 =w$ and $\lambda_2 =1 / \nu$.
Similarly, one can show that the correlation function takes the form 
\begin{eqnarray}
\label{eq:corr_scale}
C(x, h, \tau) = \xi^{-2 d_{\psi}} g_{\pm}(x/\xi, h/\xi^{-w}) \,,
\end{eqnarray}
where $d_{\psi} = (d-2+\eta)/2$.
Finally and importantly, there are also {\em corrections to scaling}.
Let us call $g_3$ the irrelevant field with the smallest eigenvalue, which scales as $g_3 \sim \xi^{\lambda_3}$, with $\lambda_3 <0$ and $|\lambda_3|$ a minimum.
This field represents the dominant correction to scaling for $\xi \to \infty$.
Therefore one has for the scaling of $\phi$, linearizing with respect to $g_3 \to 0$, 
\begin{eqnarray}
\label{eq:phi_corr}
\phi(h, \tau, g_3) = \xi^{-d} f_{\pm}(h/\xi^{-w}) 
\left[ 1 + g_3 f_{corr} + \dots \right] \,. \qquad
\end{eqnarray}
For example for $h =0$ one has for the susceptibility
\begin{eqnarray}
\label{eq:chi_corr}
\chi &=& \Gamma |\tau|^{-\gamma} 
\left[ 1 + \Gamma_{corr} \xi^{\lambda_3} + \dots \right] 
\nonumber \\
&=& \Gamma |\tau|^{-\gamma} 
\left[ 1 + \Gamma_{corr}' |\tau|^{\Delta} + \dots \right] \,,
\end{eqnarray}
where $\Delta = -\lambda_3 \nu >0$.
If $\Delta <1$, the correction becomes singular and it will dominate the regular correction terms.
%

%
%
\begin{figure}[ht]
\includegraphics[width=7.5cm]{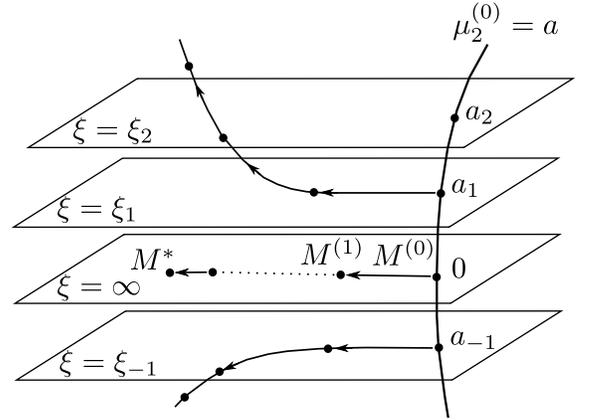}
\caption{\label{fig:muspace2}
Approaching the critical point in $\mu$-space.
The field $\mu_1^{(0)} =h$ has been set to zero.}
\end{figure}
As mentioned in the Introduction it is the great achievement of Wilson and others to have introduced the framework of renormalization group flows and fixed points to define equilibrium phases and transitions between them and to have demonstrated mathematically the mechanism for scaling and universality at the transition point.
In Fig.~\ref{fig:muspace2} a pictorial way of looking at the renormalization group in $\mu$-space is shown.
Let us take $\mu_1^{(0)} =h =0$ and consider in $\mu$-space the relevant field $\mu_2^{(0)} =a$.
The value $\mu_2^{(0)} =0$ corresponds to the critical point.
Let us draw a surface of constant $\xi$.
If it goes through $\mu_2^{(0)} =0$ (meaning $\tau =0$) then on that surface $\xi = \infty$.
As long as one stays on that surface and makes the transformation ${\cal R}$ with the length scale $s$, one will remain on that surface approaching the fixed point $\bm M^{*}$, since $\xi = \infty$ and multiplying by $s$ does not matter.
For the surface of constant $\xi$ that goes through some other $\mu_2^{(0)}$, say, $\mu_2^{(0)} <0$ (i.e., $\tau <0$) we have finite $\xi = \xi_{-1}$.
Then starting from that surface and making transformations, $\xi$ will be reduced at each step and one eventually goes out of the surface, away from the fixed point, to $T =0$.
Similarly if one starts above $T_c$ ($\mu_2^{(0)} >0$) and makes transformations, one goes eventually away from $T_c$, to $T \to \infty$.
%

%
%
\begin{figure}[ht]
\includegraphics[width=7cm]{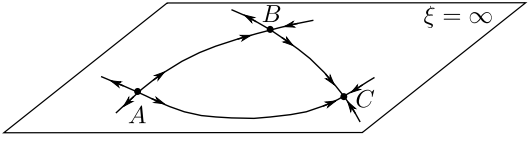}
\caption{\label{fig:muspace3}
Fixed points on the critical surface $\xi = \infty$ in $\mu$-space.
$A$ -- unstable fixed point, $B$ -- saddle point, $C$ -- stable fixed point.}
\end{figure}
On the critical surface $\xi = \infty$ one has the following picture (Fig.~\ref{fig:muspace3}).
There could be several fixed points, differing by the values of irrelevant fields.
These fixed points can be stable, unstable, and saddle node points with respect to trajectories on the critical surface, i.e., with respect to the irrelevant fields.
The only important fixed point is the one that remains stable on the critical surface.
Note that such points are always unstable with respect to the relevant fields $\mu_1 =h$ and $\mu_2 =a$ (meaning relevant directions away from the critical surface).
%

\subsection{\label{sec:5E} The $\epsilon$-expansion}
Another major achievement (Wilson and Fisher) is the $\epsilon$-expansion, which is an explicit perturbative calculation which justifies the renormalization group conjectures for spatial dimension $d$ sufficiently close to $4$.
Consider the partition function 
\begin{eqnarray}
\label{eq:Zk}
Z = \int\limits_{0 < k < k_0} {\cal D}\psi_k \exp\left\{ -\Phi[\psi_k] \right\} \,,
\end{eqnarray}
and assume that the free energy density is given by only the lowest-order terms in $\psi$
\begin{eqnarray}
\label{eq:phi_short}
\phi = a |\psi|^2 + c |\nabla \psi|^2 
= (a + c k^2) \psi_k \psi_{-k} \,.
\end{eqnarray}
Then the integral in Eq.~(\ref{eq:Zk}) is exactly solvable (each component of $k$ separates).
This is known as the Gaussian model.
The naive dimensions discussed in Sec.~\ref{sec:5C} are the scaling dimensions of this Gaussian model.
We have also seen that the dimension of $b$, the coefficient of $\psi^4$, is $d_b = 4 - d$ and for $d >4$ one has $b \to 0$ when one iterates the renormalization group.
But for $d <4$ one has $d_b >0$ and $b$ grows, so that the Gaussian model has large corrections.
The case of $b \ne 0$ is known as $\psi^4$ field theory.
In this case perturbation theory for $\phi$ has a diagrammatic form, where each element represents a certain integral in $k$-space.
For example the term $b |\psi|^4$ is represented by a $4$-vertex with strength $b$.
Note that the integrals have the form
\begin{eqnarray}
\label{eq:int_k}
&& \int d^dk_1 d^dk_2 \dots f(k_1,k_2,\dots) 
\nonumber \\
&& = \int k_1^{d-1} dk_1 f(\theta_1, \dots) 
  \int k_2^{d-1} dk_2 f(\theta_2, \dots) \dots \,, \qquad
\end{eqnarray}
and they formally depend on the spatial dimension $d$.
Wilson and Fisher proposed to make an analytic continuation of expressions such as Eq.~(\ref{eq:int_k}) from integer $d$ to continuous $d$.
They defined $\epsilon = 4-d$, which for continuous dimension $d$ can be arbitrary small, $\epsilon \ll 1$.
Then, when starting with small $b \sim \xi^{d_b}=\xi^{\epsilon}$, it remains small in the vicinity of the critical point ($\xi \to \infty$) for sufficiently small $\epsilon$.
Thus one can do perturbation theory (expansion in $b$ near $b =0$) for $\epsilon \ll 1$.
Although for fixed $\epsilon$ the perturbation expansion in $b \sim \xi^{\epsilon}$ eventually breaks down as $\xi \to \infty$, the scheme allows one to obtain a formal expansion of the eigenvalues (exponents) $\lambda_i$ as a power series in $\epsilon$, more precisely as an {\em asymptotic} expansion.
The coefficients of the Landau expansion $a(s)$, $b(s)$, etc. depend on $s$ as we iterate the renormalization group, where now the transformation factor can be written as $s = e^{l}$ with $l \to 0$ (infinitesimal transformations).
Then one can turn the transformation ${\cal R}$ into a set of differential equations, instead of discrete iterations of $s$,
\begin{eqnarray}
\label{eq:Rab}
&& \frac{d a}{d l} = 2 a + c_a b (1-a) + {\cal O}(b^2) \,,
\nonumber \\
&& \frac{d b}{d l} = \epsilon b - c_b b^2 + {\cal O}(b^3) \,,
\end{eqnarray}
with explicit expressions for $c_a$ and $c_b$ in terms of $\epsilon$ and $n$.
Let us now see if there is a self-consistent way of carrying out the renormalization group under the condition $\epsilon \ll 1$.
The fixed point is given by the condition that $a$ and $b$ should no longer vary:
\begin{eqnarray}
\label{eq:pert_fp}
\frac{d a}{d l} = \frac{d b}{d l} = 0 \,.
\end{eqnarray}
There are two fixed points: the Gaussian fixed point given by
\begin{eqnarray}
\label{eq:Gauss_fp}
a^{*} = b^{*} = 0 \;,
\end{eqnarray}
and the Wilson-Fisher fixed point
\begin{eqnarray}
\label{eq:WF_fp}
b^{*} = \epsilon / c_b \,, \;
a^{*} = -c_a b^{*} / (2 - c_a b^{*}) \,.
\end{eqnarray}
The question is, which one is stable?
Let us do a linear stability analysis of the fixed point of Eq.~(\ref{eq:Rab}), $(a^{*}, b^{*})$,
\begin{eqnarray}
\label{eq:fp_stab}
&& a = a^{*} + \delta a \,,
\nonumber \\
&& b = b^{*} + \delta b \,.
\end{eqnarray}
Linearizing Eq.~(\ref{eq:Rab}) one obtains
\begin{eqnarray}
\label{eq:dadb}
&& \frac{d \delta a}{d l} = 
2 \delta a + c_a [(1 - a^{*}) \delta b - b^{*} \delta a ] \,,
\nonumber \\
&& \frac{d \delta b}{d l} = 
\epsilon \delta b - 2 c_b b^{*} \delta b \,.
\end{eqnarray}
In the case $d >4$ one has $\epsilon <0$ and for the Gaussian fixed point, Eq.~(\ref{eq:Gauss_fp}), one finds 
\begin{eqnarray}
\label{eq:Gauss_fp_exp}
\delta b \sim e^{\epsilon l} \to 0 \,, \;
\delta a \sim e^{2 l} \,,
\end{eqnarray}
which means that the Gaussian fixed point has $\lambda_2 =2 = \nu^{-1}$ and $\lambda_3 =\epsilon <0$; it is stable on the critical surface ($a =0$).
For the Wilson-Fisher fixed point Eq.~(\ref{eq:WF_fp}), one finds
\begin{eqnarray}
\label{eq:WF_fp_exp}
\delta b \sim e^{-\epsilon l} \,,
\end{eqnarray}
which is unstable on the critical surface for $\epsilon <0$.
In the case $d <4$ one has $\epsilon >0$ and the Gaussian fixed point is unstable on the critical surface, whereas the Wilson-Fisher fixed point is stable (now $\lambda_3 =-\epsilon <0$).
For the perturbations of $a$ at the Wilson-Fisher fixed point one has
\begin{eqnarray}
\label{eq:WF_fp_exp_a}
\delta a \sim e^{(2 - \epsilon c_a/c_b) l} \,,
\end{eqnarray}
and thus $\lambda_2 =2 - \epsilon c_a/c_b = \nu^{-1}$.
Therefore one obtains the critical exponent $\nu$ as an {\em expansion in the parameter} $\epsilon$.
This can be generalized to higher orders in $\epsilon$ and in this way all critical exponents can be calculated as asymptotic series in $\epsilon$, which agree very well with experiments and other theoretical estimates.
We will discuss later on how one can verify the critical exponents and scaling functions experimentally.
An illuminating perspective on the renormalization group may be found in the review by \textcite{Fisher:1998}.
%

\subsection{\label{sec:5F} Critical dynamics}
We may generalize the Ginzburg-Landau-Wilson model to dynamics, i.e., construct dynamical models which incorporate fluctuations and have the correct hydrodynamics for $\tau >0$ and $\tau <0$.
The simplest model is relaxational with a stochastic contribution
\begin{eqnarray}
\label{eq:GLW_relax}
\partial_t \psi = 
-\Lambda_{\psi} \frac{\partial \Phi}{\partial \psi} + \theta(x,t) \,,
\end{eqnarray}
where $\Phi$ is the general Ginzburg-Landau free energy as in Eq.~(\ref{eq:Z_psi}), and $\theta$ is a noise source, a random function defined by its probability distribution.
We choose $\theta$ to be a Gaussian white noise source, such that
\begin{eqnarray}
\label{eq:noise}
&& \langle \theta(x,t) \rangle =0 \,,
\nonumber \\
&& \langle \theta(x,t)\theta(x',t') \rangle = 
2 \Lambda_{\psi} \delta(x-x') \delta(t-t') \,.
\end{eqnarray}
Since the probability distribution is Gaussian the higher correlators, e.g., $\langle \theta \theta \theta \rangle$, are expressible in terms of the second-order correlator given by Eq.~(\ref{eq:noise}).
If in the probability distribution Eq.~(\ref{eq:noise}) the coefficient $\Lambda_{\psi}$ is the same as in Eq.~(\ref{eq:GLW_relax}), then it can be shown that if $\Phi$ has no explicit time dependence the probability distribution of $\psi$ relaxes at long times to the equilibrium distribution
\begin{eqnarray}
\label{eq:Peq}
P_{eq}(\psi) = Z^{-1} \exp\left[ \Phi(\psi) \right] \,.
\end{eqnarray}
As discussed above, a model with richer hydrodynamics is the planar magnet where one has coupling of the order parameter to a conserved density
\begin{eqnarray}
\label{eq:psi_t_mz_t_noise}
&& \partial_t \psi = -2 \Lambda_{\psi}
\frac{\partial \Phi}{\partial \psi^{*}}
- i g_0 \psi \frac{\partial \Phi}{\partial m} 
+ \theta(x,t) \,,
\nonumber \\
&& \partial_t m = \lambda
\nabla^2 \frac{\partial \Phi}{\partial m}
+ 2 g_0
\textrm{Im} \left[ \psi^{*} \frac{\partial \Phi}{\partial \psi^{*}} \right] 
+ \zeta(x,t) \,, \qquad
\end{eqnarray}
where $\Phi$ is the generalization of Eq.~(\ref{eq:fe_psi_mz}) to contain high-order terms in $\psi$ and $m$, and the noise terms satisfy
\begin{eqnarray}
\label{eq:2noise}
&& \langle \theta(x,t)\theta(x',t') \rangle = 
\textrm{Re} \Lambda_{\psi} \delta(x-x') \delta(t-t') \,,
\nonumber \\
&& \langle \zeta(x,t)\zeta(x',t') \rangle = 
-\lambda \nabla^2 \delta(x-x') \delta(t-t') \,,
\nonumber \\
&& \langle \theta(x,t)\zeta(x',t') \rangle =0 \,.
\end{eqnarray}
Here again if the coefficients in Eqs.(\ref{eq:2noise}) have been chosen appropriately, the system relaxes at long times to the equilibrium distribution 
\begin{eqnarray}
\label{eq:Peq2}
P_{eq}(\psi,m) = Z^{-1} \exp\left[ \Phi(\psi,m) \right] \,.
\end{eqnarray}
As shown by Halperin and Hohenberg, the renormalization group theory of Sec.~\ref{sec:5D} may be generalized to apply to the dynamical models Eqs.~(\ref{eq:GLW_relax}) or (\ref{eq:psi_t_mz_t_noise}), and the static Wilson conjectures can be extended to the full dynamics.
The phenomenological scaling theory is recovered if one assumes that the equations of motion are transformed and reach a fixed point form upon iteration.
Linearization about the fixed point yields one more relevant exponent $z$, which controls the scaling of frequencies, and the scaling of dynamic correlation functions and critical modes as in Eq.~(\ref{eq:omega_hom}), then follows.
Just as in the static case these conjectures can then be verified in detail by carrying out an analytic $\epsilon$-expansion of the equations of motion near $4$ dimensions.
In the planar magnet [Eq.~(\ref{eq:psi_t_mz_t_noise})], for example, one now has $\Lambda_{\psi}(l)$, $\lambda(l)$, $\chi_{\psi}(l)$, $\chi_m(l)$, and $g_0(l)$.
An equation for $\Lambda_{\psi}$ for given $a(l)$, $b(l)$ has the following form 
\begin{eqnarray}
\label{eq:Lambda_l}
\frac{d \Lambda_{\psi}}{d l} = 
F[ a(l), b(l), \dots, \epsilon ] \,.
\end{eqnarray}
Solving this equation one finds dynamic fixed points and dynamic exponents in an expansion in terms of $\epsilon$.
Similar equations can be found for $\lambda(l)$ and $g_0(l)$.
For $\tau >0$ we define the characteristic frequencies
\begin{eqnarray}
\label{eq:omega_plus}
\omega_{\psi}^{+}(l) \sim \frac{\Lambda_{\psi}(l)}{\chi_{\psi}(l)} \,, \;
\omega_m^{+}(l) \sim \frac{\lambda(l)}{\chi_m(l)} k^2 \,.
\end{eqnarray}
In the ordered phase, $\tau <0$, we define
\begin{eqnarray}
\label{eq:omega_minus}
\omega_{\psi}^{-}(l) = \omega_m^{-}(l) = \pm c_s(l) k \,, \;
c_s^2 = g_0^2 \rho_s / \chi_m \,,
\end{eqnarray}
as well as the two quantities $w(l)$ and $f(l)$ given by
\begin{eqnarray}
\label{eq:wf_def}
w(l) = \frac{\omega_{\psi}^{+}(l)}{\omega_m^{+}(l)} \,, \;
f(l) = \frac{[\omega_{\psi, m}^{-}(l)]^2}{\omega_{\psi}^{+}(l) \omega_m^{+}(l)} \,.
\end{eqnarray}
Equations for $w(l)$ and $f(l)$ can be derived from the equations for $\Lambda_{\psi}(l)$, $\lambda(l)$, and the static functions $a(l)$, $b(l)$, $\rho_s(l)$, $\chi_m(l)$, and a fixed point is found, of the form
\begin{eqnarray}
\label{eq:wf_scale}
w(l) \to w^{*} = 1 + O(\epsilon) \,, \;
f(l) \to f^{*} = \epsilon + O(\epsilon^2) \,. \qquad
\end{eqnarray}
Given the existence of such a fixed point one can verify that the characteristic frequencies satisfy the dynamic scaling relation $\omega \sim k^{z} \Omega(k \xi)$, and the dynamic exponent turns out to be $z =d/2$.
In this way, the phenomenological assumptions of Sec.~\ref{sec:4B} are justified analytically to lowest order in $\epsilon$, and further terms in the $\epsilon$-expansion can also be calculated.
Similar treatments have also been carried out for the other dynamic universality classes, as described in the review of \textcite{Hohenberg:1977}.
%

\subsection{\label{sec:5G} Testing the theory experimentally}
In this section we wish to show how the detailed predictions of the renormalization group theory can be tested experimentally, thus permitting accurate estimates of the numerical values of universal exponents and amplitudes.
In the usual procedure, when measuring some physical quantity $Q(\tau)$ which has a singularity for $\tau \to 0$, one assumes the form
\begin{eqnarray}
\label{eq:Q_func}
Q(\tau) = A_Q \tau^{x_Q} \,.
\end{eqnarray}
By plotting the measured values on a $\log - \log$ scale, the exponent $x_Q$ is taken to be the best fit over a reasonably large range, especially close to $\tau =0$ (many decades).
To be more sophisticated one does a $\chi^2$-test by calculating
\begin{eqnarray}
\label{eq:chi2_test}
\chi^2 = \frac{\langle (Q_{exp} - Q_{th})^2 \rangle}{\langle (Q_{exp} + Q_{th})^2 \rangle} \,,
\end{eqnarray}
and minimizes $\chi^2=\chi^2(x_Q)$ with respect to $x_Q$.
This gives values of $x_Q$ with error bars.
However, fitting experimental data by expressions like Eq.~(\ref{eq:Q_func}) without correction terms leaves out contributions of the form $|\tau|^{\Delta}$ which are significant for $|\tau| \to 0$, $\Delta <1$.
This means that the exponents thus obtained cannot be considered to be quantitatively reliable.
Let us take as an example the superfluid transition in $^4$He ($\lambda$-transition).
The phase diagram is shown in Fig.~\ref{fig:4He_PT}.
We are interested in the transition from $^4$He-I (liquid) to $^4$He-II (superfluid) when the $\lambda$-line is crossed.
Along this line there are in fact an infinite number of $\lambda$-transitions, and the renormalization group theory predicts that universal quantities (exponents and amplitude ratios) should be the same for all those transitions (i.e., independent of $P$).
Suppose the measured quantity $Q(\tau)$ is the specific heat $C_p \sim |\tau|^{-\alpha}$ or the superfluid density $\rho_s \sim |\tau|^{\nu}$.
The usual method would give critical exponents $\alpha(P)$ and $\nu(P)$ as fit parameters for each pressure value $P$. 
How does one check that, e.g., $\alpha = 2 - 3 \nu$ holds for each $P$, or how does one account for the pressure dependence of the `best fit' exponents?
One is reminded of Einstein's statement: ``The theory decides what is measurable''.
There is another way of saying this due to Eddington: ``Never believe an experimental result until it has been confirmed by theory''.
The renormalization group theory for the superfluid transition says that there is only one transition independent of $P$, and one can write 
\begin{eqnarray}
\label{eq:Cprhos_th}
&& C_p^{\pm} = A_{\pm}(P) |\tau|^{-\alpha} \left[ 1 + B_{\pm}(P) |\tau|^{\Delta} + \dots\right] \,,
\nonumber \\
&& \rho_s = D(P) |\tau|^{\nu} \left[ 1 + E(P) |\tau|^{\Delta} + \dots \right] \,,
\end{eqnarray}
where $\alpha$, $\nu$, and $\Delta$ are universal, i.e., independent of $P$, with $\alpha = 2 - 3 \nu$.
Let us now define the amplitude ratios $R_i$ as follows: 
\begin{eqnarray}
\label{eq:Ri}
R_1 = \frac{D(P)^3}{A_{\pm}(P)} \,, \;
R_2 = \frac{A_{+}(P)}{A_{-}(P)} \,,
\nonumber \\
R_3 = \frac{B_{+}(P)}{B_{-}(P)} \,, \;
R_4 = \frac{B_{-}(P)}{E(P)} \,.
\end{eqnarray}
According to the renormalization group theory these four ratios should also be universal, i.e., independent of $P$.
Taking data for all $P$ and fitting by Eqs.~(\ref{eq:Cprhos_th}) one extracts $\alpha$, $\nu$, $\Delta$, $R_1, \dots, R_4$ and one can test the theoretical predictions.
In practice one can take $\alpha$, $\nu$, $\Delta$ from theory and fit experimental $R_i$ for all $P$.
If the $R_i$ depend on $P$ that would falsify the theory.
The main conclusion one may draw from this exercise is that no matter how good the accuracy and range of experimental data, it is not possible to determine critical exponents without some assumption about the dependence of measured quantities on temperature, say.
For example, given Eq.~(\ref{eq:Cprhos_th}) one can determine the numerical values of amplitude ratios, given the assumed values of the exponents.
In this way the consistency of the theory is directly tested and the actual values of certain quantities determined from experiments.
Systematic analysis of the experimental data in terms of Eqs.~(\ref{eq:Cprhos_th})-(\ref{eq:Ri}) has been carried out at the $\lambda$-transition by Ahlers and co-workers, where in the same experiment only the pressure varies.
The renormalization group predictions were thus rigorously tested and the agreement between experiment and theory constitutes a major triumph for both, see e.g. \textcite{Privman:1991}.
%

\section{\label{sec:6} Nonequilibrium patterns near linear instabilities}
Up to now we were interested in average quantities, averaged over the thermal noise.
Now we consider macroscopic phenomena on scale $l$ for which the scale of energies $\langle \varepsilon \rangle$ averaged over a volume $v \sim l^d$ far exceeds $ k_B T$, so we may neglect thermal noise.
Moreover we are interested in the behavior far from equilibrium.
We shall focus on systems with spontaneous symmetry breaking, so that Ginzburg-Landau theory will once again turn out to be useful.
In the phase transition theory considered up to now, the spontaneous symmetry breaking came from the phase transition.
Here we consider the bifurcation of a uniform nonequilibrium steady state, for example the instability of a horizontal fluid layer heated from below (Rayleigh-B\'enard convection).
The control parameter $R$ measures the distance from equilibrium; above a certain value $R_c$ the uniform steady state becomes linearly unstable and patterns in space and time can grow.
%

\subsection{\label{sec:6A} Classification of instabilities}
\begin{figure}[ht]
(a)\includegraphics[width=6.5cm]{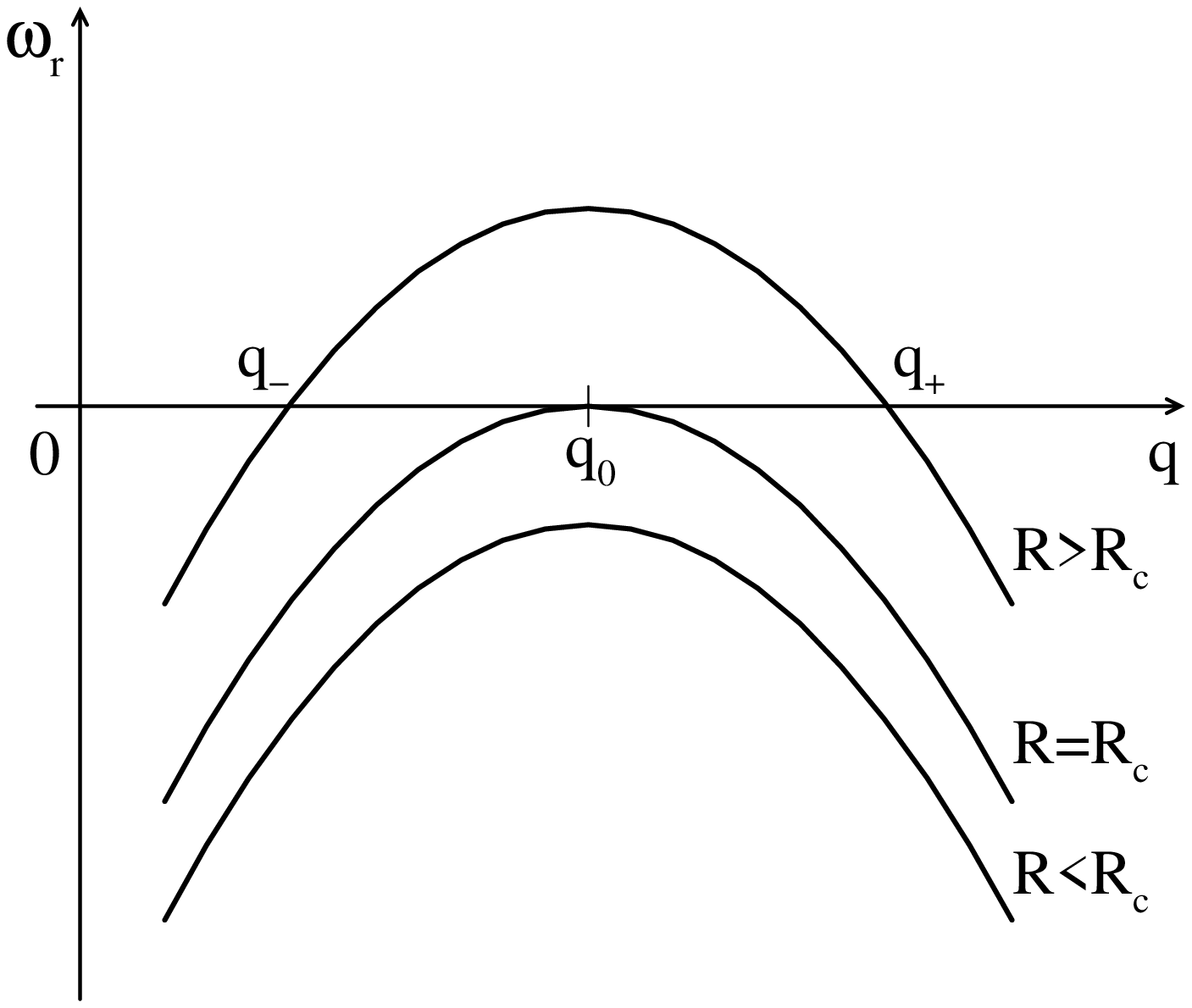}
(b)\includegraphics[width=6.5cm]{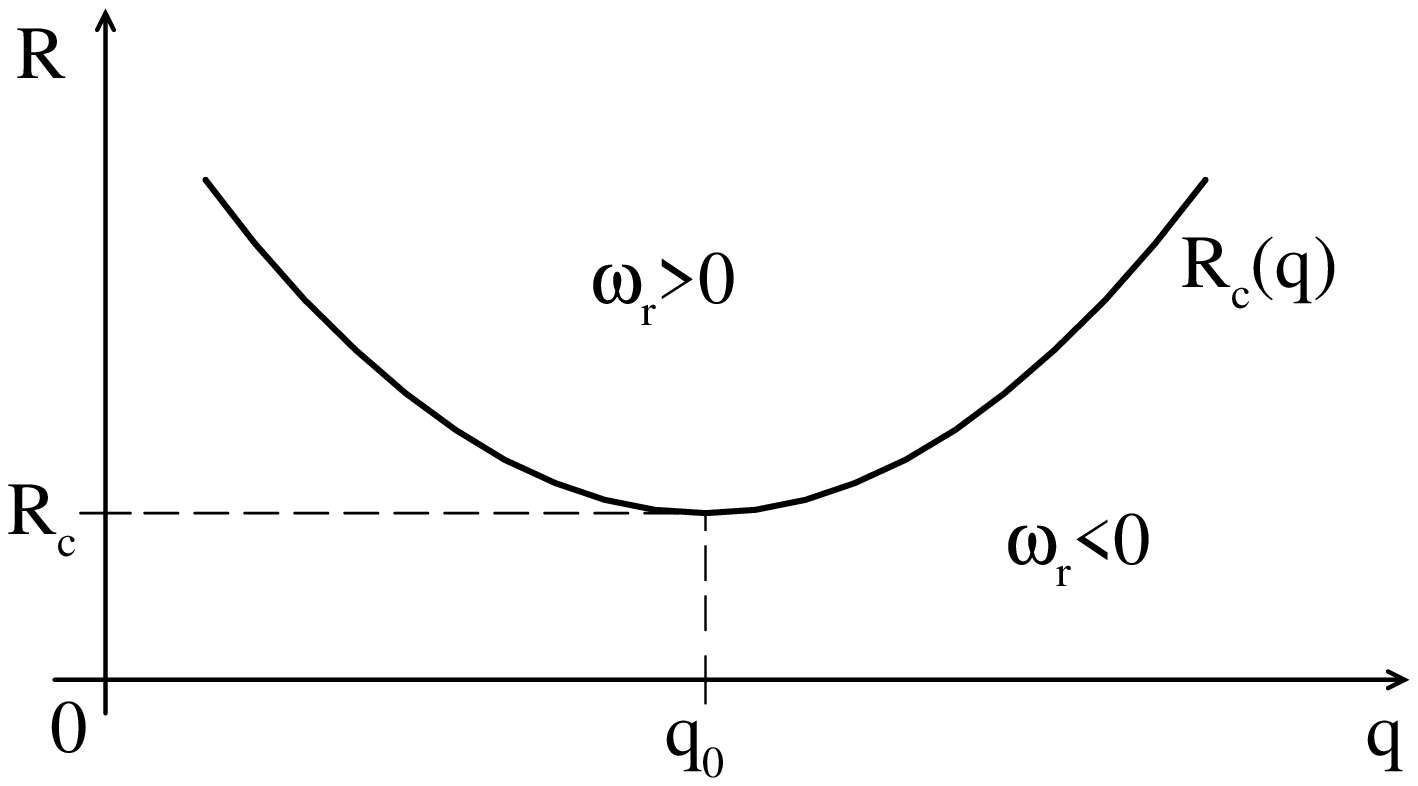}
\caption{\label{fig:growthrate}
Real part of $\omega(q,R)$ (a) and neutral curve $R_c(q)$ (b).}
\end{figure}

Consider systems described by what we will call a `microscopic model', defined by differential equations of the general form
\begin{eqnarray}
\label{eq:dudt}
\partial_t \bm u(x,t) = \bm f(R, \bm u, \nabla \bm u, \dots) \,,
\end{eqnarray}
where $\bm u = \{ u_1, u_2, \dots, u_n \}$ is an $n$-component vector and the function $\bm f = \{ f_1, f_2, \dots, f_n \}$ (also a vector) depends on the control parameter $R$.
Suppose $\bm u = \bar{\bm u}(R)$ is a uniform solution of Eq.~(\ref{eq:dudt}) with $\partial_t \bar{\bm u} =0$.
In mathematics this is referred to as an `equilibrium solution', even though the state $\bar{\bm u}(R)$ is not an equilibrium state of the physical system.
Now we ask whether $\bar{\bm u}(R)$ is linearly stable.
Linearizing Eq.~(\ref{eq:dudt}) about
$\bm u = \bar{\bm u}(R)$
\begin{eqnarray}
\label{eq:delta_u}
&& \bm u = \bar{\bm u}(R) + \delta\bm u(x,t) \,
\nonumber \\
&& \partial_t \delta u_i(x,t) =
\sum_{j} \left( \frac{\partial f_i}{\partial u_j} \right)_{\bm u = \bar{\bm u}}
\delta u_j(x,t) \,,
\end{eqnarray}
one obtains linear equations for the perturbations $\delta u_i$.
These equations can be solved by Fourier transformation
\begin{eqnarray}
\label{eq:delta_u_trans}
\delta \bm u(x,t) =
\int \delta \bm u(q,\omega) e^{i (q x - i \omega t)} dq \, d\omega \,,
\end{eqnarray}
yielding a frequency $\omega(q,R)$ for each value of the wave vector and control parameter.
Equation (\ref{eq:delta_u}) thus becomes a set of linear algebraic equations
\begin{eqnarray}
\label{eq:delta_u_sys}
\delta u_i(q,\omega) = \sum_{j} A_{i j} \delta u_j(q,\omega) \,.
\end{eqnarray}
In general $\omega(q,R) = \omega_r(q,R) + i \omega_i(q,R)$ is complex.
If $\omega_r(q,R) <0$ for all $q$, then $\delta u$ decays and $\bar{\bm u}$ is stable; if $\omega_r(q,R) >0$ then $\bar{\bm u}$ is unstable and $\omega_r(q,R) =0$ corresponds to the point of instability which occurs at $R =R_c$ [see Fig.~\ref{fig:growthrate}(a)].
%

%
%
The condition $\omega_r(q,R) =0$ defines a function $R = R_c(q)$ which is also called the neutral curve.
Let us consider the case when $R_c(q)$ has a minimum at a certain $q = q_0$ (which could also be zero).
For $q = q_0$, $R = R_c(q_0) \equiv R_c$, one has for the real part $\omega_r =0$ and for the imaginary part we define $\omega_0 \equiv \omega_i(q_0,R_c)$.
The classification of instability type in time and space is based on the values of $q_0$ and $\omega_0$, see \textcite{Cross:1993}, \textcite{Cross:2009}.
If $\omega_0 =0$ and $q_0 =0$ one has a transition from one uniform steady state to another, there is no pattern and we will not consider this case.
The length scale $q_0^{-1} = \ell_0$ is what we will (formally) consider to be the `microscale'.
The cases we consider are:
Type I$_s$: Stationary-periodic, $\omega_0 =0$, $q_0 \ne 0$.
The critical mode at $R=R_c$ is time independent, $\delta \bm u \sim e^{i q_0 x}$ and for $R >R_c$ all modes with $q_{-} < q < q_{+}$ grow [see Fig.~\ref{fig:growthrate}(a)].
The instability results in a stationary stripe pattern (in $2d$).
Type III$_o$: Oscillatory-uniform, $\omega_0 \ne 0$, $q_0 =0$.
Here the critical mode at $R=R_c$ is $\delta \bm u \sim e^{i \omega_0 t}$, i.e. it has $q =0$.
Above $R_c$ there is a band of unstable modes with $q^2 < q_{\pm}^2 \sim R - R_c$, with growth rates $\textrm{Re} \, \omega(q,R) \sim q_{\pm}^2 - q^2$.
Type I$_o$: Oscillatory-periodic, $\omega_0 \ne 0$, $q_0 \ne 0$.
The critical mode depends on space and time, $\delta \bm u \sim A_{-} e^{i (q_0 x - \omega_0 t)} + A_{+} e^{i (q_0 x + \omega_0 t)}$.
It results in traveling waves.
In general, above $R_c$ the modes within the band $q_{-} < q < q_{+}$ are unstable.
As these modes grow they interact due to nonlinearity and one mode typically emerges.
This is an ideal pattern.
The following questions arise:\\
(i) Which ones of these modes are stable?\\
(ii) They exist in a continuum. Which one is selected?\\
(iii) How do such patterns evolve as $R$ increases quasistatically?
In equilibrium steady states, the answers to these questions can be found by minimization of a free energy.
Here we have no such principle, so the problems are much more complex and less general.
We will answer some of these questions using Ginzburg-Landau equations.
%

\subsection{\label{sec:6B} Pattern forming systems}
\subsubsection{\label{sec:6B1} Experimental systems}
In this section we describe very briefly some examples of physical systems undergoing linear instabilities, according to the instability type.
Type I$_s$: \\
-~Rayleigh-B\'enard convection in a horizontal fluid layer of height $d$ heated from below.
The control parameter is proportional to the temperature difference $\Delta T$ between the lower and the upper plate.
Above a critical value of $\Delta T$ the uniform heat conduction state becomes unstable and a stationary convective flow in the form of a series of rolls (stripe pattern) with $q_0 \sim 1/d$ develops. \\
-~Taylor-Couette flow of a fluid between two coaxial cylinders with rotating inner cylinder.
The control parameter is proportional to the angular velocity $\Omega$ of the inner cylinder.
For small $\Omega$ one has a uniform velocity profile which becomes unstable above a critical value $\Omega_c$ and a system of toroidal rolls around the inner cylinder is formed (Taylor-Couette vortices).
If one also rotates the outer cylinder the instability type can be changed to I$_o$.
Type I$_o$: \\
-~Thermal convection in fluid mixtures -- traveling rolls.
Under certain conditions thermal convection in a fluid mixture can change from a stationary (type I$_s$) to an oscillatory (type I$_o$) bifurcation. \\
-~The same is true for Taylor-Couette flow in certain regimes in which both the inner and the outer cylinder are rotating.
Type III$_o$: \\
-~Belousov-Zhabotinsky chemical reaction: as first demonstrated by Belousov and further explored by Zhabotinsky, certain chemical reactions spontaneously change over from quiescent to oscillatory when concentrations and reaction rates are varied. \\
-~Electrical activity of heart muscle: the electric potentials in heart muscle can also experience bifurcations to various regimes of oscillatory behavior.
%

\subsubsection{\label{sec:6B2} Reaction-diffusion model}
Patterns developed in reaction-diffusion equations for chemical or biological media were first discovered by Turing in 1952.
Depending on the details they can be of type I$_s$, I$_o$, or III$_o$.
Traditionally the type I$_s$ (stationary) instability goes by the name `Turing pattern', and the type III$_o$ instability is called `oscillatory instability', even though Turing discussed both types in his paper in 1952.
The reaction-diffusion model is defined by linear diffusion and a nonlinear reaction function in Eq.~(\ref{eq:dudt})
\begin{eqnarray}
\label{eq:RDg}
\partial_t \bm u = f(\bm u) + D \nabla^2 \bm u \,.
\end{eqnarray}
The simplest reaction-diffusion system is the two-component model
\begin{eqnarray}
\label{eq:RD}
&& \partial_t u_1 = f_1(u_1,u_2) + D_1 \nabla^2 u_1 \,,
\nonumber \\
&& \partial_t u_2 = f_2(u_1,u_2) + D_2 \nabla^2 u_2 \,.
\end{eqnarray}
Suppose $\bar{\bm u} = (\bar{u}_1,\bar{u}_2)$ is a uniform steady solution of Eqs.~(\ref{eq:RD}).
Taking $\bm u = \bar{\bm u} + \delta \bm u$ and linearizing Eqs.~(\ref{eq:RD}) one obtains two coupled linear differential equations for perturbations
\begin{eqnarray}
\label{eq:RD_perp}
&& \partial_t \delta u_i = \sum_{j=1,2} a_{i j} \delta u_j \,, \;
i = 1, 2 \,,
\nonumber \\
&& a_{i j} = \left( \frac{\partial f_i}{\partial u_j} \right)_{\bm u = \bar{\bm u}} \,.
\end{eqnarray}
Using Fourier transformation
\begin{eqnarray}
\label{eq:RD_trans}
\delta \bm u =
\int \delta \bm u(q) e^{i (q x - i \omega t)} dq \, d\omega \,,
\end{eqnarray}
the equations reduce to an eigenvalue problem for $\omega$
\begin{eqnarray}
\label{eq:RD_ev}
&& \tilde{A}(q) \delta \bm u(q) = \omega(q) \delta \bm u(q) \,,
\nonumber \\
&& \tilde{A}(q) =
\begin{pmatrix}
a_{1 1} - D_1 q^2 & a_{1 2} \\
a_{2 1} & a_{2 2} - D_2 q^2
\end{pmatrix} \,.
\end{eqnarray}
The eigenvalues for this $2 \times 2$ matrix $\tilde{A}$ are
\begin{eqnarray}
\label{eq:RD_omega}
\omega_{\pm}(q) = && \frac{1}{2} \textrm{tr} \tilde{A}
\pm \frac{1}{2} \left[ (\textrm{tr} \tilde{A})^2 - 4 \, \textrm{det} \tilde{A} \right]^{1/2}
\nonumber \\
&& \equiv \frac{1}{2}T_q \pm \frac{1}{2} \left[ T_q^2 - 4 D_q \right]^{1/2} \,.
\end{eqnarray}
Suppose $\bar{\bm u}$ is stable at $q =0$.
This means that $\textrm{Re} \, \omega(0) <0$, which can be realized for
\begin{eqnarray}
\label{eq:DT_0}
T_0 \equiv \textrm{tr} \tilde{A}(0) <0 \,, \;
D_0 \equiv \textrm{det} \tilde{A}(0) >0 \,,
\end{eqnarray}
so that the stability condition at $q =0$ becomes
\begin{eqnarray}
\label{eq:DTA_0}
a_{1 1} + a_{2 2} <0 \,, \;
a_{1 1} a_{2 2} - a_{1 2} a_{2 1} >0 \,.
\end{eqnarray}
Let us choose $a_{1 1} >0$, then $a_{2 2}$ and $a_{1 2} a_{2 1}$ both are negative.
The question is, can one get an instability at $q \ne 0$ (due to diffusion)?
One has
\begin{eqnarray}
\label{eq:T_q}
T_q = a_{1 1} + a_{2 2}
- (D_1 + D_2) q^2 < T_0 < 0 \,,
\end{eqnarray}
because the diffusion coefficients $D_1$, $D_2$ are positive.
To get an instability we need $\textrm{Re} \, \omega(q) >0$ for some value of $q$.
Since $T_0 <0$ the only way to accomplish this is to have
\begin{eqnarray}
\label{eq:D_q}
D_q = (a_{1 1} - D_1 q^2) (a_{2 2} - D_2 q^2) - a_{1 2} a_{2 1} \le 0 \,, \qquad
\end{eqnarray}
and the threshold is given by $D_q =0$.
To find the $q$ value for which $D_q$ first becomes zero one minimizes $D_q$ with respect to $q$
\begin{eqnarray}
\label{eq:RD_q0}
\partial_{q^2} [D_q] =0 \,: \;
D_1 D_2 q^2 - D_1 a_{2 2} - D_2 a_{1 1} \,,
\end{eqnarray}
which gives
\begin{eqnarray}
\label{eq:RD_q02}
q_0^2 = \frac{1}{2} \left( \frac{a_{1 1}}{D_1} + \frac{a_{2 2}}{D_2} \right) \,,
\end{eqnarray}
a quantity that has to be positive.
Taking into account that $a_{1 1} >0$ and $a_{2 2} <0$ and defining $l_1^2 = D_1/a_{1 1}$, $l_2^2 = D_2/|a_{2 2}|$ one obtains
\begin{eqnarray}
\label{eq:RD_q02_l}
q_0^2 = \frac{1}{2} \left( \frac{1}{l_1^2} - \frac{1}{l_2^2} \right) \,,
\end{eqnarray}
which is positive for $l_1 < l_2$ or equivalently
\begin{eqnarray}
\label{eq:D1D2}
\frac{D_2}{D_1} > \frac{|a_{2 2}|}{a_{1 1}} \,.
\end{eqnarray}
In such a case one has a type I$_s$ instability.
For $a_{1 1} >0$, $a_{2 2} <0$ $u_1$ is referred as the activator and $u_2$ as the inhibitor.
The condition $l_1 < l_2$ implies short-range activation and long-range inhibition.
Typically in real systems $|a_{2 2}|/a_{1 1} \sim 5 - 10$ and most chemicals have approximately the same diffusion coefficients, so that Turing patterns (i.e., type I$_s$) were not observed for many years.
One needs sufficiently different diffusion coefficients for activation and inhibition.
After this was understood, such systems were in fact prepared experimentally and (stationary) Turing patterns have now been observed in a number of chemical systems.
As mentioned above, Turing also showed that an instability of type III$_o$ is rather easy to obtain in reaction-diffusion systems.
According to Eq.~(\ref{eq:RD_omega}) this occurs for
\begin{eqnarray}
\label{eq:DT_0_2}
T_0 \ge 0 \,, \;
D_0 >0 \,, \;
T_0^2 < 4 D_0 \,.
\end{eqnarray}
The threshold is given by $T_0 =0$ and the frequency of oscillations is $\omega_0 = 2 \sqrt{D_0}$.
Historically Turing was the first to understand how various patterns can arise spontaneously out of the instability of a nonequilibrium homogeneous steady state.
His motivation seems to have been to understand differentiation during embryo development.
It can also be shown that a type I$_o$ is impossible in this model.
One needs at least one more variable, $u_3$, to get a type I$_o$ instability.
For completeness, let us also mention natural patterns:
Excitable biological media such as nerve pulses, heart muscle, aggregation patterns of Dictyostelium (slime mold), zebra stripes, leopard spots, some patterns in developing embryos; snow flakes; sand dunes; the red spot of Jupiter; spiral galaxies.
All of these systems display spontaneous pattern formation with many similarities to patterns found in the simple models we are discussing, but under natural conditions systems rarely operate near the linear instability of a uniform state.
%

\subsection{\label{sec:6C} Amplitude equations: the real and complex Ginzburg-Landau equations: potential and non-potential dynamics}
\subsubsection{\label{sec:6C1} The real Ginzburg-Landau equation}
Let us consider a one-dimensional system (one space, one time dimension) defined by a differential equation
\begin{eqnarray}
\label{eq:1d_model}
\partial_t u(x,t) = f(R, u, \nabla u, \dots) \,,
\end{eqnarray}
which has a uniform solution $\bar{u}$ and shows a type I$_s$ instability for $R >R_c$ ($q_0 \ne 0$).
We wish to study the nonlinear states of $u(x,t)$ for $R >R_c$.
Introduce the reduced control parameter
\begin{eqnarray}
\label{eq:r_red}
r = \frac{R - R_c}{R_c} \,,
\end{eqnarray}
and assume that the growing solution has the following form near threshold ($r \ll 1$):
\begin{eqnarray}
\label{eq:du_amp}
\delta u(x,t) = \delta u_0 \left[ A(x,t) e^{i q_0 x} + \textrm{c.c.} \right] + \textrm{h.o.t.} \,,
\end{eqnarray}
where $A(x,t)$ is complex (c.c. means complex conjugate and h.o.t denotes high-order terms).
Then inserting Eq.~(\ref{eq:du_amp}) into the original equation (\ref{eq:1d_model}) and expanding in $r$, one obtains an equation for the amplitude $A(x,t)$
\begin{eqnarray}
\label{eq:ampl}
\tau_0 \partial_t A(x,t) = r A + l_0^2 \partial_x^2 A - g_0 |A|^2 A + \textrm{h.o.t.} \,, \qquad
\end{eqnarray}
where $\tau_0$ and $l_0$ are time and length scales that can be obtained from the function $\omega(q,R)$ arising from the linear instability of the system (\ref{eq:1d_model}).
We now show that the form of the equation can be inferred using symmetry arguments and the assumptions $r \ll 1$, $|A| \ll 1$, and $|\nabla A| \ll 1$.
The symmetry requirements that constrain the form of the amplitude equation arise from the need for consistency with the physical symmetries of the original system (\ref{eq:1d_model}).
These are \\
(i) translation symmetry: Eq.~(\ref{eq:ampl}) should be unchanged by the substitution $A \to A e^{i \Delta}$, since by Eq.~(\ref{eq:du_amp}) it implies a translation of the system $x \to x + \Delta/q_0$:
\begin{eqnarray}
\label{eq:u_trans}
\delta u(x,t) &\to& \delta u_0 A(x,t) e^{i \Delta} e^{i q_0 x} + \textrm{c.c.}
\nonumber \\
&\to& \delta u_0 A(x,t) e^{i q_0 (x + \Delta/q_0)} + \textrm{c.c.} \,.
\end{eqnarray}
(ii) parity symmetry: Eq.~(\ref{eq:ampl}) should be unchanged under the double substitution $A \to A^{*}$, $x \to -x$, which corresponds to an inversion of the coordinates in Eq.~(\ref{eq:1d_model}).
From translational symmetry we conclude that $A$ must be complex.
Algebraic products of $A$ and $A^{*}$ that lead to odd powers such as $A$, $|A|^2 A$, $|A|^4 A$, etc. are invariant under both symmetries and are thus allowed, whereas even powers such as $A^2$, $|A|^2$, $|A|^2 A^2$ or certain other odd powers such as $A^3$, $|A|^2 A^3$ are ruled out by translation symmetry.
Since the equation for $u(x,t)$ has a first time derivative and is dissipative, i.e., not time-reversal invariant, there must also be a first time derivative in the amplitude equation.
For $r >0$ the solution should grow, which is represented by the allowed term $r A$.
Terms like $\partial_x A$, although consistent with parity symmetry, can be eliminated by setting $\bar{A} = A e^{i x}$.
In general there should be diffusion given by $\partial_x^2 A$.
The nonlinear term, proportional to $|A|^2 A$, is allowed by symmetry and is responsible for saturation of the growing solution.
Higher powers of $A$ and $\partial_x A$ are negligible for $r \ll 1$.
In addition to the above symmetry arguments, Eq.~(\ref{eq:ampl}) can also be derived using a formal `multiple scales' perturbation theory treating $r$ and $\nabla A / A$ as expansion parameters.
The simplest solution of the amplitude equation (\ref{eq:ampl}) is a constant
\begin{eqnarray}
\label{eq:A_const}
A = a = \textrm{const} \,, \;
|a|^2 = \frac{r}{g_0} \,,
\end{eqnarray}
which exists for $g_0 >0$.
Note that if $g_0 <0$ one needs higher-order terms like $|A|^4 A$ in the amplitude equation to stabilize solutions for $r >0$.
Thus for $g_0 >0$ and for small $r \ll 1$ the amplitude $|a|$ is also small.
For $r >0$ one can rescale Eq.~(\ref{eq:ampl})
\begin{eqnarray}
\label{eq:AXT_scale}
A = (r/g_0)^{1/2} \bar{A} \,, \;
x = l_0 r^{-1/2} X \,, \;
t = \tau_0 r^{-1} T \,, \qquad
\end{eqnarray}
leading to the scaled real Ginzburg-Landau equation (RGLE)
\begin{eqnarray}
\label{eq:RGLE}
\partial_T \bar{A} = \bar{A} + \partial_X^2 \bar{A} - |\bar{A}|^2 \bar{A} \,.
\end{eqnarray}
Here $X \sim r^{1/2}$ plays the role of slow scale or mesoscale close to the instability threshold.
Similarly $T \sim r$ represents the slow time.

Consider now a plane wave solutions of Eq.~(\ref{eq:RGLE})
\begin{eqnarray}
\label{eq:pw_sol}
\bar{A}_K = a_K e^{i K X} \,.
\end{eqnarray}
Inserting Eq.~(\ref{eq:pw_sol}) into Eq.~(\ref{eq:RGLE}) one finds
\begin{eqnarray}
\label{eq:pw_disp}
a_K^2 = 1 - K^2 \,,
\end{eqnarray}
so that the solution exists for $-1 < K < 1$.
Going back to the original scaling given by Eq.~(\ref{eq:du_amp}) one finds
\begin{eqnarray}
\label{eq:pw_du_amp}
\delta u(x,t) \sim a_K e^{i (q_0 + r^{1/2} K) x} + \textrm{c.c.} \,.
\end{eqnarray}
Thus above threshold plane wave solutions of Eq.~(\ref{eq:1d_model}) exist in the range $q_{-} \le q \le q_{+}$, with $q_{\pm} = q_0 \pm r^{1/2}$.
These solutions are referred to as `rolls' or `stripes' in Rayleigh-B\'enard convection.
%

%
%
\begin{figure}[ht]
\includegraphics[width=7cm]{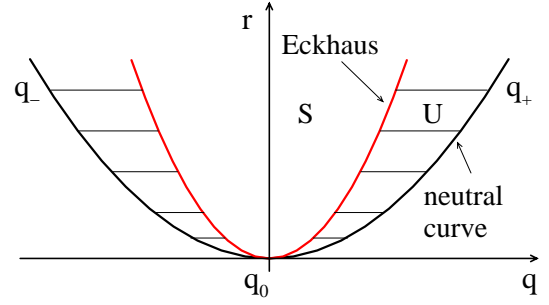}
\caption{\label{fig:eckhaus}
Stability diagram of plane wave solutions of the real Ginzburg-Landau equation.
In the region marked S there are stable stripe solutions.
These solutions still exist in the region marked U, but they are unstable.}
\end{figure}
Now we can study the linear stability of the stripe solutions, considering perturbations of the form
\begin{eqnarray}
\label{eq:pw_stab}
\bar{A}(X,T) = a_K e^{i K X} + \delta \bar{A}(X,T) =
\bar{A}_K + \delta \bar{A} \,. \qquad
\end{eqnarray}
The linearized equation for $\delta \bar{A}$ is given by
\begin{eqnarray}
\label{eq:pw_pert}
\partial_T \delta \bar{A} =
\delta \bar{A} + \partial_X^2 \delta \bar{A}
- 2 |\bar{A}_K|^2 \delta \bar{A} - \bar{A}_K^2 \delta \bar{A}^{*} \,, \qquad
\end{eqnarray}
and similarly for $\delta \bar{A}^{*}$
\begin{eqnarray}
\label{eq:pw_pert_cc}
\partial_T \delta \bar{A}^{*} =
\delta \bar{A}^{*} + \partial_X^2 \delta \bar{A}^{*}
- 2 |\bar{A}_K|^2 \delta \bar{A}^{*} - (\bar{A}_K^{*})^2 \delta \bar{A} \,. \;\;\qquad
\end{eqnarray}
These are linear partial differential equations but the coefficients are not constant, since they depend on $\bar{A}_K(X,T)$ which is periodic in space.
Since the coefficients are periodic ($\sim e^{i K X}$) a solution can be searched in the form
\begin{eqnarray}
\label{eq:pw_pert_sol}
\delta \bar{A} = e^{i K X}
\left[ \delta a_{+} e^{i Q X}
+ \delta a_{-}^{*} e^{-i Q X} \right] \,,
\end{eqnarray}
where $\delta a_{\pm}(t) \sim e^{\omega_K T}$.
In this way one gets for $\omega_K$
\begin{eqnarray}
\label{eq:pw_pert_omega}
\omega_K &=& (1 - K^2) - Q^2
+ \left( (1 - K^2)^2 + 4 K^2 Q^2 \right)^{1/2} \,,
\nonumber \\
&=& -\frac{1 - 3 K^2}{1 - K^2} Q^2 + O(Q^4) \,.
\end{eqnarray}
For $1/3 \le K^2 \le 1$ one has $\omega_K \ge 0$ and therefore the plane wave solution $\bar{A}_K$ becomes unstable.
This is the so-called Eckhaus instability.
Thus plane wave solutions $\bar{A}_K$ exist for $-1 < K < 1$ but they are only stable in the subrange $-1/\sqrt{3} < K < 1/\sqrt{3}$ (or $q_{-}/\sqrt{3} < q-q_0 < q_{+}/\sqrt{3}$, see Fig.~\ref{fig:eckhaus}).
In higher dimensions one can also consider more complicated forms of perturbations $\delta \bar{A}$, when the Ginzburg-Landau equation contains in addition gradients in the $y$ direction (perpendicular to $x$).
In this way one finds the stability domain for the plane wave solutions of Eq.~(\ref{eq:1d_model}).
Finally we discuss the dynamics of the amplitude function given by the real Ginzburg-Landau equation.
Let us define a kind of free energy
\begin{eqnarray}
\label{eq:FE_ampl}
\bar{\Phi} = \frac{1}{2} \int dx \left[ -|\bar{A}|^2 + \frac{1}{2} |\bar{A}|^4 + |\partial_X \bar{A}|^2 \right] \,.
\end{eqnarray}
If $\bar{A}$ is a solution of the real Ginzburg-Landau equation Eq.~(\ref{eq:RGLE}) then
\begin{eqnarray}
\label{eq:dPhidT}
\frac{d \bar{\Phi}}{d T} = - \int dx |\partial_T \bar{A}|^2 \, \le 0 \,.
\end{eqnarray}
Thus all dynamics makes $\bar{\Phi}$ decrease and we refer to this as `potential dynamics', analogous to the situation in equilibrium.
The system always ends up in a stationary state, a local minimum of $\bar{\Phi}$.
We note, however, that this situation is special to the real Ginzburg-Landau equation.
It is not typical for nonequilibrium systems.
%

\subsubsection{\label{sec:6C2} The complex Ginzburg-Landau equation}
\begin{figure}[ht]
(a)\includegraphics[width=6.5cm]{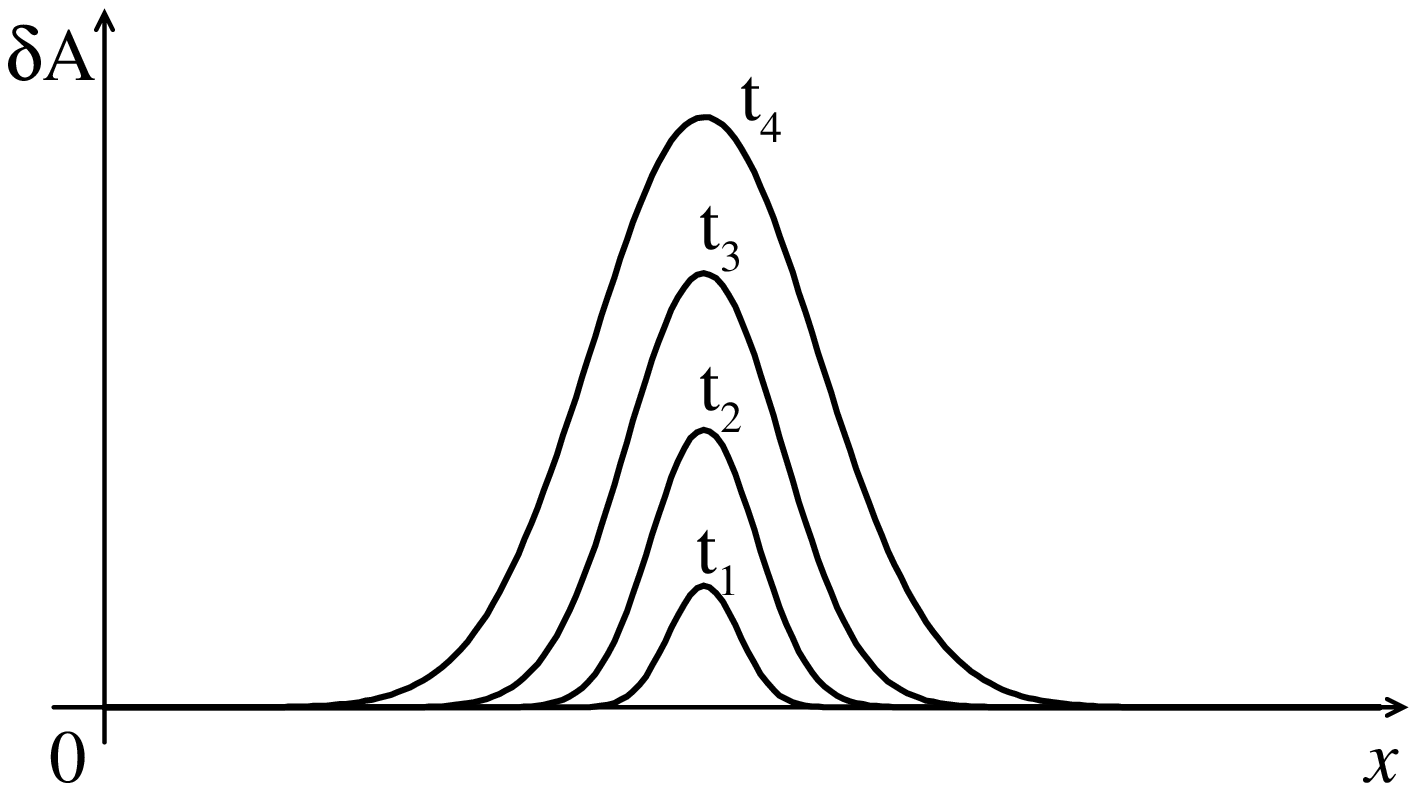}
(b)\includegraphics[width=6.5cm]{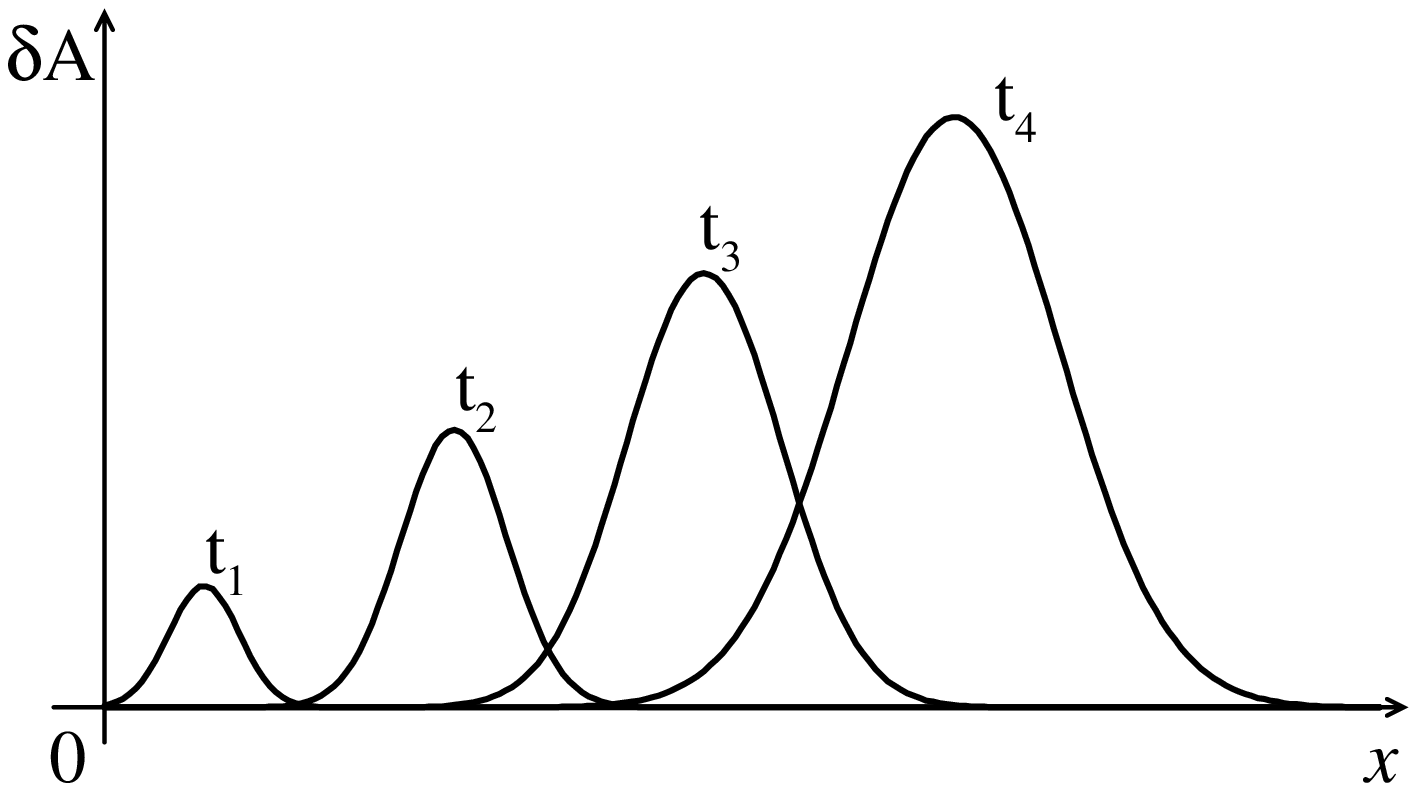}
\caption{\label{fig:absconv}
Growth of a localized perturbation for an absolute instability (a) and a convective instability (b).}
\end{figure}

Consider now the amplitude equation for a type III$_o$ instability (oscillatory-uniform) where a uniform solution of Eq.~(\ref{eq:1d_model}) becomes unstable for $R >R_c$ with $\omega_0 \ne 0$, $q_0 =0$.
Assume for the growing solution near threshold ($r \ll 1$) the form
\begin{eqnarray}
\label{eq:du_osc}
\delta u(x,t) = \delta u_0 \left[ A(x,t) e^{i \omega_0 t} +
\textrm{c.c} \right] + \textrm{h.o.t.} \,.
\end{eqnarray}
The amplitude $A(x,t)$ is again complex though its phase has a rather different significance than for the type I$_s$ system.
Here it is the local phase of the temporal oscillations, and a change of phase corresponds to a shift of the time coordinate.
The magnitude and phase of the amplitude $A$ describe slowly varying spatial and temporal modulations of the spatially uniform `fast' oscillation $e^{i {\omega_0} t}$.
The equation for $A(x,t)$ can again be inferred phenomenologically from symmetry arguments and the lowest-order result is
\begin{eqnarray}
\label{eq:CGLE1d}
\tau_0 \partial_t A = r A + (1 + i c_1) l_0^2 \partial_x^2 A
- g_0 (1 - i c_3) |A|^2 A \,. \;\;\qquad
\end{eqnarray}
The coefficients on the r.h.s. of this equation are in general complex ($c_1 \ne 0$, $c_3 \ne 0$) and as discussed below this makes a huge difference in the dynamics.
The complex coefficients arise because the amplitude $A^{*}$ describes the amplitude of the time reversed oscillation $e^{-i \omega t}$ which is different from the original oscillation $e^{i \omega t}$, due to the absence of time inversion symmetry.
In the previous case, $A^{*}$ describes the oscillation of the space reversed component $e^{-i q_0 x}$, which is related to the original component $e^{i q_0 x}$ by inversion symmetry.
For $r >0$ we can also rescale $A$, $x$, and $t$ in Eq.~(\ref{eq:CGLE1d}) as in Eq.~(\ref{eq:AXT_scale}), to obtain the so-called complex Ginzburg-Landau equation (CGLE)
\begin{eqnarray}
\label{eq:CGLE}
\partial_T \bar{A} = \bar{A} + (1 + i c_1) \nabla^2 \bar{A}
- (1 - i c_3) |\bar{A}|^2 \bar{A} \,.
\end{eqnarray}
Previously we considered the one-dimensional case, but for type III$_o$ systems the same equation is obtained in higher spatial dimensions.
The important difference with the previous case is that here there is no potential $\bar{\Phi}$, and the dynamics of $A$ is much more complicated than a simple minimization as in Eq.~(\ref{eq:dPhidT}).
Let us consider traveling wave solutions of Eq.~(\ref{eq:CGLE})
\begin{eqnarray}
\label{eq:tw}
&& \bar{A}_K = a_K e^{i (K x - \Omega_K t)} \,,
\nonumber \\
&& a_K^2 = 1 - K^2 \,, \;
\Omega_K = -c_3 + (c_1 + c_3) K^2 \,.
\end{eqnarray}
The group velocity $s$ is then given by
\begin{eqnarray}
\label{eq:groupv}
s = \partial_K \Omega_K = 2 K (c_1+c_3) \,.
\end{eqnarray}
The linear stability of the traveling waves can be studied similarly to the real case by setting $\bar{A} = \bar{A}_K + \delta \bar{A}$ with
\begin{eqnarray}
\label{eq:tw_pert}
\delta \bar{A} = e^{i (K x - \Omega_K t)}
\left[ \delta a_{+} e^{i Q X} + \delta a_{-}^{*} e^{-i Q X} \right] \,,
\end{eqnarray}
where $\delta a_{\pm} \sim e^{\Lambda_K T}$.
Solving linear equations for $\delta a_{\pm}$ one then finds $\Lambda_K(Q)$, which determines the stability of the traveling wave solution $\bar{A}_K$ with respect to perturbations with wave vector $Q$.
Consider the two-dimensional case and pick the direction of the wave vector $\bm K = K \hat{\bm x}$.
But the wave vector of the perturbations $\bm Q$ could be in any direction.
In the limit $|\bm Q| \ll 1$ one has in leading order:
\begin{eqnarray}
\label{eq:Lambda_K}
&& \Lambda_K(Q) = i s Q_x
- D_{\parallel}(K) Q_x^2 - D_{\perp}(K) Q_y^2 \,,
\nonumber \\
&& D_{\parallel}(K) = 1 - c_1 c_3 - \frac{2 (1 + c_3^2) K^2}{1 - K^2} \,,
\nonumber \\
&& D_{\perp}(K) = 1 - c_1 c_3 \,.
\end{eqnarray}
The traveling wave solution Eq.~(\ref{eq:tw}) is stable for $D_{\parallel} >0$, $D_{\perp} >0$.
The first instability one encounters has $D_{\parallel} \le 0$, which gives for the wave vector $K$
\begin{eqnarray}
\label{eq:BF}
K^2 \ge K_{BF}^2 \,, \;
K_{BF}^2 = \frac{1 - c_1 c_3}{3 - c_1 c_3 + 2 c_3^2} \,.
\end{eqnarray}
This is the so-called Benjamin-Feir instability which is the analogue of the Eckhaus instability for the real Ginzburg-Landau equation.
Setting $c_1 = c_3 =0$ in Eq.~(\ref{eq:BF}) gives $K^2 \ge 1/3$ as in Eq.~(\ref{eq:pw_pert_omega}).
Another new feature which appears in the CGLE is convective versus absolute instability.
For stationary instabilities (type I$_s$, RGLE) we did not ask what the spatial form of the perturbations $\delta u$ or $\delta \bar{A}$ was, because it did not matter.
For oscillatory instabilities (type III$_o$, CGLE) on the other hand, it matters.
We should consider spatially localized perturbations and ask if they grow.
If they do, there are 2 possibilities:
(i) they grow at a fixed location -- this is an absolute instability,
(ii) they grow but are swept away -- this is a convective instability, see Fig.~\ref{fig:absconv}.
The absolute instability is similar to the type I$_s$ situation.
Convective instabilities are a new feature of type III$_o$ systems, and they occur as well in more general instabilities such as pipe flow.
%

%
%
The criterion for absolute instability involves setting up a wave packet and seeing if growth is faster than advection.
Consider a localized perturbation of the form
\begin{eqnarray}
\label{eq:local_pert}
\delta \bar{A}(X,T) = \int dQ e^{i Q X + \Omega_Q T} \delta A(Q,0) \,,
\end{eqnarray}
where $\delta A(Q,0)$ represents the perturbation at $T =0$ and we inquire whether it grows with time.
It can be rewritten as
\begin{eqnarray}
\label{eq:local_pert2}
\delta \bar{A} &=&
\int dQ e^{i Q X + \Omega_Q T} \int dX' e^{-i Q X'} \delta A(X',0) \,,
\nonumber \\
&=& \int dX' \delta A(X',0) \int dQ e^{i Q (X-X') + \Omega_Q T} \,.
\end{eqnarray}
The integral over $Q$ can be calculated by stationary phase approximation in the complex $Q$-plane.
The major contribution to the $Q$-integral comes from the point where $\partial_Q \Omega_Q =0$.
Solving this condition one finds $Q =Q_s$ (complex) and the integral is then given by
\begin{eqnarray}
\label{eq:stat_phase}
\int dQ e^{i Q (X-X') + \Omega_Q T} \simeq e^{i Q_s (X-X') + \Omega_s T} \,,
\end{eqnarray}
where $\Omega_s = \Omega_{Q_s}$.
The perturbation Eq.~(\ref{eq:local_pert2}) is then
\begin{eqnarray}
\label{eq:local_pert3}
\delta \bar{A}(X,T) =
e^{\Omega_s T} \int dX' \delta A(X',0) e^{i Q_s (X-X')} \,, \qquad
\end{eqnarray}
and for $X =0$, carrying out the integral over $X'$ one finds
\begin{eqnarray}
\label{eq:local_pert4}
\delta \bar{A}(0,T) = e^{\Omega_s T} \delta \tilde{A}(Q_s) \,.
\end{eqnarray}
Thus, an absolute instability takes place for $\textrm{Re} \, \Omega_s >0$, where $Q_s$ is defined by the condition $\partial_Q \Omega_Q(Q=Q_s) =0$.
The criterion for convective instability is $\textrm{Re} \, \Omega_Q >0$ for some $Q$ with a nonzero group velocity ($s = \partial_Q \textrm{Im} \, \Omega_Q  \ne 0$), thus the perturbation is growing and propagating (or advected) away.
Considering perturbations of traveling waves with growth rate $\Lambda_K(Q)$ given by Eq.~(\ref{eq:Lambda_K}), the stationary phase point $Q_s$ is then given by
\begin{eqnarray}
\label{eq:Q_s}
\partial_Q \Lambda_K(Q) =0 \,.
\end{eqnarray}
The condition $\textrm{Re} \, \Lambda_K(Q_s) >0$ gives the range of $K^2 > K_A^2$ with $K_A^2 = 4(1 + c_1^2)$, where traveling waves are absolutely unstable.
The criterion for convective instability gives $K^2 > K_C^2 = K_{BF}^2$ and one finds $K_A > K_C$.
It is important that depending on $c_1$, $c_3$, one can have $K_C^2 <0$ and in this case there will be no stable traveling waves.
%

\subsection{\label{sec:6D} Defect solutions of the Ginzburg-Landau equations}
The stripe patterns and plane waves considered up to now are only the simplest `ideal' solutions of the Ginzburg-Landau equations, referring to an infinite system.
There are of course many other types of solutions, which appear under more realistic conditions.
In this section we study a class of patterns we call `defect solutions'.
These are formed by piecing together different ideal patterns, or by perturbing the patterns locally.
%

\subsubsection{\label{sec:6D1} Defects in the real Ginzburg-Landau equation}
\begin{figure}[ht]
(a)\hspace*{1.0cm}\includegraphics[width=5cm]{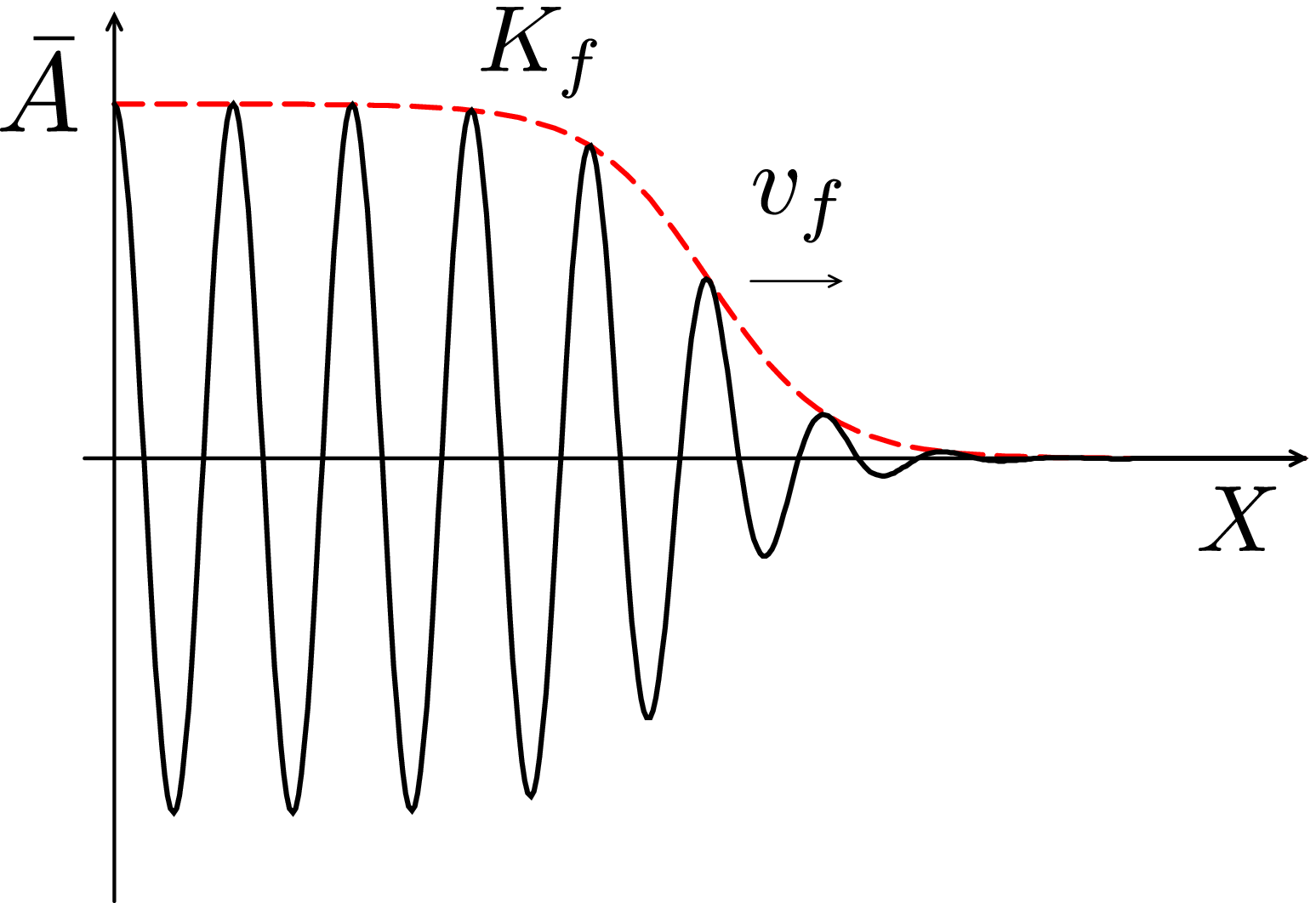}\\
\vspace*{0.5cm}
(b)\hspace*{1.0cm}\includegraphics[width=5cm]{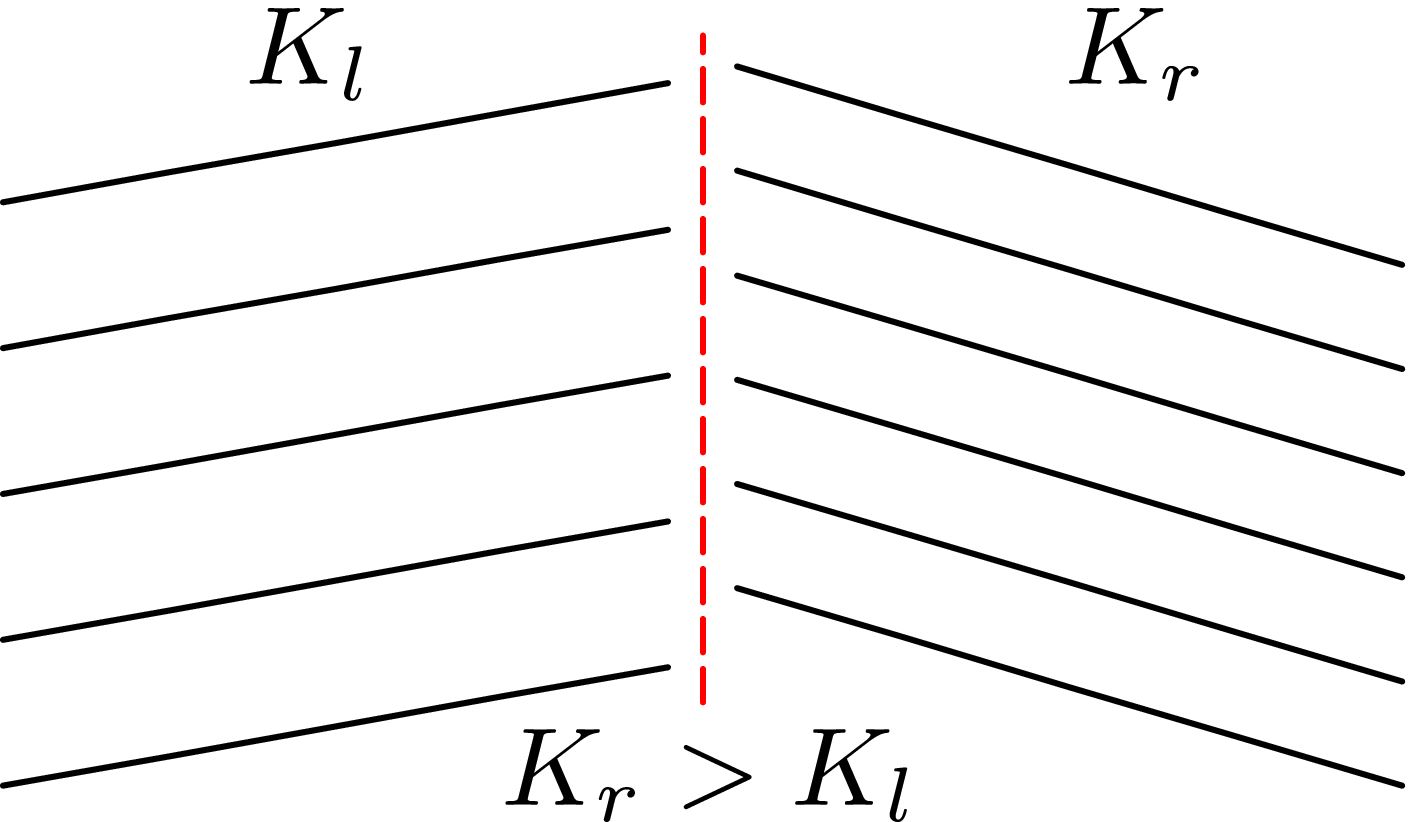}\\
\vspace*{1.0cm}
(c)\hspace*{1.0cm}\includegraphics[width=5.5cm]{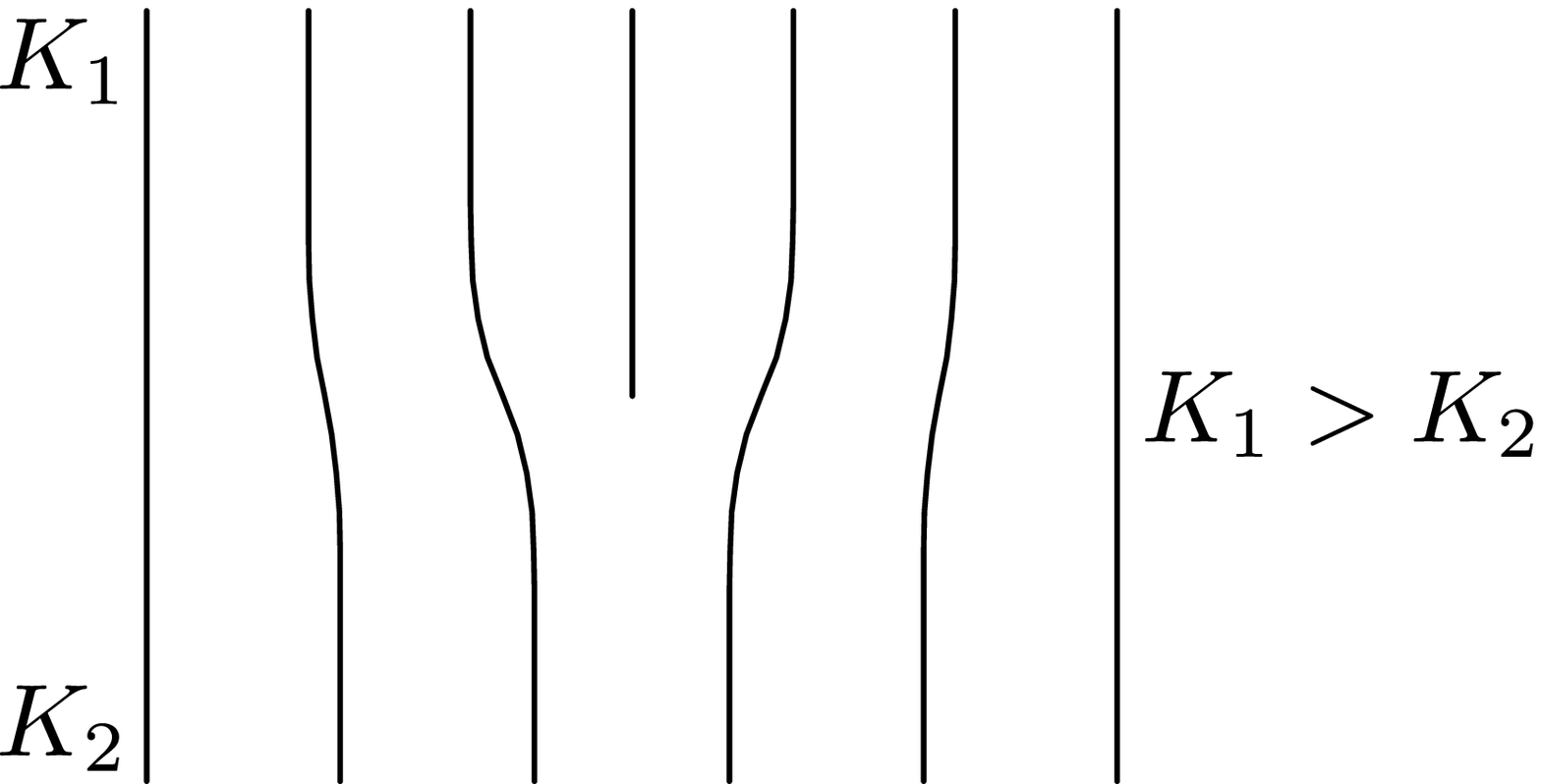}\\
\vspace*{0.5cm}
(d)\hspace*{0.5cm}\includegraphics[width=3cm]{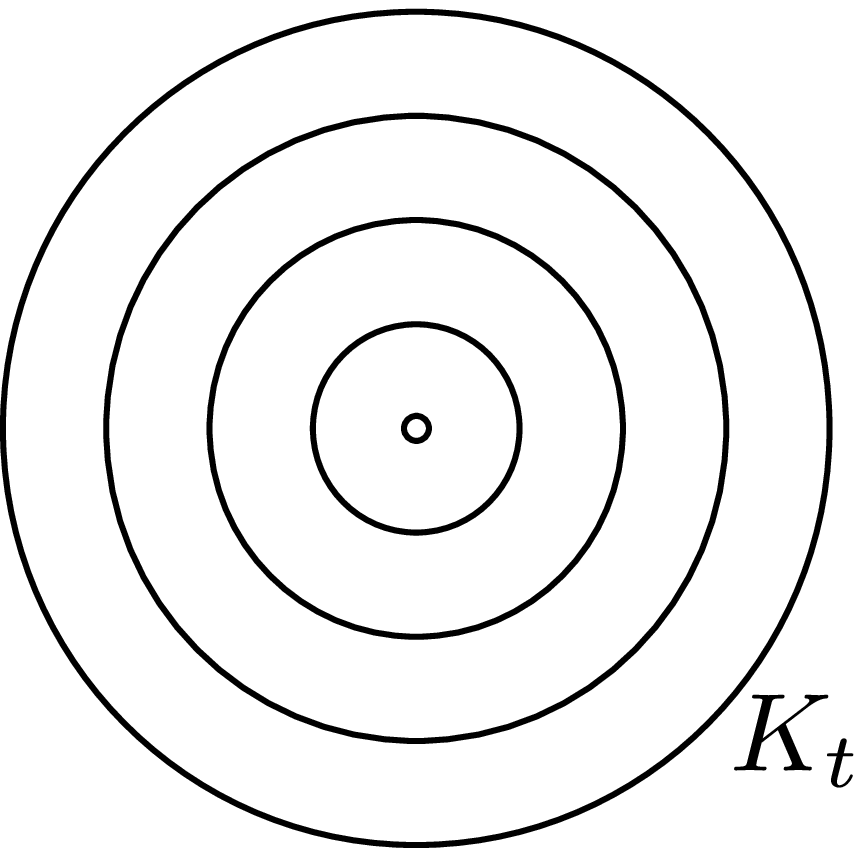}
\caption{\label{fig:pattern2d}
Defect patterns for the real Ginzburg-Landau equation: front in 1d (a), domain boundary (b), dislocation (c) and target pattern (d) in 2d.}
\end{figure}

The simplest `defect' in one dimension is to consider Eq.~(\ref{eq:RGLE}) on a semi-infinite domain, $X >0$, with the boundary conditions $\bar{A} =0$ at $X =0$, and $\bar{A} = a_K e^{i K X}$ for $X \to \infty$.
It is then found that the band $-1 \le K \le 1$ collapses to a {\em single} point $K =0$, i.e. only a constant satisfies both the Ginzburg-Landau equation (\ref{eq:RGLE}) and the boundary conditions.
This constant for $\bar{A}$ corresponds to a pattern with $q =q_0$ for $u$.
Another solution of Eq.~(\ref{eq:RGLE}) is a front in 1d.
Consider for $r >0$ the solution $\bar{A}(X,0) =0$ at $T =0$ which is unstable.
Let us add a localized perturbation $\delta \bar{A}$ at $X =0$.
According to the Ginzburg-Landau equation, $\delta \bar{A}$ will grow and eventually saturate its amplitude due to the nonlinear term.
Suppose it leaves behind a plane wave.
In this process the solution is propagating into the unstable state ($\bar{A} =0$), so there is a front velocity $v_f$ [see Fig.~\ref{fig:pattern2d}(a)].
What wave vector $K_f$ is selected behind the front?
Using the Ginzburg-Landau equation one can calculate the selected wave vector $K_f$ (see below).
Another example of a defect solution occurs in an instability of type I$_s$ in two dimensions.
The ideal pattern above the instability has the form of stationary stripes.
Let us consider so called domain boundaries between regions with different orientations of the stripes [see Fig.~\ref{fig:pattern2d}(b)].
They are characterized by a wave vector $K_l$ on the left and $K_r$ on the right.
Using the Ginzburg-Landau equation in 2d one can calculate the behavior of such configurations and analyze whether the domain boundary is stable or mobile (unstable).
%

%
%
Stripe patterns in 2d can also have dislocations [see Fig.~\ref{fig:pattern2d}(c)], in which a stripe boundary abruptly terminates, creating a pattern with wave vector $K_1$ on top (in the far field) and $K_2$ on the bottom.
If the dislocation moves upward we say the state $K_2$ is preferred over the state $K_1$ and the opposite is true if the dislocation moves downward.
We will see below that such pattern competition allows one to define a `preferred' wave vector.
Target patterns [Fig.~\ref{fig:pattern2d}(d)] represent another type of stripe pattern in two dimensions.
One can analyze the existence and stability of such solutions and calculate the wave vector $K_t$ far from the target center.
Here again we can ask, which wave vector will be selected in the far field.
%

\subsubsection{\label{sec:6D2} Defects in the complex Ginzburg-Landau equation}
\begin{figure}[ht]
\includegraphics[width=3cm]{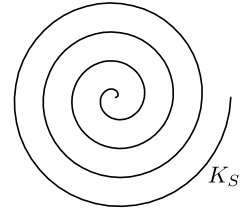}
\caption{\label{fig:spiral}
Defect pattern for the complex Ginzburg-Landau equation: spiral in 2d.}
\end{figure}

For the complex equation \eqref{eq:CGLE} in one dimension we can also create fronts [Fig.~\ref{fig:pattern2d}(a)], with wave vector $K_f$.
In two dimensions the best known example is the generalization of a target, namely a spiral pattern represented in polar coordinates $R, \theta$ as (note, here $R$ is the radial coordinate, not the control parameter!),
\begin{eqnarray}
\label{eq:spiral}
\bar{A}_S(R,\theta, T) = a(R) e^{i [-\Omega_S T + m \theta + \psi(R)]} \,,
\end{eqnarray}
where in the far field $\psi(R) \sim K_S R$ (see Fig.~\ref{fig:spiral}).
It looks like a traveling wave far away from the center of the spiral.
Inserting Eq.~(\ref{eq:spiral}) into the CGLE Eq.~(\ref{eq:CGLE}) one can find $a(R)$.
For a given $K_S$ the frequency $\Omega_S = \Omega(K_S)$ is determined by the dispersion relation of plane wave solutions, Eq.~(\ref{eq:tw}).
Depending on the sign of $m$ the spiral wave unwinds or winds.
Spiral patterns are typical for systems with type III$_o$ instability.
%

%
%
In three dimensions such solutions correspond to a vortex line
\begin{eqnarray}
\label{eq:vortex}
\bar{A}_V(R,\theta, z, T) = a(R,z) e^{i [-\Omega_V T + m \theta + \psi(R) + k_z z]} \,. \qquad
\end{eqnarray}
In this simple example the vortex core is a line in the $z$ direction and it is characterized by the wave vector $k_z$.
Deforming the vortex line one can make, e.g., a `smoke ring', typically with $k_z =0$.
It has a close analogue to classical hydrodynamic vortex rings and is also relevant to the dynamics of superfluids described by the Gross-Pitaevskii equation.
%

\subsection{\label{sec:6E} Pattern selection}
The problem of pattern selection arises quite generally because the equations we are considering have many solutions above threshold for given external conditions (fixed control parameter $R$), whereas observed patterns constitute a much more restricted set.
Thus, among the allowed (i.e. linearly stable) solutions some seem to be preferred over others and we would like to understand the selection process.
The discussion of defect solutions in the previous subsections leads to the following questions:\\
(i) Are there constraints in either space or time that reduce the multiplicity of allowed solutions?\\
(ii) In situations where a multiplicity of solutions remains after constraints have been applied, is there any ordering between the solutions such that one is preferred over the other?
As mentioned earlier, in equilibrium bulk systems the free energy provides the ordering principle, so that the solution with the lowest value is preferred.
We can thus anticipate that for the real Ginzburg-Landau equation the potential (\ref{eq:FE_ampl}) will play the same role and the solution $\bar{A} =const$ ($K =0$, corresponding to $q =q_0$ for $u$) will be preferred.
Indeed, in contrast to the ideal case, the semi-infinite system with $\bar{A}(X=0) =0$ has only the constant as a solution.
In addition, for each one of the defects we considered (front, dislocation, target), the $K =0$ solution, corresponding to $q =q_0$ for $u$, is preferred.
We expect that any type I$_s$ system with potential dynamics will favor the ideal pattern that minimizes the potential.
Such systems, however, are the exception, valid only very near threshold, and many other type I$_s$ cases with \emph{nonpotential} dynamics have been studied both theoretically and experimentally.
In such systems it is found, for example, that dislocations select a unique wave vector $q_d$ for each value of $R$ at which the dislocation is stationary.
Similarly, targets select a unique $q_t$ in the far field, and fronts leave behind a definite $q_f$.
The interesting result, which has been confirmed both experimentally and theoretically, is that in general $q_d(R)$, $q_f(R)$ and $q_t(R)$ are all different for $R > R_c$, thus falsifying the claims of universal selection principles for nonequilibrium steady states.
On the other hand, in any type I$_s$ system with potential dynamics (i.e. satisfying an equation like (\ref{eq:RGLE}) with $\tilde{\Phi}$ more general than (\ref{eq:FE_ampl}), (\ref{eq:dPhidT})), different selection mechanisms represented by different types of defects, all select the same wave vector.
As mentioned above, however, potential dynamics is the exception rather than the rule for nonequilibrium steady states.
For example, it no longer applies at the next order in the expansion in Eq.~(\ref{eq:du_amp}) for familiar examples such as Rayleigh-B\'enard convection.
The simplest example of nonpotential dynamics is the complex Ginzburg-Landau equation (\ref{eq:CGLE}), for which the wave vectors $K_f$ and $K_S$ selected by the front and the spiral, respectively, depend on $c_1$ and $c_3$ are are in general different.
There is thus no universal selection principle in this case even at lowest order in this Type III$_o$ case.
We conclude this brief discussion of pattern selection by considering the effects of noise, either in the form of random initial conditions or as external forcing throughout the time dependence.
In the first case it has been conjectured that the fastest growing linear mode will dominate the evolution at later times, but this idea is too simplistic and no general rule has emerged.
As regards external forcing, thermal noise, which was considered explicitly in Sec.~\ref{sec:5} above for the study of phase transitions, can be argued to have a negligible effect on patterns at the macroscopic scales usually studied.
Instrumental noise, on the other hand, can certainly be important, especially in situations where deterministic constraints are insufficient to define a unique pattern, but here again no general laws are known.
The reader interested in further information on pattern selection is referred to Chapter~8 of \textcite{Cross:2009} and Sec.~VI of \textcite{Cross:1993}.
%

\subsection{\label{sec:6F} Solutions of the Ginzburg-Landau equations: temporal and spatiotemporal chaos}
\subsubsection{\label{sec:6F1} Temporal chaos}
We first briefly discuss the Lorenz model with three degrees of freedom, $x(t)$, $y(t)$ and $z(t)$
\begin{eqnarray}
\label{eq:Lorenz}
&& \dot{x}(t) = -\sigma (x - y) \,,
\nonumber \\
&& \dot{y}(t) = r x - y - x z \,,
\nonumber \\
&& \dot{z}(t) = b (x y - z) \,,
\end{eqnarray}
where $r$ is the control parameter and $\sigma$, $b$ some fixed numbers.
Due to nonlinearity there is no analytic solution of this model.
It is found numerically that for $r <1$ the solution is uniform,  $x=y=z=0$ at long time, and for $1 < r < r_1$ there exist nonzero (fixed point) solutions $x = \bar{x}$, $y = \bar{y}$, $z = \bar{z}$, where $r_1(b,\sigma)$ is some constant.
For the standard values $b =8/3$, $\sigma =10$ used by Lorenz we have $r_1 =24.74$, and for $r >r_1$ the fixed point is unstable.
Coexisting with the fixed point solution, in the range $r_2 < r < r_1$ with $r_2 = 13.9$, there exists a periodic solution, called a limit cycle.
This type of behavior is standard for ordinary differential equations.
What Lorenz found in addition, however was a great surprise: in a domain $r > r_3$, with $r_3 =24.06$ there was another solution that was neither constant nor periodic in time, but irregular, with continuous Fourier spectrum.
Irregular solutions from deterministic equations were called chaotic.
This was a great discovery by Lorenz in 1963.
Let us now consider a geometrical representation of the dynamics of the Lorenz model Eq.~(\ref{eq:Lorenz}) in terms of its phase space.
The dimension of the phase space is $D =3$, which is the number of dynamical variables.
The initial conditions are represented by points in this phase space.
The time evolution of the solution is represented by a trajectory in the phase space.
If the trajectories all go to some fixed point, this point is an attractor $A^{*}$ with a dimension $D^A =0$.
With the parameter values $b =8/3$, $\sigma =10$ chosen by Lorenz, for $r <1$ there is one fixed-point attractor at $x=y=z=0$.
For $1 < r < r_1$ there exists another fixed-point attractor $A^{*} = (\bar{x},\bar{y},\bar{z})$ with the dimension $D^{A^{*}} =0$.
The limit cycle with $x(t)$, $y(t)$ and $z(t)$ periodic in time, which appears for $r_2 < r < r_1$, is represented by a loop in the phase space and the dimension of this attractor is $D^A =1$.
For $r > r_3$ a so-called chaotic or {\em strange attractor} was found in another region of phase space.
It is a complicated object in phase space, which looks like a composition of butterfly wings.
The dimension of this attractor can be estimated and it was found to be $D^A = 2.06 < D$, which is not an integer and it is smaller than $D$ but more than for a plane ($2 < D^A < D$).
Such attractors are called fractal (Mandelbrot).
In general, if the dynamics is dissipative then there exists an attractor with a dimension $D^A < D$, where $D$ is the dimension of the phase space.
If the dynamics is regular, then $D^A$ is an integer.
If $D^A$ is non-integer then the dynamics is chaotic.
Thus chaotic dynamics is characterized by continuous spectra of dynamical variables and by non-integer attractor dimension in the phase space.
Another feature of chaotic dynamics, which in many ways is even more basic, is sensitive dependence on initial conditions.
If one chooses two arbitrarily close points near the attractor then the trajectories emanating from those points diverge arbitrarily far from each other after long times.
This is the idea of unpredictability.
It can be quantified by the Lyapunov exponent which is defined in the following way.
Let us take $\bm \Delta(t) = P_1(t) - P_2(t)$ as the distance between two points emanating from close initial points $P_1(0)$ and $P_2(0)$ in the phase space.
At long time the distance $\bm \Delta(t)$ grows and one can calculate
\begin{eqnarray}
\label{eq:Lyapunov}
\lim\limits_{t \to \infty ,\, \bm \Delta(0) \to 0} \bm \Delta(t) \sim e^{\lambda_1 t} \,,
\end{eqnarray}
where $\lambda_1$ is called a Lyapunov exponent.
If $\lambda_1 <0$ the solution is regular (fixed point or limit cycle), but if $\lambda_1 >0$ the solution is chaotic.
By generalizing $\bm \Delta(t)$ to $n$-dimensional volumes in phase space (with $n \le D$), one obtains a spectrum of Lyapunov exponents $\{ \lambda_i \}$ with $1 \le i \le D$ and $\lambda_i < \lambda_1$ for $i>1$.
Most directions have $\lambda_i <0$ but if the largest exponent $\lambda_1$ is positive then the system is chaotic.
Note the double limit in Eq.~(\ref{eq:Lyapunov}).
The attractor dimension $D^A$ can be evaluated once the positive Lyapunov exponents are known.
Consider now partial differential equations (PDEs).
By definition the dimension of its phase space is infinite, $D =\infty$, since we are dealing with a continuum model.
We will see that for dissipative PDEs the dimension of the attractors in the phase space remains finite in a system of finite size, and thus $D^A\ll D$.
Let us consider in particular the RGLE or the CGLE in 1d.
In the scaling of Eqs.~(\ref{eq:RGLE}) and (\ref{eq:CGLE}) the mesoscale is $\xi = \ell_0 r^{-1/2} =1$ and the only remaining scale is given by the system size $L$ ($0 \le X \le L$).
It is the number of meso-units in the system.
We consider first a `small' system where $L = O(1)$.
If the dynamics of $\bar{A}$ is given by the RGLE, Eq.~(\ref{eq:RGLE}), which is potential, the attractor is a fixed point with dimension $D^A =0$.
For the CGLE in 1d, on the other hand, we have
\begin{eqnarray}
\label{eq:scaledCGLE1d}
\partial_T \bar{A}  = \bar{A} + (1 + i c_1) \partial_X^2 \bar{A} - (1 - i c_3) |\bar{A}|^2 \bar{A} \,, \qquad
\end{eqnarray}
and if $c_1 c_3 > 1$ there are no stable plane waves [see Eq.~(\ref{eq:Lambda_K})].
Taking $L=O(1)$ with periodic boundary conditions one indeed finds chaotic dynamics in numerical simulations.
What is the nature of this chaos and what is the dimension of the attractor?
For $L =O(1)$ it has been shown that there exists a solution of the form
\begin{eqnarray}
\label{eq:twomodes}
\bar{A}(X,T) = \sum_{n =1}^3 a_n(T) \phi_n(X) \,,
\end{eqnarray}
where the $\phi_n(X)$ are suitably defined basis functions, and the complex coefficients $a_1(T)$, $a_2(T)$, $a_3(T)$ satisfy Lorenz-like ordinary differential equations.
This reduced form with phase space dimension $D =6$ gives results very close numerically to what one finds from full simulations of Eq.~(\ref{eq:scaledCGLE1d}), where $D =\infty$.
The reduced model with 6 real modes identifies the so-called active modes in the system and it has an attractor dimension $D^A < 6$.
We will call such Lorenz-like chaotic dynamics `temporal chaos'.
In real experiments on Rayleigh-B\'enard convection in cells with lateral size close to the distance between plates, which means $L =O(1)$, Ahlers was the first to find chaotic behavior analogous to that of a Lorenz model.
%

\subsubsection{\label{sec:6F2} Spatiotemporal chaos}
\begin{figure}[ht]
(a)\includegraphics[width=3.5cm]{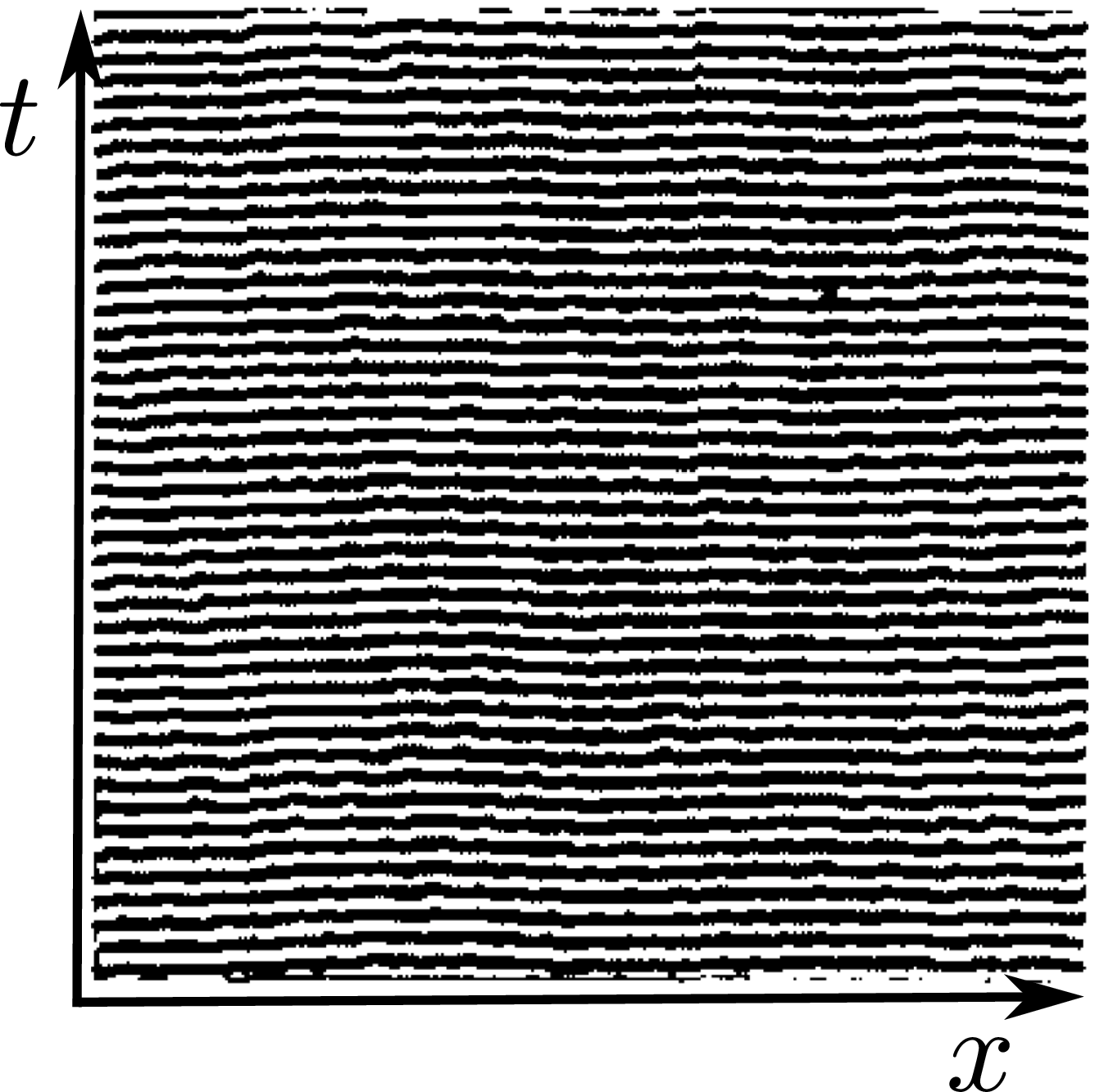}
(b)\includegraphics[width=3.5cm]{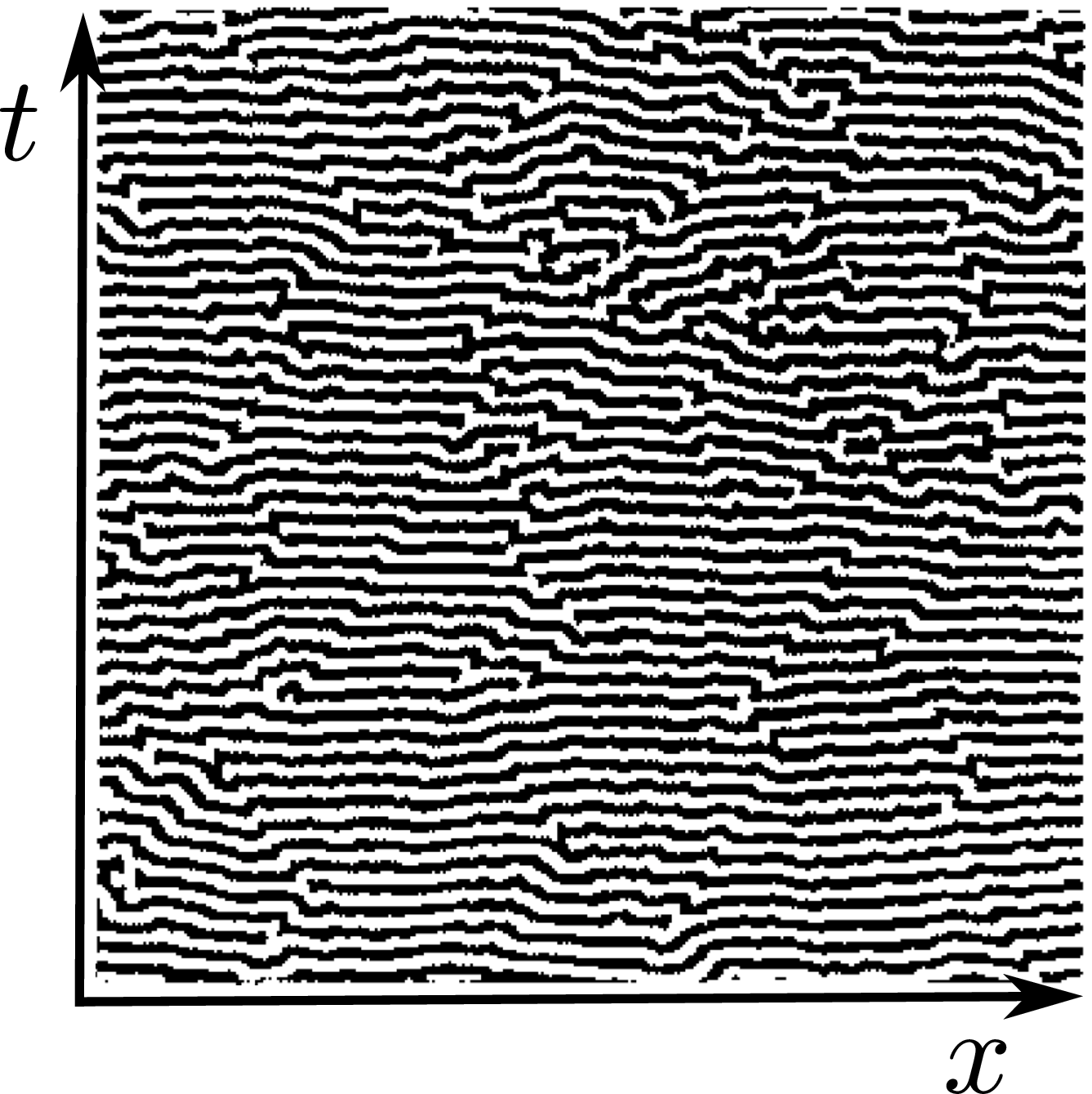}
\caption{\label{fig:chaos}
The phase $\phi(x,t)$ is plotted as a function of $x$, for different times.
Part (a) shows phase chaos, in which $\phi(x)$ is continuous for each value of $t$.
Part (b) shows defect chaos, in which discontinuities (space-time defects) appear whenever the amplitude $a(x,t)$ vanishes.}
\end{figure}

Now we consider large systems, i.e., $L \gg 1$.
These can be considered as consisting of small subsystems with size $L_i =O(1)$ interacting with each other in space and time.
For each subsystem $L_i$ the dimension of the attractor $D_i^A \lesssim D_i = O(1)$.
Thus for the whole system the dimension of the attractor $D^A(L) \sim L D_i^A$, i.e., $D^A(L)$ scales with system size $L$.
More generally, in a system with physical dimension $d$, and volume $L^d$, we define the dimension density of the attractor as
\begin{eqnarray}
\label{eq:rho_A}
\rho_A = D^A(L) / L^d \,,
\end{eqnarray}
and if $\rho_A$ remains finite as $L \to \infty$ we call this `extensive chaos' or `spatiotemporal chaos'.
Another way to define extensive chaos is to say that the number of positive Lyapunov exponents increases linearly with the system volume.
The CGLE in 1d for $c_1 c_3 >1$ has been simulated for $L= O(1000)$ for long times, and one finds two regimes depending on $c_1$, $c_3$: (i) phase chaos and (ii) defect chaos.
In the regime of phase chaos we write
\begin{eqnarray}
\label{eq:phase_ini}
\bar{A}(X,T) = a(X,T) e^{i \phi(X,T)} \,,
\end{eqnarray}
and start with $a(X,0) \approx 0.5$.
Then the amplitude $a(X,T)$ remains nonzero at all times.
The phase $\phi(X,T)$ has variations but as long as $a \ne 0$ we can define a winding number
\begin{eqnarray}
\label{eq:winding}
\nu = \int\limits_{0}^{L} dX \phi(X,T) \,,
\end{eqnarray}
which is independent of $T$ [Fig.~\ref{fig:chaos}(a)].
%

%
%
\begin{figure}[ht]
\includegraphics[width=7.5cm]{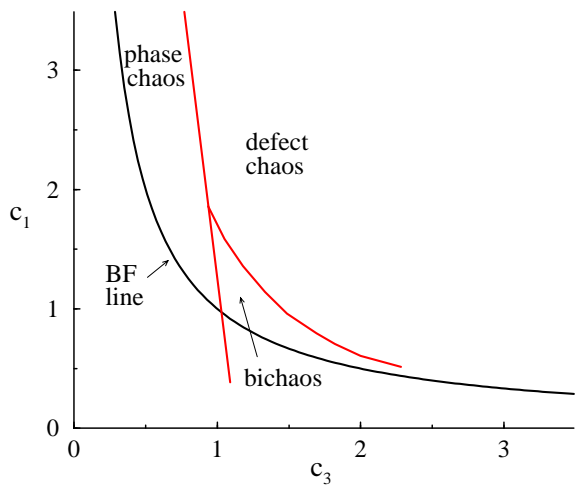}
\caption{\label{fig:phasediag}
Phase diagram for CGLE in 1d in the $c_1-c_3$ plane.
To the left of the BF line ($c_1 c_3 =1$) plane wave modes are linearly stable.
The region marked PC is phase chaos ($n_D =0$), DC is defect chaos ($n_D \ne 0$) and the regime marked BC has at least two chaotic attractors (bichaos), depending on initial conditions.}
\end{figure}

In the defect chaos regime $a(X,T)$ vanishes at some values of $X$ and $T$.
At those points $\phi$ jumps by a finite amount [Fig.~\ref{fig:chaos}(b)].
These points are referred to as `space-time defects' in the phase $\phi$.
The density of defects $n_D$ in the domain $0 < T < T_1$, $0 < X < L$, quantifies the regime for large $L$ and $T$.
If $n_D >0$ we have defect chaos, whereas $n_D =0$ signifies phase chaos.
In Fig.~\ref{fig:phasediag} the phase diagram for the CGLE in 1d is shown in the $c_1-c_3$ plane.
%

%
%
Thus the CGLE can be used to illustrate the passage from temporal (Lorenz-like) chaos to spatiotemporal (extensive) chaos, simply by increasing $L$.
Many questions remain about the precise behavior of this nonequilibrium system with many degrees of freedom, but the essential difference between temporal and spatiotemporal chaos is already illustrated by the CGLE model.
Similar behavior, with even richer structure, is found for the CGLE in two and three dimensions, where extensive spiral chaos appears for suitable choice of parameters $c_1$ and $c_3$.
%


\begin{acknowledgments}
The authors wish to thank Michael Cross, Daniel Fisher and Bert Halperin for constructive
comments and criticisms on a previous version of this paper.
\end{acknowledgments}


\bibliography{GL_theory_archive}

\begin{thebibliography}{12}%
\makeatletter
\providecommand \@ifxundefined [1]{%
 \@ifx{#1\undefined}
}%
\providecommand \@ifnum [1]{%
 \ifnum #1\expandafter \@firstoftwo
 \else \expandafter \@secondoftwo
 \fi
}%
\providecommand \@ifx [1]{%
 \ifx #1\expandafter \@firstoftwo
 \else \expandafter \@secondoftwo
 \fi
}%
\providecommand \natexlab [1]{#1}%
\providecommand \enquote  [1]{``#1''}%
\providecommand \bibnamefont  [1]{#1}%
\providecommand \bibfnamefont [1]{#1}%
\providecommand \citenamefont [1]{#1}%
\providecommand \href@noop [0]{\@secondoftwo}%
\providecommand \href [0]{\begingroup \@sanitize@url \@href}%
\providecommand \@href[1]{\@@startlink{#1}\@@href}%
\providecommand \@@href[1]{\endgroup#1\@@endlink}%
\providecommand \@sanitize@url [0]{\catcode `\\12\catcode `\$12\catcode
  `\&12\catcode `\#12\catcode `\^12\catcode `\_12\catcode `\%12\relax}%
\providecommand \@@startlink[1]{}%
\providecommand \@@endlink[0]{}%
\providecommand \url  [0]{\begingroup\@sanitize@url \@url }%
\providecommand \@url [1]{\endgroup\@href {#1}{\urlprefix }}%
\providecommand \urlprefix  [0]{URL }%
\providecommand \Eprint [0]{\href }%
\providecommand \doibase [0]{http://dx.doi.org/}%
\providecommand \selectlanguage [0]{\@gobble}%
\providecommand \bibinfo  [0]{\@secondoftwo}%
\providecommand \bibfield  [0]{\@secondoftwo}%
\providecommand \translation [1]{[#1]}%
\providecommand \BibitemOpen [0]{}%
\providecommand \bibitemStop [0]{}%
\providecommand \bibitemNoStop [0]{.\EOS\space}%
\providecommand \EOS [0]{\spacefactor3000\relax}%
\providecommand \BibitemShut  [1]{\csname bibitem#1\endcsname}%
\let\auto@bib@innerbib\@empty
\bibitem [{\citenamefont {Cross}\ and\ \citenamefont
  {Greenside}(2009)}]{Cross:2009}%
  \BibitemOpen
  \bibfield  {author} {\bibinfo {author} {\bibnamefont {Cross}, \bibfnamefont
  {M}}, \ and\ \bibinfo {author} {\bibfnamefont {H.}~\bibnamefont {Greenside}}}
  (\bibinfo {year} {2009}),\ \href@noop {} {\emph {\bibinfo {title} {{Pattern
  Formation and Dynamics in Nonequilibrium Systems}}}}\ (\bibinfo  {publisher}
  {Cambridge University Press},\ \bibinfo {address} {New York})\BibitemShut
  {NoStop}%
\bibitem [{\citenamefont {Cross}\ and\ \citenamefont
  {Hohenberg}(1993)}]{Cross:1993}%
  \BibitemOpen
  \bibfield  {author} {\bibinfo {author} {\bibnamefont {Cross}, \bibfnamefont
  {M~C}}, \ and\ \bibinfo {author} {\bibfnamefont {P.~C.}\ \bibnamefont
  {Hohenberg}}} (\bibinfo {year} {1993}),\ \bibfield  {title} {\enquote
  {\bibinfo {title} {{Pattern formation outside of equilibrium}},}\ }\href
  {\doibase 10.1103/RevModPhys.65.851} {\bibfield  {journal} {\bibinfo
  {journal} {Rev. Mod. Phys.}\ }\textbf {\bibinfo {volume} {65}}~(\bibinfo
  {number} {3}),\ \bibinfo {pages} {851--1112}}\BibitemShut {NoStop}%
\bibitem [{\citenamefont {Fisher}(1998)}]{Fisher:1998}%
  \BibitemOpen
  \bibfield  {author} {\bibinfo {author} {\bibnamefont {Fisher}, \bibfnamefont
  {M~E}}} (\bibinfo {year} {1998}),\ \bibfield  {title} {\enquote {\bibinfo
  {title} {{Renormalization group theory: Its basis and formulation in
  statistical physics}},}\ }\href {\doibase 10.1103/RevModPhys.70.653}
  {\bibfield  {journal} {\bibinfo  {journal} {Rev. Mod. Phys.}\ }\textbf
  {\bibinfo {volume} {70}}~(\bibinfo {number} {2}),\ \bibinfo {pages}
  {653--681}}\BibitemShut {NoStop}%
\bibitem [{\citenamefont {Ginzburg}(1960)}]{Ginzburg:1960}%
  \BibitemOpen
  \bibfield  {author} {\bibinfo {author} {\bibnamefont {Ginzburg},
  \bibfnamefont {V~L}}} (\bibinfo {year} {1960}),\ \bibfield  {title} {\enquote
  {\bibinfo {title} {{Some remarks on phase transitions of the second kind and
  the microscopic theory of ferroelectric materials}},}\ }\href@noop {}
  {\bibfield  {journal} {\bibinfo  {journal} {Fiz. Tverd. Tela}\ }\textbf
  {\bibinfo {volume} {2}}~(\bibinfo {number} {9}),\ \bibinfo {pages}
  {2031--2043}},\ \bibinfo {note} {[Sov. Phys. Solid State {\bf 2} (9),
  1824--1834 (1961)]}\BibitemShut {NoStop}%
\bibitem [{\citenamefont {Goldenfeld}(1992)}]{Goldenfeld:1992}%
  \BibitemOpen
  \bibfield  {author} {\bibinfo {author} {\bibnamefont {Goldenfeld},
  \bibfnamefont {N}}} (\bibinfo {year} {1992}),\ \href@noop {} {\emph {\bibinfo
  {title} {{Lectures on Phase Transitions and the Renormalization Group}}}}\
  (\bibinfo  {publisher} {Perseus Books},\ \bibinfo {address} {Reading,
  Massachusetts})\BibitemShut {NoStop}%
\bibitem [{\citenamefont {Hohenberg}\ and\ \citenamefont
  {Halperin}(1977)}]{Hohenberg:1977}%
  \BibitemOpen
  \bibfield  {author} {\bibinfo {author} {\bibnamefont {Hohenberg},
  \bibfnamefont {P~C}}, \ and\ \bibinfo {author} {\bibfnamefont {B.~I.}\
  \bibnamefont {Halperin}}} (\bibinfo {year} {1977}),\ \bibfield  {title}
  {\enquote {\bibinfo {title} {{Theory of dynamic critical phenomena}},}\
  }\href {\doibase 10.1103/RevModPhys.49.435} {\bibfield  {journal} {\bibinfo
  {journal} {Rev. Mod. Phys.}\ }\textbf {\bibinfo {volume} {49}}~(\bibinfo
  {number} {3}),\ \bibinfo {pages} {435--479}}\BibitemShut {NoStop}%
\bibitem [{\citenamefont {Landau}\ and\ \citenamefont
  {Lifshitz}(1987)}]{FluidMech:1987}%
  \BibitemOpen
  \bibfield  {author} {\bibinfo {author} {\bibnamefont {Landau}, \bibfnamefont
  {L~D}}, \ and\ \bibinfo {author} {\bibfnamefont {E.~M.}\ \bibnamefont
  {Lifshitz}}} (\bibinfo {year} {1987}),\ \href@noop {} {\emph {\bibinfo
  {title} {{Fluid Mechanics}}}},\ Vol.~\bibinfo {volume} {6}\ (\bibinfo
  {publisher} {Pergamon Press},\ \bibinfo {address} {Oxford})\BibitemShut
  {NoStop}%
\bibitem [{\citenamefont {Landau}\ \emph {et~al.}(1994)\citenamefont {Landau},
  \citenamefont {Lifshitz},\ and\ \citenamefont {Pitaevskii}}]{StatPhys:1994}%
  \BibitemOpen
  \bibfield  {author} {\bibinfo {author} {\bibnamefont {Landau}, \bibfnamefont
  {L~D}}, \bibinfo {author} {\bibfnamefont {E.~M.}\ \bibnamefont {Lifshitz}}, \
  and\ \bibinfo {author} {\bibfnamefont {L.~P.}\ \bibnamefont {Pitaevskii}}}
  (\bibinfo {year} {1994}),\ \href@noop {} {\emph {\bibinfo {title}
  {{Statistical Physics Part 1}}}},\ Vol.~\bibinfo {volume} {5}\ (\bibinfo
  {publisher} {Pergamon Press},\ \bibinfo {address} {Oxford})\ Chap.\ \bibinfo
  {chapter} {XIV}\BibitemShut {NoStop}%
\bibitem [{\citenamefont {Levanyuk}(1959)}]{Levanyuk:1959}%
  \BibitemOpen
  \bibfield  {author} {\bibinfo {author} {\bibnamefont {Levanyuk},
  \bibfnamefont {A~P}}} (\bibinfo {year} {1959}),\ \bibfield  {title} {\enquote
  {\bibinfo {title} {{Contribution to the theory of light scattering near the
  second-order phase-transition points}},}\ }\href@noop {} {\bibfield
  {journal} {\bibinfo  {journal} {Zh. Eksp. Teor. Fiz.}\ }\textbf {\bibinfo
  {volume} {36}}~(\bibinfo {number} {3}),\ \bibinfo {pages} {810--818}},\
  \bibinfo {note} {[Sov. Phys. JETP {\bf 9} (3), 571--576 (1959)]}\BibitemShut
  {NoStop}%
\bibitem [{\citenamefont {Patashinskii}\ and\ \citenamefont
  {Pokrovskii}(1964)}]{Patashinskii:1964}%
  \BibitemOpen
  \bibfield  {author} {\bibinfo {author} {\bibnamefont {Patashinskii},
  \bibfnamefont {A~Z}}, \ and\ \bibinfo {author} {\bibfnamefont {V.~L.}\
  \bibnamefont {Pokrovskii}}} (\bibinfo {year} {1964}),\ \bibfield  {title}
  {\enquote {\bibinfo {title} {{Phase transitions of second kind in a Bose
  fluid}},}\ }\href@noop {} {\bibfield  {journal} {\bibinfo  {journal} {Zh.
  Eksp. Teor. Fiz.}\ }\textbf {\bibinfo {volume} {46}}~(\bibinfo {number}
  {3}),\ \bibinfo {pages} {994--1017}},\ \bibinfo {note} {[Sov. Phys. JETP {\bf
  19} (3), 677--691 (1964)]}\BibitemShut {NoStop}%
\bibitem [{\citenamefont {Pfeuty}\ and\ \citenamefont
  {Toulouse}(1977)}]{Pfeuty:1977}%
  \BibitemOpen
  \bibfield  {author} {\bibinfo {author} {\bibnamefont {Pfeuty}, \bibfnamefont
  {P}}, \ and\ \bibinfo {author} {\bibfnamefont {G.}~\bibnamefont {Toulouse}}}
  (\bibinfo {year} {1977}),\ \href@noop {} {\emph {\bibinfo {title}
  {{Introduction to the Renormalization Group and to Critical Phenomena}}}}\
  (\bibinfo  {publisher} {John Wiley \& Sons},\ \bibinfo {address}
  {London})\BibitemShut {NoStop}%
\bibitem [{\citenamefont {Privman}\ \emph {et~al.}(1991)\citenamefont
  {Privman}, \citenamefont {Hohenberg},\ and\ \citenamefont
  {Aharony}}]{Privman:1991}%
  \BibitemOpen
  \bibfield  {author} {\bibinfo {author} {\bibnamefont {Privman}, \bibfnamefont
  {V}}, \bibinfo {author} {\bibfnamefont {P.~C.}\ \bibnamefont {Hohenberg}}, \
  and\ \bibinfo {author} {\bibfnamefont {A.}~\bibnamefont {Aharony}}} (\bibinfo
  {year} {1991}),\ \bibfield  {title} {\enquote {\bibinfo {title} {{Universal
  Critical-Point Amplitude Relations}},}\ }in\ \href@noop {} {\emph {\bibinfo
  {booktitle} {{Phase Transitions and Critical Phenomena}}}},\ Vol.~\bibinfo
  {volume} {14},\ \bibinfo {editor} {edited by\ \bibinfo {editor}
  {\bibfnamefont {C.}~\bibnamefont {Domb}}\ and\ \bibinfo {editor}
  {\bibfnamefont {J.~L.}\ \bibnamefont {Lebowitz}}}\ (\bibinfo  {publisher}
  {Academic Press},\ \bibinfo {address} {London})\ pp.\ \bibinfo {pages}
  {1--134}\BibitemShut {NoStop}%
\end{thebibliography}%

\end{document}